%% file: garse.tex
\documentclass{elsart}

\usepackage{epsfig}

\usepackage{amssymb}

\usepackage{amsmath}
\journal{Physics Reports}

\newcommand{\al}{\alpha}
\newcommand{\bt}{\beta}
\newcommand{\gm}{\gamma}
\newcommand{\Gm}{\Gamma}
\newcommand{\dl}{\delta}
\newcommand{\Dl}{\Delta}
\newcommand{\vf}{\varphi}
\newcommand{\ve}{\varepsilon}
\newcommand{\tht}{\theta}

\newcommand{\lb}{\lambda}
\newcommand{\sg}{\sigma}
\newcommand{\vth}{\vartheta}
\newcommand{\om}{\omega}

\newcommand{\cL}{\mathcal{L}}
\newcommand{\cV}{\mathcal{V}}
\newcommand{\cT}{\mathcal{T}}

\newcommand{\la}{\langle}
\newcommand{\ra}{\rangle}
\newcommand{\wt}{\widetilde}
\newcommand{\wh}{\widehat}
\newcommand{\us}{\underset}
\newcommand{\pa}{\partial}
\newcommand{\ol}{\overline}
\newcommand{\ul}{\underline}

\newcommand{\IIm}{\operatorname{Im}}
\newcommand{\arctg}{\operatorname{arctg}}

\begin{document}

\begin{frontmatter}



\title{Light Front Formalism for Composite Systems and Some of Its Applications in Particle and Relativistic Nuclear Physics}


\author[rmi,geneva]{V. R. Garsevanishvili\corauthref{cor}},
\corauth[cor]{Corresponding author.}
\ead{garse@hepi.edu.ge}
\author[energy]{A. A. Khelashvili},
\author[energy]{Z. R. Menteshashvili},
\author[energy]{M. S. Nioradze}

\address[rmi]{A. Razmadze Mathematical Institute, 1,~M. Aleksidze St., Tbilisi 0193, Georgia}
\address[geneva]{TH Division, CERN, CH-1211 Geneva, 23, Switzerland}
\address[energy]{Institute of High Energy Physics and Informatization, Tbilisi State University, 9,~University St., Tbilisi 0186, Georgia}

\begin{abstract}
Light front formalism for composite systems is presented. Derivation of equations for bound state and scattering problems are given. Methods of constructing of elastic form factors and scattering amplitudes of composite particles are reviewed. Elastic form factors in the impulse approximation are calculated. Scattering amplitudes for relativistic bound states are constructed. Some model cases for transition amplitudes are considered. Deep inelastic form factors (structure functions) are expressed through light front wave functions. It is shown that taking into account of transverse motion of partons leads to the violation of Bjorken scaling and structure functions become square of transverse momentum dependent. Possible explanation of the EMC-effect is given. Problem of light front relativization of wave functions of lightest nuclei is considered. Scaling properties of deuteron, ${}^3He$ and ${}^4He$ light front wave functions are checked in a rather wide energy range.
\end{abstract}

\end{frontmatter}

\newpage 

\input contents.tex

\newpage
\input garse0.tex

\input garse1.tex

\input garse2.tex

\input garse3.tex

\input garse4.tex

\section*{Acknowledgements}

Some results presented here have been obtained in collaboration with Rudolf Faustov, Alexander Kvinikhidze, Victor Matveev, Albert Tavkhelidze. The experimental situation has been discussed in groups of Victor Glagolev, Pierre Juillot, Teodor Siemiarczuk. One of the authors (V.R.G.) expresses his deep gratitude to Alvaro De Rujula for the warm hospitality at the CERN TH Division and helpful discussions. Friendly atmosphere of discussions at the CERN TH Division with Vladimir Kadyshevsky, Vadim Kuzmin, Matey Mateev, Igor Tkachev, Urs Wiedemann and meetings with Stanley Brodsky, Vladimir Karmanov, Jean-Francois Mathiot, Gerald Miller, Hans-Christian Pauli were highly stimulating. The authors are indebted to Gerald Brown for his kind attention to this work.

\input ref.tex
\end{document}

%% file: contents.tex
\section*{Contents}

\begin{enumerate}
\item[\bf I.] {\bf Introduction }
\dotfill{ 3}

\begin{enumerate}
\item[1.] Notations 
\dotfill{ 6}

\item[2.] Generators of the Poincare Group 
\dotfill{ 6}

\item[3.] Light Front Quantization 
\dotfill{ 9}

\end{enumerate}

\medskip

\item[\bf II.] {\bf Light Front Formalism for Bound States }
\dotfill{ 11}

\begin{enumerate}
\item[1.] Equation for the Two-Body Bound State Wave Function 
\dotfill{ 11}

\item[2.] Equation for the Scattering Amplitude and Relation to \\the Equation in the Infinite Momentum Frame 
\dotfill{ 14}

\item[3.] The Case of Two Spin-$1/2$ Particles 
\dotfill{ 15}

\item[4.] Equation for the Many-Body Bound State Wave Function 
\dotfill{ 20}

\end{enumerate}

\medskip
\item[\bf III.] {\bf Relativistic Elastic Form Factors and Scattering Amplitudes \\for Composite systems} 
\dotfill{ 23}

\begin{enumerate}
\item[1.] Formulation of the Method 
\dotfill{ 23}

\item[2.] Elastic Form Factor in the Impulse Approximation 
\dotfill{ 27}

\item[3.] Relativistic Form Factor for the Many-Body System 
\dotfill{ 28}

\item[4.] Asymptotic Behaviour of the Pion Form Factor at Large \\Momentum Transfer 
\dotfill{ 31}

\item[5.] Scattering of Relativistic Composite Systems 
\dotfill{ 34}

\item[6.] Particle Exchange in the Intermediate State 
\dotfill{ 36}

\item[7.] Constituent Interchange Mechanism 
\dotfill{ 37}

\end{enumerate}

\medskip
\item[\bf IV.] {\bf Deep Inelastic Form Factors of Composite Systems and \\Multiquark States in Nuclei} 
\dotfill{ 39}

\begin{enumerate}
\item[1.] Construction of Tensor $W_{\mu\nu} $ 
\dotfill{ 39}

\item[2.] Lowest Order in the Electromagnetic Interaction 
\dotfill{ 43}

\item[3.] Model Parametrizations of Wave Functions 
\dotfill{ 48}

\item[4.] Quark Degrees of Freedom in Nuclei and the EMC-Effect 
\dotfill{ 50}

\end{enumerate}

\medskip
\item[\bf V.] {\bf Processes Involving High Energy Nuclei and Problem \\of Relativisation of Nuclear Wave Functions }
\dotfill{ 58}

\begin{enumerate}
\item[1.] Scale-Invariant Parametrization of the Deuteron Relativistic  \\Wave Function 
\dotfill{ 58}

\item[2.] Break-up of the Relativistic Deuteron and the Verification \\of Scaling Properties of Its Wave Function 
\dotfill{ 60}

\item[3.] The Break-up of More Complicated Nuclei 
\dotfill{ 69}

\item[4.] Scaling Properties of the Relativistic Wave Functions of \\${}^4He$ and ${}^3He$ Nuclei 
\dotfill{ 76}

\end{enumerate}

\item[] References 
\dotfill{ 82}
\end{enumerate}

%% file: garse0.tex
\section*{\large I. Introduction}

\def\theequation{1.\arabic{equation}}

Traditional way of describing bound state in quantum field theory is the Bethe--Salpeter formalism \cite{1}. However, the dependence of the Bethe--Salpeter wave function on the relative time of two particles makes difficult the probabilistic interpretation of the wave function and gives rise to a number of mathematical problems. On the other hand, the relativistic quasipotential approach of Logunov--Tavkhelidze \cite{2} based on the two-time formalism for Green functions enables one to avoid the difficulty connected with the relative time and is much simpler in applications. Similar three dimensional approach for bound state and scattering problems in Hamiltonian formalism of quantum field theory has been developed in Ref. \cite{3}. 

We give here the derivation of the three-dimensional equations in the light front formalism. This form of the three-dimensional approach turned out to be effective in the treatment of high energy interactions of composite systems. In particular, it gives a number of scale invariant predictions for observable quantities which can be checked experimentally. 
In this approach the relativistic composite system with the total 4-momentum $P$ is described by means of the relativistic wave function $\Phi_P([x_i,\vec{p}_{i,\bot}])$ \cite{4}--\cite{10}, where the ``longitudinal motion'' of constituents is parametrized by means of the scale-invariant variables
$$
    x_i=\frac{p_{i,0}+p_{i,z}}{P_0+P_z}\,,
$$
where $p_{i,\mu}$ $(\mu=0,1,2,3$ is the Lorentz index) and $P_\mu$ are an individual 4-momentum of the $i$-th constituent and the total momentum of the system, respectively. Variables $x_i$ are ratios of the light front variables. In terms of these variables the wave function of the composite system reflects, in particular, the dependence of the internal motion of constituents on the total momentum of the system.

In what follows upper-case letters will denote the characteristics (momenta, masses, etc.) of the composite systems and lower-case letters the characteristics of constituents. Square brackets in the argument of the wave function $\Phi_P$ denote the set of the variables $x_i$ and $\vec{p}_{i,\bot}$ which satisfy the conditions:
$$
    \sum_{i=1}^A x_i=1; \quad 0<x_i<1; \quad \sum_{i=1}^A \vec{p}_{i,\bot}=\vec{P}_\bot.
$$

Recent research has given an insight into the complicated relativistic structure of hadrons. Besides, the study of high-energy nucleus interactions has shown that idea of nuclei as systems of quasi-independent nonrelativistic nucleons is incomplete. Hence it has became necessary to take into account the relativistic character of intrinsic motion of nucleons and to consider the quark degrees of freedom in nuclei. 

The above-stated reasons have motivated the subject of research considered in this review.

According to modern concepts nucleons and nuclei are bound states of more elementary constituents. Interactions between them or their interactions with particles which at the present stage can be considered as elementary are principal sources of information on bound states. 

Investigations with beams of high-energy nuclei carried out in Dubna, Berkeley and other laboratories since the beginning of the seventies have stimulated interest in the study of nuclei as relativistic composite systems and the processes of interaction between them (see, e.g., review papers \cite{7}, \cite{9}, \cite{11}--\cite{35} and the references therein). Presently, there are beams of high-energy nuclei with various atomic number in a wide energy range from several GeV/nucleon up to several hundred GeV/nucleon at CERN  \cite{36}--\cite{38}. Note that this range corresponds to relativistic physics and the non-relativistic theory of nuclear reactions (see, e.g., the review papers \cite{39}, \cite{40} and references therein) based on the Schroedinger formalism of quantum mechanics does not work in this case. Relativistic effects have to be taken into account in this energy region and high-energy phenomena can be adequately described only using the relativistically invariant formalism. It seems more convenient to plot also the data in terms of variables which have the relativistic origin.

In the experiments with beams of high-energy nuclei the situations often occur when small internucleon distances must be taken into account. This corresponds to large relative momenta of constituents of the nucleus. The problem arises to describe adequately nuclei at arbitrary momenta of their constituents. That is why it is more convenient to describe the processes involving relativistic nuclei in terms of wave functions, in which the relativistic character of nucleon motion is taken into account, instead of the ordinary quantum-mechanical nuclear wave functions, which correspond to the motion of nucleons with small internal momenta. 

Note that the idea to use light front formalism in the relativistic nuclear physics has been put forward in Ref. \cite{7}. At present there are a number of interesting reviews on the light front dynamics and its applications (see, e.g., Refs. \cite{41}--\cite{52}). Some interesting applications of light front variables in the high energy hadron-hadron and nucleus-nucleus phenomenology can be found in Refs. \cite{53}, \cite{54} and references therein.

The review is organized as follows:

Chapter II is devoted to the formulation of the three-dimensional formalism for composite systems in terms of light front variables. Equations are derived for bound states and scattering problems. Cases of spin-0 and spin-$1/2$ constituents are considered. It is shown how equations of this approach are related with or differ from the Weinberg equation obtained in the framework of the old-fashioned perturbation theory in the infinite momentum frame.

Chapter III deals with the method of constructing relativistic elastic form-factors and scattering amplitudes of composite systems in the light front formalism. A general expression is obtained for the matrix element of the current of composite system in terms of light front wave functions and the generalized vertex operator $\wt\Gm_\mu$. The explicit form of the vertex operator is found in the impulse approximation and the electromagnetic form-factor is calculated for a system, consisting of two or arbitrary number of constituents.

Asymptotic behaviour of the pion form-factor is investigated at large momentum transfer. 

Problems of the interaction of relativistic composite systems are also discussed in this chapter. The scattering amplitude is expressed in general form using relativistic wave functions and the transition operator. The constituent interchange mechanism is considered.

Chapter IV is devoted to the study of deep inelastic from-factors of composite systems. Like the case of elastic form-factors, the general expression for the deep inelastic tensor $W_{\mu\nu}$ is calculated in terms of the relativistic wave functions and the generalized two-photon vertex $\wt\Gm_{\mu\nu}$. The explicit form of this operator is calculated in the lowest order in the electromagnetic interaction and expressions are given for the structure functions $W_1$ and $\nu W_2$. It is shown that if the transverse motion of quarks is taken into account, the Bjorken scaling is violated and the structure functions become the square of the momentum transfer dependent.  The problem of explaining of the  so-called EMC-effect is stated. It is shown that one possible way to explain this effect is the scattering of electrons on the colourless multiquark configurations in nuclei.

In Chapter V problems of the relativization of nuclear wave functions are posed and solved. The case of the simplest nucleus -- deuteron is considered in detail. A way to the relativization of the well-known nonrelativistic wave functions of the deuteron is indicated. The scaling properties of the light front wave function of the deuteron are checked by studying spectator-nucleon distributions in the deuteron break-up at various incident energies. It is shown that the scaling properties of the relativistic wave function hold with good accuracy in a wide energy range. Calculations are performed using the relativistic analogue of the well-known Hulten wave function of the deuteron.

Problems of relativization of wave functions of more complex nuclei are also considered. The differential cross section of the break-up of the relativistic nucleus on the hydrogen target is calculated. Theoretical calculations are compared with experimental data on the ${}^4He\to {}^3He$, ${}^3He \to d$ fragmentation. The analysis of data shows that the scaling properties of the relativistic wave functions of the ${}^4He$ and ${}^3He$ nuclei hold with a good accuracy  in the energy range considered. 
\section{Notations}

Let us introduce the light front coordinates. Instead of usual Lorentz coordinates $x_\mu(t,x,y,z)\equiv (x_0,x_1,x_2,x_3)$ we consider new light front coordinates $x_\mu= (x_+,x_1,x_2,x_-)$, where $x_\pm=\frac{1}{\sqrt{2}}\,(x_0\pm x_3)$. If $\wh x_\mu$ are the Lorentz coordinates, then transformation to light front coordinates is performed by the matrix $C_{\mu\nu}$:
$$
    C_{\mu\nu} =\begin{pmatrix}
                    \frac{1}{\sqrt{2}} & 0 & 0 & \frac{1}{\sqrt{2}} \\
                    0 & 1 & 0 & 0 \\
                    0 & 0 & 1 & 0 \\
                    \frac{1}{\sqrt{2}} & 0 & 0 & -\frac{1}{\sqrt{2}} 
                \end{pmatrix}, \qquad x_\mu=C_{\mu\nu}\wh x_\mu.
$$

Metric tensors are related to each other by the relation:
\begin{gather*}
    \wh g_{\mu\nu}=C_{\al\mu} \wh g_{\al\bt} C_{\bt\nu}^{-1}, \quad
    g_{\al\bt}=\begin{pmatrix} 
                    1, & 0, & 0, & 0 \\
                    0, & -1, & 0, & 0 \\
                    0, & 0, & -1, & 0 \\
                    0, & 0, & 0, & -1 
                \end{pmatrix}, \qquad 
    g_{\mu\nu}=\begin{pmatrix} 
                    0, & 0, & 0, & 1 \\
                    0, & -1, & 0, & 0 \\
                    0, & 0, & -1, & 0 \\
                    1, & 0, & 0, &  0
                \end{pmatrix}.
\end{gather*}

It is obvious that
\begin{gather*}
x^2=2x_+x_--\vec{x}{\,}_\bot^2 \\
xy=x_+y_-+x_-y_+-\vec{x}_\bot \vec{y}_\bot.
\end{gather*}

\section{Generators of the Poincare Group}

Let's consider generators of the Poincare group. In usual notations $\wh P_\mu$, $\wh M_{\mu\nu}$ obey the algebra:
\allowdisplaybreaks
\begin{align*}
    [\wh P_\mu,\wh P_\nu] & = 0,\\
    [\wh M_{\mu\nu},\wh P_\rho] & =i(\wh g_{\nu\rho} \wh P_\mu -\wh g_{\mu\rho} \wh P_\nu), \\
    [\wh M_{\mu\nu},\wh M_{\rho\lb}] & =i(\wh g_{\mu\lb} \wh M_{\nu\rho} +
        \wh g_{\nu\rho} \wh M_{\mu\lb}-\wh g_{\mu\rho} \wh M_{\nu\lb}-
        \wh g_{\nu\lb} \wh M_{\mu\rho}).
\end{align*}

Generators of three dimensional rotations and pure Lorentz transformations (boosts) are related to $M_{\mu\nu}$ as follows:
\begin{align*}
    J_i & =\frac{1}{2}\, \ve_{ijk}\,\wh M_{jk} =
        (\wh M_{23}, \wh M_{31}, \wh M_{12}), \\
    K_i& =\wh M_{0i} \\
\intertext{or}
    \wh M_{\mu\nu}& =\begin{pmatrix}
                        0, & -K_1, & -K_2, & -K_3 \\
                        K_1, & 0, & J_3, & -J_2 \\
                        K_2, & -J_3, & 0, & J_1 \\
                        K_3, & J_2, & -J_1, & 0
                    \end{pmatrix}.
\end{align*}

Performing the transformation by means of $C$-matrix, we obtain for the light front  generators:
\begin{gather*}
    P_\mu=(P_+,P_1,P_2,P_-), \quad P_\pm=\frac{1}{\sqrt{2}}\,(P_0\pm P_3), \\
    M_{\mu\nu} =\begin{pmatrix}
                    0, & -S_1, & -S_2, & K_3 \\
                    S_1, & 0, & J_3, & B_1 \\
                    S_2, & -J_3, & 0, & B_2 \\
                    -K_3, & -B_1, & -B_2, & 0
                \end{pmatrix},
\end{gather*}
where 
\begin{align*}
    B_i & =\frac{1}{\sqrt{2}}\,(\wh M_{i0}+\wh M_{3i}), \\
    S_i & =\frac{1}{\sqrt{2}}\,(\wh M_{i0}-\wh M_{3i}),
\end{align*}
i.e.
\begin{align*}
    B_1 & =\frac{1}{\sqrt{2}}\,(K_1+J_2), \quad 
        B_2 =\frac{1}{\sqrt{2}}\,(K_2-J_1), \\
    S_1 & =\frac{1}{\sqrt{2}}\,(K_1-J_2), \quad 
        S_2 =\frac{1}{\sqrt{2}}\,(K_2+J_1).
\end{align*}

In the usual covariant theory initial state of the system is given on the hyperplane $t=const$. In quantum mechanics and quantum field theory canonical commutation relations between canonically conjugated quantities are given on this hyperplane. The plane $t=const$ remains invariant under three-transformations generated by momentum operators $\wt P_i$ and under three dimensional rotations generated by angular momentum operators $\wh M_{ij}=-\wh M_{ji}$, e.i. the stability group of the hyperplane $t=const$ is $O(3) \times T(3)$ (or $SU(2) \times T(3)$). 

Corresponding operators $J_k$ and $\wh P_i$ are called kinematic operators (according to Dirac). If the system of particles is considered, eigenvalues of kinematical operators are equal to the sum of eigenvalues of kinematical operators of constituents. 

Evolution of the system proceeds in the direction orthogonal to the initial hyperplane, or in the direction of the time. The operator, which generates this evolution is Hamiltonian $H=\wh P_0$, $H$ is dynamical operator. For the system of particles it is not equal to the sum of Hamiltonians of individual particles, but depends on the interaction between them. Besides $\wh P_0$ Lorentz boost operators $K_i$ are also dynamical operators. So we have 6 kinematical operators $(\wh P_i, \wh M_{ij})$ and 4 dynamical operators $(\wh P_0,K_i)$. 

Another possibility for postulating commutation relations was pointed out by Dirac in 1949 \cite{55}. In particular, he proposed to use light front hyperplane $x_+=const$ for postulating commutation relations. The hyperplane $x_+=const$ is not changed under translations in the orthogonal directions $\vec{x}_\bot$, which are generated by $\vec{P}_\bot$ and in $x_-$ direction, which is generated by $P_+$. Besides it is invariant under two-dimensional rotations generated by $M_{12}=J_3$ and boosts generated by $B_i$ $(i=1,2)$. Corresponding stability group  is $E(2)\times T(3)$. Evolution of the system proceeds in the $x_+$-direction and the role of shift generators in this direction is played by $P^-=P_+$. Thus we have 6 kinematic operators $(P^+, \vec{P}_\bot, J_3,B_i)$ and 4 dynamical operators $(P^-,S_i,K_3)$. 

From commutation relations between generators:
\begin{gather*}
    [P^-,P^i]=[P^-,P^+]=[J_3,P^-]=0, \\
    [B_i,P^-]=-iP^-,
\end{gather*}
it follows that an arbitrary 4-vector $A^\mu\,(A^+,\vec{A}_\bot,A^-)$ is transformed according to the rule:
$$
    e^{i\vec{v}_\bot\vec{B}_\bot}A^\mu e^{-i\vec{v}_\bot\vec{B}_\bot}=
    \big( A^+,\vec{A}_\bot+\vec{v}_\bot A^+, A^-+\vec{v}_\bot\vec{A}_\bot+
        \tfrac{1}{2}\,\vec{v}{\,}_\bot^2 A^+\big).
$$
It is obvious that $B_{1,2}$ generate Galilei boosts in the directions $x$ and $y$, $P^+$ plays the role of mass, $\vec{P}_\bot$ plays the role of momentum, $J_3$ plays the role of angular momentum and $P^-$ plays the role of the Hamiltonian.

Consider now commutation relations between dynamical operators:
\begin{align*}
    [S_i,S_j] & =0, \quad [S_i,P^-]=0, \\
    [J_3,S_i] & =i\ve_{ijk} S_k, \\
    [S_i,P^+] & =-iP^i, \\
    [S_i,P_j] & =-i\dl_{ij} P^+.
\end{align*}

From these commutation relations it follows that transformation rule for arbitrary 4-vector $A^\mu$ is the following:
$$
    e^{i\vec{u}_\bot\vec{S}_\bot}A^\mu e^{-i\vec{u}_\bot\vec{S}_\bot}=
    \big( A^++\vec{u}_\bot\vec{A}_\bot + \tfrac{1}{2}\,\vec{u}{\,}_\bot^2 A^-, 
        \vec{A}_\bot+\vec{u}_\bot A^-,A^-\big).
$$

Comparison shows that the system of operators $(P_-,P^i,J_3,S_i)$ allows the same interpretation as $(P^+,P^i,J_3,B_i)$, but in relation to the hyperplane $x^-=const$, i.e. if the initial conditions are given on the plane $x^-=const$.

Let us consider now commutation relations:
\begin{align*}
    [S_i,B_j] & =-i\ve_{ij3}J_3+i\dl_{ij}K_3, \\
    [K_3,P^i] & =[K_3,J_3]=0, \\
    [K_3,P^\pm] & =\mp iP^\pm.
\end{align*}

From these relations it follows that the transformation rule for arbitrary 4-vector $A^\mu$ looks as follows:
$$
    e^{i\om K_3}A^\mu e^{-i\om K_3} =\big(e^\om A^+,\vec{A}_\bot, e^{-\om} A^-\big).
$$
Besides, from commutation relations:
\begin{align*}
    & [K_3,B_i] = -iB_i, \\
    & [K_3,S_i] = iS_i
\end{align*}
it follows that
\begin{align*}
    e^{i\om K_3} B_i e^{-i\om K_3} & =e^w B_i, \\
    e^{i\om K_3} S_i e^{-i\om K_3} & =e^{-w} S_i. 
\end{align*}
Thus, in the light front system the boost operator along the third axis acts as a scale transformation. This means that the light front is well suited for the treatment of high energy problems.

\section{Light Front Quantization}

For simplicity let's consider first the free scalar field with mass $m$. In covariant formulation the commutator of fields looks as follows:
$$
    [\vf(x),\vf(y)] =\Dl(x-y),
$$
where \cite{56}:
\begin{equation}\label{1}
    \Dl (x)=\frac{1}{2\pi}\,\ve(x_0)\dl(x^2)-\frac{m}{4\pi\sqrt{x^2}}\,
        \ve(x_0) \tht(x^2) J_1(m\sqrt{x^2}).
\end{equation}
Using formulae:
$$
    \ve(x_0)\tht(x^2)|_{x_+=0}=0, \quad 
            \ve(x_0)\dl(x^2)|_{x_+=0}=\frac{\pi}{2}\,\ve(x_-)\dl(x^2),
$$
one obtains from \eqref{1}
\begin{equation}\label{2}
    [\vf(x),\vf(y)]_{x_+=y_+}=\frac{i}{4}\,\ve(x_--y_-) \dl^{(2)}(x_\bot-y_\bot).
\end{equation}

As it can be seen from here the commutator does not vanish at $x_-\to \pm\infty$. Thus, the problem of boundary conditions at infinity arises. It was shown \cite{57} that the limit $x_-\to \pm \infty$ corresponds to the contribution of zero modes $P_+=0$. It was shown in Refs. \cite{58}, \cite{59} that the commutator on the light front depends on the interaction. Various methods of excluding zero modes based on the periodic boundary conditions have been proposed. Some authors assume that it is necessary to work with two Hamiltonians $P_-$ and $P_+$, or, if one uses only one Hamiltonian, say $P_-$, it is necessary to require fulfilment of periodic boundary conditions with respect to the variable $x_-$. It is interesting  to note that if zero modes $P_+=0$ are excluded, then the vacuum, as a state with $P_+=0$, is at the same time the lowest state of the Hamiltonian $P_-$.

The scattering matrix in the light front quantization is defined as a generalization  of Dyson formula:
\begin{equation}\label{2a}
    S=T_+ \exp\Big[ -i\int d^4 x H_I(x)\Big],
\end{equation}
where $H_I(x)$ is the the interaction Hamiltonian in the interaction representation, $T_+$ is the ordering operator with respect to the variable $x_+$. If the interaction Lagrangian does not contain derivatives, it can be shown that $H_I(x)=-\cL_I(x)$ as in the usual case. For the equivalence with usual case it is necessary to show equivalence of $T_+$-product to the usual $T$-product.

Consider
\begin{align*}
    T_+(\vf(x)\vf(y)) & = \tht(x_+-y_+)[\vf(x),\vf(y)]+\vf(y)\vf(x), \\
    T(\vf(x)\vf(y)) & = \tht(x_0-y_0)[\vf(x),\vf(y)]+\vf(y)\vf(x).
\end{align*}

These two expressions may differ only in the cases: when $x_0>y_0$, we have $x_+<y_+$ or when $x_0<y_0$, we have $x_+>y_+$. It is easy to see that in both cases $(x-y)^2<0$, i.e. interval is space-like and commutator is identically zero. 

Thus, both orderings coincide and the light front $S$-matrix coincides with the covariant $S$-matrix. In the spin case additional  noncovariant terms arise, but they cancel with corresponding noncovariant terms in propagators \cite{60}--\cite{64}.

%% file: garse1.tex
\section*{\large II. Light Front Formalism for Bound States}

\def\theequation{2.\arabic{equation}}
\setcounter{equation}{0}
\setcounter{section}{0}

\section{Equation for the Two-Body Bound State Wave Function}

In this section we develop the three-dimensional formalism for composite systems in light front variables. These variables were introduced by Dirac \cite{55} with the aim to construct the quantum theory with canonical communication relations on the light front hyperplane (instead of traditionally used $t=0$ hyperplane). We note that the real progress in this direction has been achieved much later \cite{60}--\cite{65}.

Let us consider the Bethe--Salpeter amplitude (wave function)
\begin{equation}\label{1.1}
    \chi_{P,\al}=\left\la 0\mid T(\vf_1(x_1)\vf_2(x_2)) \mid P,\al\right\ra=
        e^{-iPX}\chi_{P,\al}(x).
\end{equation}
Here $|\, P,\al\ra$ is the state vector of two particles with total 4-momentum $P$ and quantum numbers $\al$, $X=(x_1+x_2)/2$ is the centre-of-mass coordinate, $P=p_1+p_2$. Relative coordinate and momentum are defined as
\begin{equation}\label{1.2}
    x=x_1-x_2, \qquad p=\frac{p_1-p_2}{2}\,,
\end{equation}
and the light front variables can be introduced
\begin{equation}\label{1.3}
    x_\pm=\frac{x_0\pm x_3}{2}\,, \quad p_\pm=p_0\pm p_3, \quad P_\pm=P_0\pm P_3.
\end{equation}

Let us introduce the Fourier transform $\chi_{P,\al}(p)=\chi_{P,\al}(p_-,p_+,\vec{p}_\bot)$ of the Bethe--Salpeter amplitude
\begin{align}
    \chi_{P,\al}(x) &= \chi_{P,\al}(x_+,x_-,\vec{x}_\bot) =
        \int d^4 p e^{-ipx} \chi_{P,\al}(p) \notag  \\
    & =\frac{1}{2} \int dp_+\,dp_-\,d\vec{p}_\bot\,
        e^{-i(p_+x_-+p_-x_+-\vec{p}_\bot\vec{x}_\bot)}\chi_{P,\al}(p) \label{1.4}
\end{align}
and define the function
\begin{equation}\label{1.5}
    \Psi_{P,\al}(p_+,\vec{p}_\bot) =\int_{-\infty}^\infty dp_-\chi_{P,\al}
        (p_-,p_+,\vec{p}_\bot).
\end{equation}
It can be shown that the  function $\Psi_{P,\al}(p_+,\vec{p}_\bot)$ depends on the values of the Bethe--Salpeter amplitude on the light front  hyperplane
$$
    x_0+x_3=0.
$$
In fact, using definition (\ref{1.5}) and Fourier transformation (\ref{1.4}), we get:
\begin{equation}\label{1.6}
    \Psi_{P,\al}(p_+,\vec{p}_\bot)\!=\!\frac{2}{(2\pi)^3} \int dx_+\,dx_-\,d\vec{x}_\bot\,
        \dl(x_+) e^{-i(p_+x_--\vec{p}_\bot\vec{x}_\bot)} \chi_{P,\al}(x_+,x_-,\vec{x}_\bot).
\end{equation}

Let us consider now the two-particle Green function 
\begin{align}
    G(x_1,x_2;x_1',x_2') & =G(X-X';x,x')=
        \left\la 0\mid T(\vf_1(x_1)\vf_2(x_2)\vf_1^+(x_1')\vf_2^+(x_2')) 
            \mid 0\right\ra \notag \\
    & =\frac{1}{(2\pi)^3} \int dP\,dp\,dp'\,e^{-iP(X-X')-i(px-p'x')}
        G(P;p,p'). \label{1.7}
\end{align}
Here the total and relative 4-momenta and 4-coordinates in the initial and final states are introduced as follows:
\begin{align}
    & P=p_1+p_2, \quad p=\frac{p_1-p_2}{2}\,, \quad 
        X=\frac{x_1+x_2}{2}\,, \quad x=x_1-x_2, \label{1.8} \\
    & P'=p_1'+p_2', \quad p'=\frac{p_1'-p_2'}{2}\,, \quad 
        X'=\frac{x_1'+x_2'}{2}\,, \quad x'=x_1'-x_2'. \label{1.9}
\end{align}

Fourier transform of the ``two-time'' Green function can be defined as
\begin{equation}\label{1.10}
    \wt G(P;p_+,\vec{p}_\bot;p_+'\vec{p}{\,}_\bot')=
        \int_{-\infty}^\infty dp_-\,dp_-'\, G(P;p,p').
\end{equation}

For free particles we have
\begin{equation}\label{1.11}
    G^{(0)}(P;p,p')= \frac{-\dl^{(4)}(p-p')}
        {\big[\big(\frac{P}{2}+p\big)^2 -m_1^2+i\ve\big]\,
        \big[\big(\frac{P}{2}-p\big)^2 -m_2^2+i\ve\big]}\,.
\end{equation}
Performing the integration according to the definition (\ref{1.10}), we obtain \cite{4}:
\begin{align}
    \wt G(P;p_+,\vec{p}_\bot;p_+'\vec{p}{\,}_\bot') 
    & =\frac{4\pi i\dl(p_+-p_+')\dl^{(2)}(\vec{p}_\bot-\vec{p}{\,}_\bot')\tht(x)\tht(1-x)}
        {P_+x(1-x) \big[P^2+\vec{P}{\,}_\bot^2- 
            \frac{(\vec{P}/2+\vec{p})_\bot^2+m_1^2}{x} -
            \frac{(\vec{P}/2-\vec{p})_\bot^2+m_2^2}{1-x}\big]} \notag \\
    & =\wt G^{(0)}(P;p_+,\vec{p}_\bot)\dl(p_+-p_+')\dl^{(2)}(\vec{p}_\bot-\vec{p}{\,}_\bot').
        \label{1.12}
\end{align}

In this expression the variable $x$ is introduced in the following way
\begin{equation}\label{1.13}
    x=\frac{1}{2}+\frac{p_+}{P_+}\,.
\end{equation}
It is obvious that when the variable $x$ varies in the limits
\begin{equation}\label{1.14}
    0<x<1,
\end{equation}
the variable $p_+$ varies in the interval $[-P_+/2,\,P_+/2]$.

Inverse operator can be defined by the relation 
\begin{gather}
    \int_{-P_+/2}^{P_+/2} dp_+'' \int d\vec{p}{\,}_\bot'' 
        \wt G^{-1}(P;p_+,\vec{p}_\bot; p_+'',\vec{p}{\,}_\bot'')
        \wt G(P;p_+'',\vec{p}{\,}_\bot''; p_+',\vec{p}{\,}_\bot') \notag \\
    =\dl(p_+-p_+')\dl^{(2)}(\vec{p}_\bot-\vec{p}{\,}_\bot'). \label{1.15}
\end{gather}

If we introduce the interaction kernel $V$ (quasipotential):
\begin{align}
    \wt G^{-1}(P;p_+,\vec{p}_\bot; p_+',\vec{p}{\,}_\bot')& =
        {\wt G}^{(0)^{-1}}(P;p_+,\vec{p}{\,}_\bot)
        \dl(p_+-p_+')\dl^{(2)}(\vec{p}_\bot-\vec{p}{\,}_\bot') \notag \\
    & \quad -\frac{1}{4\pi i} \,V(P;p_+,\vec{p}_\bot;p_+',\vec{p}{\,}_\bot')  \label{1.16}
\end{align}
after simple transformations the equation for the wave function
\begin{equation}\label{1.17}
    \Phi_{P,\al}(x,\vec{p}_\bot)=P_+x (1-x)\Psi_{P,\al}(p_+,\vec{p}_\bot)
\end{equation}
takes the form 
\begin{multline}\label{1.18}
    \bigg[ P^2-\frac{(\vec{p}_\bot+(1/2-x)\vec{P}_\bot)^2+m_1^2}{x}-
        \frac{(\vec{p}_\bot+(1/2-x)\vec{P}_\bot)^2+m_2^2}{1-x}\bigg]
        \Phi_{P,\al}(x,\vec{p}_\bot) \\
    =\int_0^1 \frac{dx'}{x'(1-x')}\, \int d\vec{p}{\,}_\bot' 
        V(P;p_+,\vec{p}_\bot;p_+',\vec{p}{\,}_\bot')\Phi_{P,\al}(x',\vec{p}{\,}_\bot').
\end{multline}

The equation obtained gives the wave function of a bound state in an arbitrary Lorentz reference frame. Comparing it with the equation in the frame where $\vec{P}_\bot=0$ we get the transformation property for the wave function from the arbitrary frame to the frame in which the total transverse momentum of two-particle bound state is equal to zero:
\begin{equation}\label{1.19}
    \Phi_P(x,\vec{p}_\bot)=\Phi_{\vec{P}_\bot=0}(x,\vec{p}_\bot+(1/2-x)\vec{P}_\bot).
\end{equation}

\section{Equation for the Scattering Amplitude and Relation to the Equation in the Infinite Momentum Frame}

Let us derive now the equation for the two-body scattering amplitude. The definition of the scattering amplitude $T(P;p,p')$ in the 4-dimensional covariant Bethe--Salpeter formalism looks as follows:
\begin{align}
    G(P;p,p') & =G^{(0)}(P;p,p') \notag \\
    & \quad +\int d^4p''d^4p''' G^{(0)}(P;p,p'')T(P;p'',p''') G^{(0)}(P;p''',p') \notag \\
    & =G^{(0)}(P;p) \dl^{(4)}(p-p')+G^{(0)}(P;p) T(P;p,p') G^{(0)}(P,p'). \label{1.20}
\end{align}

We define the quantity $\wt T(P;p_+,\vec{p}_\bot;p_+',\vec{p}{\,}_\bot')$ by the similar expression
\begin{multline}\label{1.21}
    \wt G(P;p_+,\vec{p}_\bot;p_+',\vec{p}{\,}_\bot') =
        \wt G^{(0)}(P;p_+,\vec{p}_\bot)\dl(p_+-p_+')
        \dl^{(2)}(\vec{p}_\bot-\vec{p}{\,}_\bot') \\
    +\wt G^{(0)}(P;p_+,\vec{p}_\bot) 
        \wt T(P;p_+,\vec{p}_\bot;p_+',\vec{p}{\,}_\bot')
        \wt G^{(0)}(P;p_+',\vec{p}{\,}_\bot'). 
\end{multline} 

Integrating (\ref{1.20}) according to (\ref{1.10}), we get: 
\begin{equation}\label{1.22}
    \wt G=\wt G^{(0)}+\wt{G^{(0)} TG^{(0)}}. 
\end{equation}
Comparing formulas (\ref{1.22}) and (\ref{1.21}), we obtain:
\begin{equation}\label{1.23}
    \wt T=\wt G^{(0)}{\,}^{-1} \cdot \wt{G^{(0)} TG^{(0)}} \cdot \wt G^{(0)^{-1}}.
\end{equation}
It can be shown that on mass-shell the following equality holds: 
$$
    \wt T=T.
$$

Let us derive now the equation for the amplitude $\wt T$. Using the definition \eqref{1.16}, we get the equation for the Fourier transform of the ``two-time'' Green function
\begin{equation}\label{1.24}
    \wt G=\wt G^{(0)}+\wt G^{(0)}V \wt G.
\end{equation}

In \eqref{1.24} the multiplication is understood as a three-dimensional integration over the corresponding variables $x$ and $\vec{p}_\bot$. Comparing \eqref{1.24} and \eqref{1.21} one can see that
\begin{equation}\label{1.25}
    \wt T\wt G^{(0)}=V \wt G
\end{equation}
from which the equation for scattering amplitude $\wt T$ follows:
\begin{equation}\label{1.26}
    \wt T=V+V\wt G^{(0)}\wt T.
\end{equation}
In the frame, where total transverse momentum is zero $\vec{P}_\bot=0$, the equation \eqref{1.26} looks as follows:
\begin{gather}
    \wt T(P;x,\vec{p}_\bot;x',\vec{p}{\,}_\bot')= V(P;x,\vec{p}_\bot;x',\vec{p}{\,}_\bot') \notag \\
    +\int_0^1 \frac{dx''}{x''(1-x'')} \int d\vec{p}{\,}_\bot''\,
        \frac{V(P;x,\vec{p}_\bot;x'',\vec{p}{\,}_\bot'') 
        \wt T(P;x'',\vec{p}{\,}_\bot'';x',\vec{p}{\,}_\bot')}
            {\big[ \frac{m_1^2+\vec{p}{\,}_\bot^{\prime\prime2}}{x''}+
                \frac{m_2^2+\vec{p}{\,}_\bot^{\prime\prime2}}{1-x''}-P^2-i\ve\big]}\,. \label{1.27}
\end{gather}
In a number of papers (see, e.g., \cite{66}--\cite{69}) the composite systems have been described on the basis of the called old-fashioned three-dimensional perturbation theory in the infinite momentum frame, which has been used by Weinberg \cite{70} in the relativistic quantum field theory. The equation \eqref{1.27} is the close analogue of the equation derived 
in \cite{70}. In the lowest order or perturbation theory equation (\ref{1.27}) reproduces the equation from \cite{70} and at the same time contains the regular method of constructing the interaction kernel in the higher orders of perturbation theory. We will not discuss this point here, but recall that as in the canonical three-dimensional approach \cite{2} there exist two methods of constructing of the interaction kernel (by means of the ''two-time'' Green function and by means of the scattering amplitude on the mass-shell).

We note, however, that there exist one substantial difference between the equation derived here and equation of \cite{70}. In the light front approach the equation is written in an arbitrary Lorentz frame and ``longitudinal motion'' of constituents is parameterized in terms of scale invariant and Lorentz invariant variable $x=(P/2+p)_+/ P_+$. The Weinberg equation is written in the infinite momentum frame and ``longitudinal motion'' is parameterized in terms of the variable $x=(P/2+p)_3/P_3$, which is not Lorentz invariant.

\section{The Case of Two Spin-$1/2$ Particles}

Let us consider the problem of two spin-$1/2$ particles interacting with scalar or vector gluon fields. 

Let us start with definition of Green function in the interaction representation in terms of $x_+$-ordered product:
$$
    G_+(x_1,x_2;x_3,x_4)= \frac{\la 0 \mid T_+(\psi_1(x_1)\psi_2(x_2) 
        \ol{\psi}_2(x_3) \ol{\psi}_4(x_4)S)\mid 0\ra}{\la 0 \mid S \mid 0\ra},
$$
where $S$-matrix is defined by the relation \eqref{2a}. It was noticed that in models like $g\ol{\psi}\Gm\psi\vf$, $g\ol{\psi}\gm_\mu\psi B_\mu$, etc. noncovariant expressions in vertices and propagators cancel in $S$-matrix. That is why we can use the covariant $S$-operator.

Using usual consideration \cite{1} one can derive the light front Bethe--Salpeter equation:
\begin{gather*}
    G_+(x_1,x_2;x_3,x_4)=\ol{S}{\,}_F^{(1)}(x_1-x_3)\ol{S}{\,}_F^{(2)}(x_2-x_4) \\
    -\!\int \!\!dx_5\, dx_6\, dx_7\,dx_8\, \ol{S}{\,}_F^{(1)}(x_1\!-\!x_5)
        \ol{S}{\,}_F^{(2)}(x_2\!-\!x_6) K(x_5,x_6;x_7,x_8) G_+(x_7,x_8;x_3,x_4),
\end{gather*}
where the kernel $K$ contains the sum of all two particle irreducible diagrams in the light front formalism.
$$
    \ol{S}_F(x)=S_F(x)-\frac{i}{4}\,\gm_+\ve(x_-) \dl(x_+)\dl^{(2)}(\vec{x}_\bot).
$$

In the momentum representation the free fermion propagator is of the form:
$$
    \ol{S}_F(p)=\frac{\wh p+m}{p^2-m^2+i\ve}-\frac{1}{2}\,\frac{\gm_+}{p_+}=
        \frac{\wh{\ol{p}}+m}{p^2-m^2+i\ve},
$$
where in the nominator the momentum $p$ is on the mass shell:
$$
    \wh{\ol{p}}=\frac{1}{2}\,\gm_+ p_-+\frac{1}{2}\,\gm_- p_+-\vec{\gm}_\bot\vec{p}_\bot,
        \quad p_-=\frac{\vec{p}{\,}_\bot^2+m^2}{p_+}\,.
$$

In the momentum representation the Bethe--Salpeter equation looks as follows:
\begin{multline*}
    G_+(P;p,q)=\ol{S}{\,}_F^{(1)}(\mu_1P+p) \ol{S}{\,}_F^{(2)}(\mu_2P-q)
        \dl^{(4)}(p-q) \\
    -\ol{S}{\,}_F^{(1)}(\mu_1 P+p) \ol{S}{\,}_F^{(2)}(\mu_2 P-p) 
        \int d^4 q' K(P;p,q') G_+(P;q',q).
\end{multline*}
Here the usual notations are used:
$$
    P=p_1+p_2; \quad p=\mu_2p_1-\mu_1p_2; \quad \mu_1+\mu_2=1; \quad 
        \mu_i=\frac{m_i}{m_1+m_2}\,.
$$

Solving this equation by iterations we get the resolvent representation:
\begin{multline*}
    G_+(P;p,q)=\ol{S}{\,}_F^{(1)}(\mu_1 P+p) \ol{S}{\,}_F^{(2)}(\mu_2 P-q)
        \dl^{(4)}(p-q) \\
    -\ol{S}{\,}_F^{(1)}(\mu_1 P+p) \ol{S}{\,}_F^{(2)}(\mu_2 P-p) T_+(P;p,q)
        \ol{S}{\,}_F^{(1)}(\mu_1 P+q) \ol{S}{\,}_F^{(2)}(\mu_2 P-q).
\end{multline*}
Here the resolvent $T_+$ coincides with the covariant $T$-matrix. Thus, the only difference consists in the external propagators.

The contribution of the two-body bound state can be evaluated in the usual way. Let $x_{1+},x_{2+}>x_{3+},x_{4+}$. Then:
$$
    G_+(x_1,x_2;x_3,x_4)=-\sum_{P,\al} \chi_{P,\al}(x_1,x_2) 
        \ol{\chi}_{P,\al}(x_3,x_4),
$$
where bound state wave function is defined as follows:
$$
    \chi_{P,\al}(x_1,x_2) =\la 0 \mid T_+(\psi(x_1)\psi(x_2))\mid P,\al\ra.
$$
For the contribution of bound state $M_B$ we have:
\begin{gather*}
    -\int \frac{d^4P}{(2\pi)^3} \,\tht(P_+) \dl(P^2-M_B^2) 
        \chi_{P,\al}(x_1,x_2) \ol{\chi}_{P,\al}(x_3,x_4) \\
    \times \tht\Big(X_+-X_+'-\frac{1}{2}\, |x_+|-\frac{1}{2} |x_+'|\Big).
\end{gather*}
Here usual notation of centre of mass coordinates $(X,X')$ and relative coordinates $(x,x')$ are introduced. This expression can be rewritten as:
\allowdisplaybreaks
\begin{gather*}
    -\frac{i}{(2\pi)^4} \int d^4 P \tht(P^+) e^{iP(X-X')}\,
        \frac{\chi_{P,\al}(x) \ol{\chi}_{P,\al}(x')}{P^+(P^--P_B^-+i\ve)}
        e^{\frac{i}{2}(P^--P_B^-)(|x_+|+|x_+'|)}, \\
    P_B^-=\frac{\vec{P}{\,}_\bot^2+M_B^2}{P^+}\,.
\end{gather*} 
Thus in the momentum representation we have:
$$
    G_+(P;p,q)\simeq -\frac{1}{(2\pi)^4}\,
        \frac{\chi_{P,\al}(p) \ol{\chi}_{P,\al}(q)}{P^+(P^--P_B^-+i\ve)}+
        \text{regular terms at} P_-=P_B^-.
$$

Equation for bound state wave function looks as follows \cite{6}:
$$
    \chi_{P,\al}(p) =-\ol{S}{\,}_F^{(1)}(\mu_1P+p) \ol{S}{\,}_F^{(2)}(\mu_2P-p) 
        \int d^4 q \,K(P;p,q)\chi_{P,\al}(q).
$$

Let us proceed now to the three dimensional (quasipotential) light front formulation of the bound state problem.

Bethe--Salpeter amplitude (wave function) is defined as:
$$
    \chi_{P,\al}(x_1,x_2)=\la 0 \mid T_+(\ol{\psi}(x_1)\psi(x_2)) \mid P,\al \ra=
        e^{iPx}\chi_{P,\al}(x).
$$

Let us derive the equation for the three-dimensional wave function $\chi_{P,\al}$ $(x_+=0,\ul{x})$, $\ul{x}=(x_-,\vec{x}_\bot)$.

In the momentum representation we have:
$$
    \psi_P(\ul{p}) =\psi_P(p_+,\vec{p}_\bot)=\int_{-\infty}^\infty \chi_p(p)\, dp_-.
$$
It is necessary to investigate properties of the ``two-time'' Green function:
$$
    \wt G_+(P;\ul{p}\,,\ul{q})=\int_{-\infty}^\infty G_+(P;p,q)\,dp_-\,dq_-.
$$
Let us consider first the free Green function:
$$
    \wt G_0(P;\ul{p}\,,\ul{q})=(\wh{\ol{p}}_1+m_1)(\wh{\ol{p}}_2+m_2) \dl^{(3)}(\ul{p}-\ul{q})I_0,
$$
where 
\begin{align*}
    I_0 & =\int_{-\infty}^\infty dp_-\,dq_- \; \Dl_0^{(1)}(\mu_1P+p)\Dl_0^{(2)}(\mu_1P-p)
        \dl(p_--q_-) \\
    & =-\frac{1}{xP_+(1-x)P_+}\, \int_{-\infty}^\infty 
        \frac{dp_-}{(p_--\mu_1P_--\ol{p}{\,}_1^-+\frac{i\ve}{x})
            (p_--\mu_2P_-+\ol{p}{\,}_2^-+\frac{i\ve}{1-x})}\,, \\
    & \qquad x=\frac{p_{1,+}}{P_+}\,, \quad 1-x=\frac{p_{2,+}}{P_+}\,, \quad 
        0\leq x\leq 1.
\end{align*}

Special property of the spin case is that numerators of free propagators do not participate in the integration, because they do not depend on $p_-$ and $q_-$. After integration we obtain:
$$
    \wt G_0(P;\ul{p}\,,\ul{q})=(\wh{\ol{p}}_1+m_1)(\wh{\ol{p}}_2+m_2)\,
        \frac{-2\pi i}{x(1-x)P_+}\,\frac{\tht(x)\tht(1-x) \dl^{(3)}(\ul{p}-\ul{q})}
            {(P^2-\frac{\vec{p}_\bot^2+m_1^2}{x}-\frac{\vec{p}_\bot^2+m_2^2}{1-x}+i\ve)}\,.
$$

The reference frame with $\vec{P}_\bot=0$ is used in this integration.

Since $G_+$ $(\wt G_+)$ is defined on the subspace with positive frequencies, it is necessary to project onto this subspace. Define:
\begin{gather*}
    \ol{u}{\,}_1^{\lb_1}(\ol{p}_1) \ol{u}{\,}_2^{\lb_2}(\ol{p}_2)\wt G_+
        \ol{u}{\,}_1^{\lb_1'}(\ol{q}_1)\ol{u}{\,}_2^{\lb_2'}(\ol{q}_2)\equiv 
        \wt g_+^{\lb_1\lb_2,\lb_1'\lb_2'}\,, \\
    \ol{u}{\,}_1^{\lb_1}(\ol{p}_1) \ol{u}{\,}_2^{\lb_2}(\ol{p}_2)\wt T
        \ol{u}{\,}_1^{\lb_1'}(\ol{q}_1)\ol{u}{\,}_2^{\lb_2'}(\ol{q}_2)\equiv 
        \wt \cT^{\lb_1\lb_2,\lb_1'\lb_2'}
\end{gather*}
and obtain the matrix equation
$$
    \wt g_+=\wt g_0\times \wt\cT\times \wt g_0,
$$
where
$$
    \wt g\equiv \wt g_0(P;x,\vec{p}_\bot)=
        \frac{-2\pi i}{x(1-x)P_+}\,\frac{2m_1 2m_2\tht(x)\tht(1-x)}
            {(P^2-\frac{\vec{p}_\bot^2+m_1^2}{x}-\frac{\vec{p}_\bot^2+m_2^2}{1-x}+i\ve)}\,.
$$
Inverting the operator $\wt g_+$ we get:
$$
    \wt g_+^{-1}=\wt g_0^{-1}-V.
$$
Here the quasipotential $V$ is represented by the series, which can be written in the following formal form:
$$
    V=\wt\cT\times (1+\wt g_0 \wt\cT)^{-1}=(1+\wt\cT\wt g_0)^{-1}\times \wt\cT.
$$

The equation for the bound state wave function is of the form \cite{6}:
\begin{multline*}
    \bigg( P^2-\frac{\vec{p}_\bot^2+m_1^2}{x}-\frac{\vec{p}_\bot^2+m_2^2}{1-x}\bigg) 
        \Phi_P(x,\vec{p}_\bot) \\
    =\frac{2m_1\,2m_2}{2(2\pi)^3x(1-x)} \int_0^1 dx' \int d\vec{p}{\,}_\bot' 
        V_P(x,\vec{p}_\bot; x',\vec{p}{\,}_\bot')\Phi(x',\vec{p}{\,}_\bot').
\end{multline*}

The wave function $\Phi_P$ is related to the initial wave function $\psi$ by the relation
$$
    \Phi_P(x,\vec{p}_\bot) =\ol{u}_1(\ol{p}_1) \ol{u}_2(\ol{p}_2) \psi_P(x,\vec{p}_\bot).
$$

This equation coincides with corresponding equation for scalar particles \cite{4}.

\begin{figure}
\begin{center}
\includegraphics*{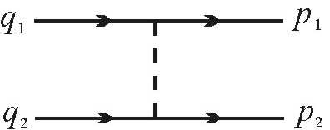}
\end{center}
\caption{}
\label{fig0}
\end{figure}

Let us construct now one meson exchange quasipotential in the theory $g\ol{\psi}\psi\vf$, which corresponds to the diagram of Fig. 1 and looks as follows:
$$
    K=\frac{-ig^2}{(2\pi)^4}\,\frac{1}{(p-q)^2-\mu^2+i\ve}\,.
$$

According to the quasipotential constructing rule it is necessary to consider the integral:
\begin{gather}
    I_1=\int_{-\infty}^\infty dp_-\,dq_- \; \Dl_0^{(1)}(\mu_1P+p)\Dl_0^{(2)}(\mu_2P-p)
        K(P;p,q)\dl(p_--q_-) \notag \\
    \times \Dl_0^{(1)}(\mu_1P+q)\Dl_0^{(2)}(\mu_2P-q), \label{*}
\end{gather}
where
$$
    K(P;p,q)=\frac{-ig^2}{(2\pi)^4}\, \frac{\ol{u}_1(\ol{p}_1)\ol{u}_2(\ol{p}_2) 
        u_1(\ol{q}_1) u_2(\ol{q}_2)}{(x-x') P_+ [p^--q^--\frac{(\vec{p}_\bot-\vec{q}_\bot)^2+\mu^2}
            {P_+(x-x')}+\frac{i\ve}{x-x'}]}\,.
$$

Integrating the expression \eqref{*} we obtain for the quasipotential:
$$
    V_P=g^2 \ol{u}_1(\ol{p}_1) u_1(\ol{q}_1) \ol{u}_2(\ol{p}_2) u_2(\ol{q}_2)
        \tht(x)\tht(1-x) \tht(x')\tht(1-x') \cV_P,
$$
where
\begin{gather}
    |x-x'| \cV(x,\vec{p}_\bot,x',\vec{p}{\,}_\bot') \notag \\
    =\tht(x-x') \bigg[ P^2-\frac{\vec{p}_\bot^2+m_1^2}{1-x}-
        \frac{\vec{p}_\bot^{\prime2}+m_2^2}{1-x'}-
        \frac{(\vec{p}_\bot-\vec{p}{\,}_\bot')^2+\mu^2}{x-x'}+i\ve \bigg]^{-1} \notag \\
    \quad +\tht(x'-x) \bigg[ P^2-\frac{\vec{p}_\bot^2+m_1^2}{x}-
        \frac{\vec{p}_\bot^{\prime 2}+m_2^2}{x'}-
        \frac{(\vec{p}_\bot-\vec{p}{\,}_\bot')^2+\mu^2}{x'-x}+i\ve \bigg]^{-1}. \label{2.29n}
\end{gather}

Note that light front formulation of the quasipotential approach for spin particles is free from difficulties, which are characteristic to the usual formulation with equal times \cite{71}.

\section{Equation for the Many-Body Bound State Wave Function}

Formalism developed in the previous sections can be generalized to the case of $N$ relativistic interacting particles. The way of this generalization can be seen, if instead of the variable $x$, defined by the relative momentum, two variables $x^{(1)}$ and $x^{(2)}$, defined by the individual momentum of particles are used:
\begin{equation}\label{1.28}
    x^{(i)}=\frac{p_+^{(i)}}{P_+}\,, \quad i=1,2,
\end{equation}
The variables $x^{(i)}$ vary in the interval $0<x^{(i)}<1$.

Define the Fourier transform of many-body Bethe--Salpeter amplitude (wave function):
$$
    \chi_{P,\al}([x^{(i)}]) =\big\la 0\mid T(\vf_1(x_\mu^{(1)}) \vf_2(x_\mu^{(2)})
        \cdots \vf_N(x_\mu^{(N)})) \mid P,\al\big\ra
$$
by the following equation
\begin{multline}\label{1.29}
    \quad \dl^{(4)}\bigg( P-\sum_{i=1}^N p^{(i)}\bigg) \chi_{P,\al}([p^{(i)}]) \\
    =\int \prod_{i=1}^N d^4 x^{(i)} \exp \bigg[ i\sum_{i=1}^N p^{(i)}x^{(i)}\bigg]
        \chi_{P,\al}([x_\mu^{(i)}]), \quad 
\end{multline}
where 
$$
    [p^{(i)}]=p^{(1)},\dots,p^{(N)}; \quad 
        [x_\mu^{(i)}]=x_\mu^{(1)},\dots,x_\mu^{(N)}.
$$
Here we have ascribed the Lorentz index $\mu$ to the 4-coordinates $x^{(i)}$ in order to distinguish them from the scale-invariant variables $x^{(i)}$, which will be introduced later.

If we introduce the light front variables
\begin{equation}\label{1.30}
    P_\pm=P_0\pm P_3; \quad 
        p_\pm^{(i)}=p_0^{(i)}\pm p_3^{(i)}; \quad 
        x_\pm^{(i)}=\frac{x_0^{(i)}\pm x_3^{(i)}}{2}
\end{equation}
and integrate (\ref{1.29}) over $\prod\limits_{i=1}^N d p_-^{(i)}$. We obtain
\begin{multline}
    2\dl \bigg( P_+-\sum_{i=1}^N p_+^{(i)}\bigg)\dl^{(2)}
        \bigg(\vec{P}_\bot -\sum_{i=1}^N \vec{p}_\bot^{(i)}\bigg)
        \Psi_{P,\al} ([p_+^{(i)},\vec{p}{\,}_\bot^{(i)}]) \\
    =(2\pi)^N \int \prod_{i=1}^N d^4 x^{(i)}\dl(x_+^{(i)})
        \exp \bigg[ i\sum_{i=1}^N (p_+^{(i)}x_-^{(i)}-
            \vec{p}{\,}_\bot^{(i)}\vec{x}{\,}_\bot^{(i)})\bigg]
            \chi_{P,\al} ([x_\mu^{(i)}]).
\end{multline}

The function $\Psi_{P,\al}([p_+^{(i)},\vec{p}{\,}_\bot^{(i)}])$ is related to the Bethe--Salpeter amplitude in the following way 
\begin{equation}\label{1.32}
    \Psi_{P,\al}([p_+^{(i)},\vec{p}{\,}_\bot^{(i)}])=\int_{-\infty}^\infty
        \prod_{i=1}^N d p_-^{(i)} \dl \bigg( P_- -
        \sum_{i=1}^N p_-^{(i)} \bigg) \chi_{P,\al}([p^{(i)}]).
\end{equation}

Let us introduce now the Fourier transform of the ``two-time'' Green function 
\begin{gather}
    \wt G(P;[p_+^{(i)},\vec{p}{\,}_\bot^{(i)}];
        [p_+^{(i)'},\vec{p}{\,}_\bot^{(i)'}]) \notag \\
    =\int_{-\infty}^\infty \prod_{i=1}^N d p_-^{(i)}\,dp_-^{(i)'} 
        \dl\bigg( P_-\!-\!\sum_{i=1}^N p_-^{(i)}\bigg) 
        \dl\bigg( P_-\!-\!\sum_{i=1}^N p_-^{(i)'}\bigg)
        G(P;[p^{(i)}]; [p^{(i)'}]).\label{1.33}
\end{gather}
The function $G(P;[p^{(i)}]; [p^{(i)'}])$ is defined by the Fourier transformation
\begin{align}
    G([x_\mu^{(i)}]; [x_\mu^{(i)'}]) & =
    \big\la 0\mid T(\vf_1(x_\mu^{(1)}) \cdots \vf_N(x_\mu^{(N)}),
        \vf_1^+(x_\mu^{(1)'}) \cdots \vf_N^+(x_\mu^{(N)'})) \mid 0\big\ra
            \notag \\
    &=(2\pi)^{-4N} \int \prod_{i=1}^N d^4 p^{(i)} d^4p^{(i)'} 
        \exp \bigg[ -i\sum_{i=1}^N (p^{(i)}x^{(i)}-
            \vec{p}{\,}^{(i)'}\vec{x}{\,}^{(i)'})\bigg] \notag \\
    & \quad \times G(P;[p^{(i)}];[p^{(i)'}]). \label{1.34}
\end{align}
For the case of free particles we have
\begin{equation}\label{1.35}
    G^{(0)}(P;[p^{(i)}];[p^{(i)'}])=
        \frac{i^N\prod\limits_{i=1}^N \dl^{(4)}(p^{(i)}-p^{(i)'})}
            {\prod\limits_{i=1}^N ({p^{(i)}}^2-{m^{(i)}}^2+i\ve)}\,.
\end{equation}
Integrating both sides of (\ref{1.35}) according to the definition (\ref{1.33}) and omitting the $\dl$-function corresponding to the total 4-momentum conservation, we get~\cite{5}:
\begin{multline}\label{1.36}
    \wt G^{(0)} (P;[p_+^{(i)},\vec{p}{\,}_\bot^{(i)}];
            [p_+^{(i)'},\vec{p}{\,}_\bot^{(i)'}]) \\
    =\frac{(2i)^N (2\pi i)^{N-1} \prod\limits_{i=1}^N 
            \dl(p_+^{(i)}-p_+^{(i)'})
            \dl^{(2)}(\vec{p}{\,}_\bot^{(i)}-\vec{p}{\,}_\bot^{(i)'})
            \prod\limits_{i=1}^N \tht(x^{(i)}) \tht(1-x^{(i)})}
        {P_+^{N-1} \prod\limits_{i=1}^N x^{(i)} 
            \Big[ P^2-\sum\limits_{i=1}^N \frac{(\vec{p}{\,}_\bot^{(i)}-x^{(i)}\vec{P}_\bot)^2+
                m_i^2}{x^{(i)}}\Big]} \\
    =\wt G^{(0)}(P;[p_+^{(i)},\vec{p}{\,}_\bot^{(i)}]) 
        \prod_{i=1}^N \dl(p_+^{(i)}-p_+^{(i)'})
        \dl^{(2)}(\vec{p}{\,}_\bot^{(i)}-\vec{p}{\,}_\bot^{(i)'}).
\end{multline}

The variables $x^{(i)}$ are defined in the following way 
\begin{equation}\label{1.37}
    x^{(i)}=\frac{p_+^{(i)}}{P_+}\,, \quad i=1,2,\dots,N.
\end{equation}
Thus, the function $\wt G^{(0)}(P;[p_+^{(i)},\vec{p}{\,}_\bot^{(i)}])$ is defined under the conditions
\begin{equation}\label{1.38}
    \sum_{i=1}^N x^{(i)}=1; \qquad 0<x^{(i)}<1; \qquad 
        \sum_{i=1}^N \vec{p}{\,}_\bot^{(i)}=\vec{P}_\bot.
\end{equation}

Let us introduce now the inverse operator $\wt G^{-1}$ by means of the relation
\begin{multline}\label{1.39}
    \int_0^{P_+} \prod_{i=1}^N dp_+^{(i)''} 
        \int \prod_{i=1}^N d \vec{p}{\,}^{(i)''} 
        \wt G^{-1}(P;[p_+^{(i)},\vec{p}{\,}_\bot^{(i)}];
            [p_+^{(i)''},\vec{p}{\,}_\bot^{(i)''}]) \\
    \times \wt G(P;[p_+^{(i)''}, \vec{p}{\,}_\bot^{(i)''}];
            [p_+^{(i)'}, \vec{p}{\,}_\bot^{(i)'}])= 
        \prod_{i=1}^N \dl(p_+^{(i)}-p_+^{(i)'})
            \dl^{(2)}(\vec{p}{\,}_\bot^{(i)}-\vec{p}{\,}_\bot^{(i)'})
\end{multline}
and define the interaction kernel $V$:
\begin{multline}\label{1.40}
    \wt G^{-1}(P;[p_+^{(i)},\vec{p}{\,}_\bot^{(i)}];
            [p_+^{(i)'}, \vec{p}{\,}_\bot^{(i)'}])=
        {\wt{G}}^{(0)}{\,}^{-1}(P;[p_+^{(i)},\vec{p}{\,}_\bot^{(i)}];
            [p_+^{(i)'}, \vec{p}{\,}_\bot^{(i)'}]) \\
    -\frac{\dl\Big(P_+-\sum\limits_{i=1}^N p_+^{(i)}\Big)
            \dl^{(2)}\Big(\vec{P}_\bot-\sum\limits_{i=1}^N \vec{p}{\,}_\bot^{(i)}\Big)}
            {(2i)^N (2\pi i)^{N-1}}\,
        V(P;[p_+^{(i)},\vec{p}{\,}_\bot^{(i)}];
            [p_+^{(i)'}, \vec{p}{\,}_\bot^{(i)'}]).
\end{multline}

The equation for the wave function 
\begin{equation}\label{1.41}
    \Phi_{P,\al} ([x^{(i)},\vec{p}{\,}_\bot^{(i)}])=
        P_+^{N-1} \prod_{i=1}^N x^{(i)}\Psi_{P,\al} ([p_+^{(i)},\vec{p}{\,}_\bot^{(i)}])
\end{equation}
looks as follows
\allowdisplaybreaks
\begin{multline}\label{1.42}
    \bigg[ P^2-\sum_{i=1}^N 
        \frac{(\vec{p}{\,}_\bot^{(i)}-x^{(i)}\vec{P}_\bot)^2+m_i^2}{x^{(i)}}\bigg] 
        \Phi_{P,\al} ([x^{(i)},\vec{p}{\,}_\bot^{(i)}]) \\
    =\int_0^1 \prod_{i=1}^N \frac{dx^{(i)'}}{x^{(i)'}}\,
        \dl\bigg( 1-\sum_{i=1}^N x^{(i)'}\bigg)
        \int \prod_{i=1}^N d\vec{p}{\,}_\bot^{(i)'}
            \dl\bigg( \vec{P}_\bot-\sum_{i=1}^N \vec{p}{\,}_\bot^{(i)'}\bigg) \\
    \times V(P;[p_+^{(i)},\vec{p}{\,}_\bot^{(i)}]; 
            [p_+^{(i)'}, \vec{p}{\,}_\bot^{(i)'}]) 
        \Phi_{P,\al}([x^{(i)'},\vec{p}{\,}^{(i)'}]).
\end{multline}

Corresponding equation for the case of spin-$1/2$ constituents looks as follows \cite{72}, \cite{73}:
\begin{gather*}
    \bigg[ P^2 -\sum_{i=1}^N \frac{(\vec{p}{\,}_\bot^{(i)}-x^{(i)} \vec{P}_\bot)^2+
            m_i^2}{x^{(i)}}\bigg]
        \prod_{i=1}^N (\wh{\ol{p}}{\,}^{(i)}+m_i) \Phi_{P,\al}([x^{(i)},\vec{p}{\,}_\bot^{(i)}]) \\
    =\int \prod_{i=1}^N \frac{dx^{(i)'}}{x^{(i)'}} \, \dl\bigg( 1-\sum_{i=1}^N x^{(i)'}\bigg)
        \int \prod_{i=1}^N d \vec{p}{\,}_\bot^{(i)'} 
        \dl\bigg( P_\bot-\sum_{i=1}^N \vec{p}{\,}_\bot^{(i)'}\bigg) \\
    \times \prod_{i=1}^N (\wh{\ol{p}}{\,}^{(i)}+m_i) V_p([x^{(i)},\vec{p}{\,}_\bot^{(i)}];
        [x^{(i)'},p_\bot^{(i)'}]) 
        \prod_{i=1}^N (\wh{\vec{p}}{\,}^{(i)'}+m_i) \Phi_{P,\al}([x^{(i)'},\vec{p}{\,}_\bot^{(i)'}]). \end{gather*}

The formalism developed can be used for the treatment of a wide class of elementary particle and nuclear physics problems (see, e.g., \cite{74}--\cite{76}).

Other forms of the light front description of particle dynamics can be found in Refs. \cite{77}--\cite{80}. 

%% file: garse2.tex
\section*{\large III. Relativistic Elastic Form Factors and Scattering Amplitudes for Composite systems}

\def\theequation{3.\arabic{equation}}
\setcounter{equation}{0}
\setcounter{section}{0}
\section{Formulation of the Method}

The reaction of a system on a weak external perturbation corresponding to the local field $A(x)$ is described in quantum field theory by the expression \cite{56}
\begin{equation}\label{2.1}
    \Big\la P,\al \mid \frac{\dl S}{\dl A(k)} \mid Q,\bt \Big\ra \bigg|_{A=0} =
        (2\pi)^4 \dl^{(4)}(P-Q-k) \la P,\al \mid J(0) \mid Q,\bt\ra.
\end{equation}
Here $J(x)$ is the local current of the system 
\begin{equation}\label{2.2}
    J(x)=i\,\frac{\dl S}{\dl A(x)}\,S^+,
\end{equation}
$| P,\al\ra$ and $|Q,\bt\ra$ are the state vectors of composite particles with momenta $P$ and $Q$ and the sets of additional quantum numbers $\al$ and $\bt$, respectively, normalized in a relativistically invariant manner
\begin{equation}\label{2.3}
    \la P,\al \mid Q,\bt\ra =2P_0(2\pi)^3 \dl^{(3)}(\vec{P}-\vec{Q})\dl_{\al\bt}.
\end{equation}

Below we suggest a method of constructing relativistically covariant form factors of composite systems in terms of ``equal time'' three-dimensional wave function $\Phi_{P,\al} ([x^{(i)}, \vec{p}{\,}_\bot^{(i)}])$.

Let us consider first the case of two-particle system and introduce the quantity $R$ defined by the vacuum expectation value of the chronologically ordered product of Heisenberg field operators of the scalar particles $\vf_i(x_i)$ and a some local current $J(x)$:
\begin{align}
    R(x_1,x_2;y_1,y_2) & = 
        \la 0 \mid T(\vf_1(x_1) \vf_2(x_2) J(0)\vf_1^+(y_1) \vf_2^+(y_2)) \mid 0 \ra \notag \\
    & =(2\pi)^{-16} \int d^4p_1\,d^4p_2\,d^4q_1\,d^4q_2\,
        \exp \bigg[ -i\sum_{j=1}^2 (p_jx_j-q_jy_j)\bigg] \notag \\
    & \quad \times R(p_1,p_2;q_1,q_2). \label{2.4}
\end{align}

Introducing, as above, the relative 4-coordinates and 4-momenta
\begin{equation}\label{2.5}
\begin{aligned}
    & X=\frac{x_1+x_2}{2}\,, \quad x=x_1-x_2, \quad 
        Y=\frac{y_1+y_2}{2}\,, \quad y=y_1-y_2; \\
    & P=p_1+p_2, \quad p=\frac{p_1-p_2}{2}\,, \quad Q=q_1+q_2, \quad 
        q=\frac{q_1-q_2}{2}\,,
\end{aligned}
\end{equation}
we rewrite expression \eqref{2.4} in the form 
\begin{gather}
    R(X,x;Y,y) \notag \\
    =\!(2\pi)^{-16} \int\!\! d^4 P\,d^4p\,d^4Q\,d^4q\,
        \exp \big[ -i(Px\!-\!QY\!+\!px\!-\!qy)\big] R(P,p;Q,q).  \label{2.6}
\end{gather}

As is known \cite{81}, the $R$ quantity can be presented in the form
\begin{equation}\label{2.7}
    R=G\Gm G,
\end{equation}
or in the detailed form:
\begin{align}
    R(X,x;Y,y) & =\int d^4 X'\, d^4 x'\, d^4Y'\, d^4 y'\, G(X-X';x,x') \notag \\
    & \quad \times \Gm(X',x';Y'y') G(Y'-Y;y,y'). \label{2.8}
\end{align}

\begin{figure}
\begin{center}
\includegraphics*{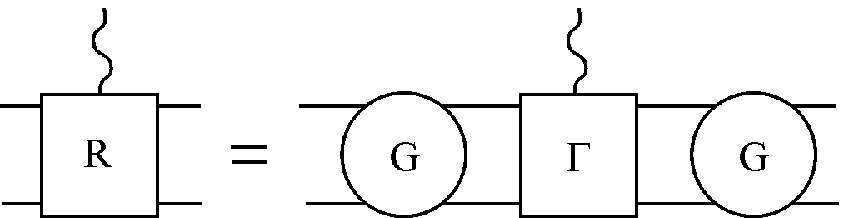}
\end{center}
\caption{}
\label{fig1}
\end{figure}

In the momentum space we get 
\begin{equation}\label{2.9}
    R(P,p;Q,q) =\int d^4 p'\,d^4q'\,G(P;p,p') \Gm(P,p';Q,q') G(Q;q',q).
\end{equation}
Here $G$ is the two-particle Green function of scalar fields $\vf_i(x_i)$ and the vertex function $\Gm$ is the sum of all two-particle irreducible diagrams for 5-point Green function \eqref{2.6} (see Fig. \ref{fig1}).

Passing to the ``two-time'' description in terms of light front variables, we define the quantity
\begin{equation}\label{2.10}
    \wt R(P;p_+,\vec{p}_\bot;Q;q_+,\vec{q}_\bot)=\int_{-\infty}^{+\infty} 
        dp_-\, dq_-\, R(P,p;Q,q).
\end{equation}

The quantity $\wt R$ can be presented in the form 
\begin{equation}\label{2.11}
    \wt R=\wt G\wt \Gm \wt G
\end{equation}
or in the detailed form 
\begin{multline}\label{2.12}
    \wt R(P;p_+,\vec{p}_\bot; Q; q_+,\vec{q}_\bot) = 
        \int_{-P_+/2}^{P_+/2} dp_+' \; \int d\vec{p}{\,}_\bot' \; 
        \int_{-Q_+/2}^{Q_+/2} dp_+''\; \int d\vec{p}{\,}_\bot'' \\
    \times \wt G(P;p_+,\vec{p}_\bot; p_+',\vec{p}{\,}_\bot')
        \wt \Gm(P;p_+',\vec{p}{\,}_\bot'; p_+'',\vec{p}{\,}_\bot'')
        \wt G(Q;p_+'',\vec{p}{\,}_\bot''; q_+,\vec{q}_\bot). 
\end{multline}
Here $\wt \Gm$ is the vertex integral operator. Let us show that the quantity $\wt\Gm$ defines the form factor of composite system. Starting from the spectral properties \cite{82} of the 5-point Green function \eqref{2.10}, it is possible to show that the quantity $\wt R$ has the pole singularities near the points corresponding to the masses $M_\al$ and $M_\bt$ of composite systems:
\begin{gather}
    \us{p^2\to M_\al^2; \;\; Q^2\to M_\bt^2}{\wt R(P;p_+,\vec{p}_\bot;Q;q_+,\vec{q}_\bot)} \notag \\
    \simeq [i(2\pi)^4]^2 \,
        \frac{\Psi_{P,\al}(p_+,\vec{p}_\bot) \la P,\al \mid J(0)\mid 
            Q,\bt\ra\Psi_{Q,\bt}^+(q_+,\vec{q}_\bot)}{(P^2-M_\al^2)(Q^2-M_\bt^2)}\,. \label{2.13}
\end{gather}
On the other hand, taking into account the pole singularities of the two-particle ``two-time'' Green function
\begin{equation}\label{2.14}
    \us{P^2\to M_\al^2}{\wt G(P;p_+,\vec{p}_\bot;p_+',\vec{p}{\,}_\bot')} \simeq
        \frac{\Psi_{P,\al}(p_+,\vec{p}_\bot)\Psi_{P,\al}^+(p_+',\vec{p}{\,}_\bot')}{P^2-M_\al^2}\,,
\end{equation}
we find from \eqref{2.12}
\begin{gather}
    \wt R(P;p_+,\vec{p}_\bot;Q,q_+,\vec{q}_\bot) \notag \\
    \simeq [i(2\pi)^4]^2\; 
        \frac{\Psi_{P,\al}(p_+,\vec{p}_\bot)\Psi_{Q,\bt}^+(q_+,\vec{q}_\bot)}
            {(P^2-M_\al^2)(Q^2-M_\bt^2)}
        \int_{-P_+/2}^{P_+/2} dp_+' \; \int d\vec{p}{\,}_\bot' \; 
        \int_{-Q_+/2}^{Q_+/2} dq_+' \; \int d\vec{q}{\,}_\bot' \notag \\
    \times \Psi_{P,\al}^+(p_+',\vec{p}{\,}_\bot') \wt \Gm_{\al\bt} (P;p_+',\vec{p}{\,}_\bot';Q;q_+',\vec{q}{\,}_\bot')
        \Psi_{Q,\bt}(q_+',\vec{q}{\,}_\bot'), \label{2.15}
\end{gather}
where 
\begin{equation}\label{2.16}
    \wt \Gm_{\al\bt} (P;p_+,\vec{p}_\bot;Q;q_+,\vec{q}_\bot)=
        \wt \Gm(P;p_+,\vec{p}_\bot;Q;q_+,\vec{q}_\bot)
            \Big|_{\substack{P^2=M_\al^2 \\Q^2=M_\bt^2}}\,.
\end{equation}
Comparing equations \eqref{2.13} and \eqref{2.15} we get the following expression for the matrix element of the local current $J$:
\begin{multline}\label{2.17}
    \la P,\al\mid J(0)\mid Q,\bt\ra =
        \int_{-P_+/2}^{P_+/2} dp_+ \; \int d\vec{p}_\bot \; 
        \int_{-Q_+/2}^{Q_+/2} dq_+ \; \int d\vec{q}_\bot  \\
    \times \Psi_{P,\al}^+(p_+,\vec{p}_\bot) \wt \Gm_{\al\bt} (P;p_+,\vec{p}_\bot;Q;q_+,\vec{q}_\bot)
        \Psi_{Q,\bt}(q_+,\vec{q}_\bot). 
\end{multline}
Equations \eqref{2.15} and \eqref{2.17} give an exact expression for the vertex operator of the composite system in terms of 4- and 5-point Green functions $G$ and $\Gm$
\begin{gather}
    \wt \Gm_{\al\bt} (P;p_+,\vec{p}_\bot;Q;q_+,\vec{q}_\bot)= 
        \lim_{\substack{P^2\to M_\al^2 \\ Q^2\to M_\bt2}} 
        \int_{-P_+/2}^{P_+/2} dp_+' \; \int d\vec{p}{\,}_\bot' \; 
        \int_{-Q_+/2}^{Q_+/2} dq_+' \; \int d\vec{q}{\,}_\bot' \notag  \\
    \times {\wt G}^{-1}(P;p_+,\vec{p}_\bot;p_+',\vec{p}{\,}_\bot') 
        [\wt{G\Gm G}](P;p_+',\vec{p}{\,}_\bot';Q;q_+',\vec{q}{\,}_\bot') \notag \\
    \times {\wt G}^{-1}(Q;q_+',\vec{q}{\,}_\bot'; q_+,\vec{q}_\bot) \label{2.18}
\end{gather}
Using the perturbation theory methods for these functions one can construct the coupling constant expansion for the vertex function of composite system.

In the case of spin-$1/2$ constituents the matrix element of the local cyrrent operator looks as follows \cite{73}:
\begin{gather*}
    (2\pi)^2 \la Q,\bt \mid J_\mu(0) \mid  P,\al \ra \\
    =\int d\ul{q}\, d\ul{p} \ol{\psi}_{Q,\bt}(\ul{q}) \,
        \frac{\wh{\ol{q}}_1+m_1}{2m_1}\,\frac{\wh{\ol{q}}_2+m_2}{2m_2} \,
        \wt\Gm_\mu(Q,\ul{q};P,\ul{p}) 
        \frac{\wh{\ol{p}}_1+m_1}{2m_1} \,\frac{\wh{\ol{p}}_2+m_2}{2m_2}\,
        \psi_{P,\al}(\ul{p}),
\end{gather*}
where, for instance, $d\ul{q}=dq_+\, d\vec{q}_\bot$, $\ul{q}=(q_+,\vec{q}_\bot)$.

\section{Elastic Form Factor in the Impulse Approximation}

\begin{figure}  
\begin{center}
\includegraphics*{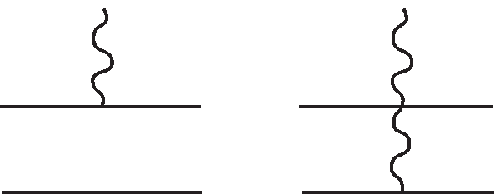}
\end{center}
\caption{}
\label{fig2}
\end{figure}

Let us consider the so-called impulse approximation for the vertex operator $\wt\Gm$ which corresponds to the limit of ``weakly bound'' (noninteracting) particles (see Fig. \ref{fig2}).

For the vertex operator, corresponding to the conserved vector current, we find
\begin{gather}
    \wt\Gm_\mu=\wt\Gm_{1\mu}^{(0)}+\wt\Gm_{2\mu}^{(0)}, \label{2.19} \\
    \wt\Gm_{i\mu}^{(0)}=[\wt G^{(0)}]^{-1} [\wt{G^{(0)} \wt\Gm_{i\mu}^{(0)} G^{(0)}}]\,
        [\wt G^{(0)}]^{-1}, \label{2.20}
\end{gather}
where 
\begin{equation}\label{2.21}
    \wt\Gm_{i\mu}^{(0)}=(2\pi)^4 e_i(p_i+q_i)_\mu \dl^{(4)}(p_j-q_j)
        [G_j^{(0)}(p_j)]^{-1} \big|_{i\neq j}.
\end{equation}
Here 
\begin{align}
    & G^{(0)}(p_j) =G_1^{(0)}(p_1) G_2^{(0)}(p_2)=
        i^2 \prod_{i=1}^2 (p_i^2-m_i^2)^{-1}, \label{2.22} \\
    & G^{(0)}(q_j) =G_1^{(0)}(q_1) G_2^{(0)}(q_2)=
        i^2 \prod_{i=1}^2 (q_i^2-m_i^2)^{-1} \label{2.23}
\end{align}
are the two-particle Green function for free particles with masses $m_i$ and charges $e_i$.

Then for the invariant form vector of the composite system defined by the relation
\begin{equation}\label{2.24}
    \la P,\al \mid J_\mu(0) \mid Q,\bt\ra =(P+Q)_\mu F(\Dl^2), \quad \Dl=P-Q,
\end{equation}
in the reference frame, in which
\begin{equation}\label{2.25}
    P_+=Q_+\,, \quad (P-Q)^2=\Dl^2=-\vec{\Dl}{\,}_\bot^2=-(\vec{P}_\bot-\vec{Q}_\bot)^2,
\end{equation}
we have \cite{4}:
\begin{gather}
    F\!=\!(-\vec{\Dl}{\,}_\bot^2)\!=\!\frac{e_1(2\pi)^3}{2}  \int_0^1 \frac{dx}{x(1-x)}\notag \\
    \times \int d\vec{p}_\bot \Phi_{\vec{P}_\bot=0}(x,\vec{p}_\bot+(1-x)\vec{\Dl}_\bot)
        \Phi_{\vec{P}_\bot=0}(x,\vec{p}_\bot) \notag \\
    +\, \mbox{similar term with} \;\; e_2. \label{2.26}
\end{gather}

Note that construction of relativistic form factors of composite systems in other versions of relativistic description of bound states is considered in Refs. \cite{74}, \cite{82}--\cite{84}. 

\section{Relativistic Form Factor for the Many-Body System}

Let us construct now a form factor for the relativistic many-body system in terms of light front many-body wave functions $\Phi_P([x^{(i)},\vec{p}{\,}_\bot^{(i)}])$. 

Consider, as in the case of two constituents, the quantity $R$, which is defined by the vacuum expectation value of the chronologically ordered product of the Heisenberg field operators $\vf_i(x_\mu^{(i)})$ and a local current $J(x)$
\begin{align}
    R([x_\mu^{(i)}]; [y_\mu^{(i)}]) & = 
        \la 0 \mid T(\vf_1(x_\mu^{(1)}) \cdots \vf_N(x_\mu^{(N)})J(0)\vf_1^+(y_\mu^{(1)}) 
            \cdots \vf_N^+(y_\mu^{(N)})) \mid 0 \ra \notag \\
    & =(2\pi)^{-4N} \int \prod_{i=1}^N  d^4p^{(i)}\,d^4 q^{(i)}\,
        \exp \bigg[ -i\sum_{i=1}^N (p^{(i)}x^{(i)}-q^{(i)}y^{(i)})\bigg] \notag \\
    & \quad \times R([p^{(i)}],][q^{(i)}]). \label{2.27}
\end{align}

The quantity $R$ can be presented in the form \cite{81}
\begin{equation}\label{2.28}
    R=G\Gm G.
\end{equation}

Multiplication in \eqref{2.28} has to be understood as an integration over the 4-coordinates of particles. $G$ is the many-body Green function of the fields $\vf_i(x_\mu^{(i)})$ and the vertex function $\Gm$ is defined by the sum of the irreducible diagrams of the $(2N+1)$-point function \eqref{2.27}.

Proceeding now to the light front description, we introduce the quantity \linebreak  $\wt R([p_+^{(i)},\vec{p}{\,}_\bot^{(i)}]; [q_+^{(i)},\vec{q}{\,}_\bot^{(i)}])$ by the relation
\begin{gather}
    \wt R([p_+^{(i)},\vec{p}{\,}_\bot^{(i)}]; [q_+^{(i)},\vec{q}{\,}_\bot^{(i)}]) \notag \\
    =\int_{-\infty}^\infty \prod_{i=1}^N d p_-^{(i)}\, dq_-^{(i)}\, 
        \dl\bigg(P_--\sum_{i=1}^N p_-^{(i)}\bigg)
        \dl\bigg(Q_--\sum_{i=1}^N q_-^{(i)}\bigg)
        R([p^{(i)}]; [q^{(i)}]) \label{2.29}
\end{gather}
and write it in the form 
\begin{equation}\label{2.30}
    \wt R=\wt G\wt\Gm \wt G.
\end{equation}
Multiplication in Eq. \eqref{2.30} has to be understood in the operator sense: 
\begin{multline}\label{2.31}
    \wt A \wt B=\int_0^{Q_+} \prod_{i=1}^N d q_+^{(i)'} 
        \dl\bigg(Q_+-\sum_{i=1}^N q_+^{(i)'}\bigg)
        \int \prod_{i=1}^N d \vec{q}{\,}_\bot^{(i)'} 
        \dl\bigg(\vec{Q}_\bot-\sum_{i=1}^N \vec{q}{\,}_\bot^{(i)'}\bigg) \\
    \times \wt A([p_+^{(i)}, \vec{p}{\,}_\bot^{(i)}]; [q_+^{(i)'}, \vec{q}{\,}_\bot^{(i)'}])
        \wt B([q_+^{(i)'}, \vec{q}{\,}_\bot^{(i)'}]; [q_+^{(i)}, \vec{q}{\,}_\bot^{(i)}]).
\end{multline}

From the spectral properties  of the function $\wt G$ \cite{82} it follows that $\wt R$ possesses the double pole singularities  
\begin{gather} 
    \wt R([p_+^{(i)},\vec{p}_\bot^{(i)}]; [q_+^{(i)},\vec{q}_\bot^{(i)}]) \notag \\
    \simeq [i(2\pi)^4]^2 \, 
       \frac{\Psi_{P,\al}([p_+^{(i)},\vec{p}{\,}_\bot^{(i)}]) \la P,\al \mid J(0)\mid 
                Q,\bt\ra\Psi_{Q,\bt}^+([q_+^{(i)},\vec{q}{\,}_\bot^{(i)}])}
            {(P^2-M_\al^2)(Q^2-M_\bt^2)} \label{2.32}
\end{gather}
in the vicinity of the points, where $N$-particle system forms the bound states with masses $M_\al$ and $M_\bt$ and a set of other quantum number $\al$ and $\bt$, respectively. 

On the other hand, knowing the pole singularities of the Green function 
\begin{equation}\label{2.33}
    \us{P^2\to M_\al^2}{\wt G([p_+^{(i)},\vec{p}{\,}_\bot^{(i)}];[q_+^{(i)},
            \vec{q}{\,}_\bot^{(i)}])}
        \simeq 2(2\pi)^4\, \frac{\Psi_{P,\al}([p_+^{(i)},\vec{p}{\,}_\bot^{(i)}])
                 \Psi_{P,\al}^+([q_+^{(i)},\vec{q}{\,}_\bot^{(i)}])}
            {(P^2-M_\al^2)}, 
\end{equation}
one can reduce the Eq. \eqref{2.32} to the form
\begin{multline}\label{2.34}
    \wt R([p_+^{(i)},\vec{p}_\bot^{(i)}]; [q_+^{(i)},\vec{q}_\bot^{(i)}])  
    \simeq [i(2\pi)^4]^2 \, 
       \frac{\Psi_{P,\al}([p_+^{(i)},\vec{p}{\,}_\bot^{(i)}]) 
                \Psi_{Q,\bt}^+([q_+^{(i)},\vec{q}{\,}_\bot^{(i)}])}
            {(P^2-M_\al^2)(Q^2-M_\bt^2)} \\
    \times \int_0^{P_+} \prod_{i=1}^N d p_+^{(i)'} 
        \dl\bigg(P_+-\sum_{i=1}^N p_+^{(i)'}\bigg)
        \int \prod_{i=1}^N d \vec{p}{\,}_\bot^{(i)'} 
        \dl^{(2)}\bigg(\vec{P}_\bot-\sum_{i=1}^N \vec{p}{\,}_\bot^{(i)'}\bigg) \\
    \times \int_0^{Q_+} \prod_{i=1}^N d q_+^{(i)'} 
        \dl\bigg(Q_+-\sum_{i=1}^N q_+^{(i)'}\bigg)
        \int \prod_{i=1}^N d \vec{q}{\,}_\bot^{(i)'} 
        \dl^{(2)}\bigg(\vec{Q}_\bot-\sum_{i=1}^N \vec{q}{\,}_\bot^{(i)'}\bigg) \\
    \times \Psi_{P,\al}^+([p_+^{(i)'},\vec{p}{\,}_\bot^{(i)'}])
        \wt \Gm_{\al\bt}([p_+^{(i)'},\vec{p}{\,}_\bot^{(i)'}];[q_+^{(i)'},\vec{q}{\,}_\bot^{(i)'}]) 
        \Psi_{Q,\bt}([q_+^{(i)'},\vec{q}{\,}_\bot^{(i)'}]).
\end{multline}
Here 
\begin{multline}\label{2.35}
    \wt \Gm_{\al\bt}([p_+^{(i)},\vec{p}{\,}_\bot^{(i)}];[q_+^{(i)},\vec{q}{\,}_\bot^{(i)}]) \\
    =\lim_{P^2\to M_\al^2, \; Q^2\to M_\bt^2} 
        \int_0^{P_+} \prod_{i=1}^N d p_+^{(i)'} 
        \dl\bigg(P_+-\sum_{i=1}^N p_+^{(i)'}\bigg)
        \int \prod_{i=1}^N d \vec{p}{\,}_\bot^{(i)'} 
        \dl^{(2)}\bigg(\vec{P}_\bot-\sum_{i=1}^N \vec{p}{\,}_\bot^{(i)'}\bigg) \\
    \times \int_0^{Q_+} \prod_{i=1}^N d q_+^{(i)'} 
        \dl\bigg(Q_+-\sum_{i=1}^N q_+^{(i)'}\bigg)
        \int \prod_{i=1}^N d \vec{q}{\,}_\bot^{(i)'} 
        \dl^{(2)}\bigg(\vec{Q}_\bot-\sum_{i=1}^N \vec{q}{\,}_\bot^{(i)'}\bigg) \\
    \times \wt G^{-1}([p_+^{(i)},\vec{p}{\,}_\bot^{(i)}];[p_+^{(i)'},\vec{p}{\,}_\bot^{(i)'}]) \\
    \times [\wt{G\Gm G}] ([p_+^{(i)'},\vec{p}{\,}_\bot^{(i)'}];[q_+^{(i)'},\vec{q}{\,}_\bot^{(i)'}])
        \wt G^{-1} ([q_+^{(i)'},\vec{q}{\,}_\bot^{(i)'}];[q_+^{(i)},\vec{q}{\,}_\bot^{(i)}]). 
\end{multline}

Comparing \eqref{2.32} with \eqref{2.34}, we get the following expression for the matrix element of the bound state current:
\begin{multline}\label{2.36}
    \la P,\al \mid J(0)\mid Q,\bt\ra \\
    =\int_0^{P_+} \prod_{i=1}^N d p_+^{(i)} 
        \dl\bigg(P_+-\sum_{i=1}^N p_+^{(i)}\bigg)
        \int \prod_{i=1}^N d \vec{p}{\,}_\bot^{(i)} 
        \dl^{(2)}\bigg(\vec{P}_\bot-\sum_{i=1}^N \vec{p}{\,}_\bot^{(i)}\bigg) \\
    \times \int_0^{Q_+} \prod_{i=1}^N d q_+^{(i)} 
        \dl\bigg(Q_+-\sum_{i=1}^N q_+^{(i)}\bigg)
        \int \prod_{i=1}^N d \vec{q}{\,}_\bot^{(i)} 
        \dl^{(2)}\bigg(\vec{Q}_\bot-\sum_{i=1}^N \vec{q}{\,}_\bot^{(i)}\bigg) \\
    \times \Psi_{P,\al}^+([p_+^{(i)},\vec{p}{\,}_\bot^{(i)}])
        \wt \Gm_{\al\bt}([p_+^{(i)},\vec{p}{\,}_\bot^{(i)}];[q_+^{(i)},\vec{q}{\,}_\bot^{(i)}]) 
        \Psi_{Q,\bt}([q_+^{(i)},\vec{q}{\,}_\bot^{(i)}]).
\end{multline}

The vertex operator $\wt\Gm_{\al\bt}$ can be constructed, using, for instance, perturbation theory methods of quantum field theory. Phenomenological vertex operators can also be used. Here we consider the so-called ``impulse approximation''. In this case 
\begin{gather}
    \wt\Gm_\mu=\sum_{i=1}^N (\wt\Gm_\mu^{(0)})_i, \label{2.37} \\
    (\wt\Gm_\mu^{(0)})_i=[\wt G^{(0)}]^{-1} [\wt{G^{(0)} (\wt\Gm_\mu^{(0)})_i G^{(0)}}]
        \, [\wt G^{(0)}]^{-1}\,. \label{2.38}
\end{gather}
Here 
\begin{gather}
    (\wt\Gm_\mu^{(0)})_i=(2\pi)^4 e_i(p^{(i)}+q^{(i)})_\mu 
        \prod_{\substack{j=1 \\ i\neq j}} \dl (p^{(j)}-q^{(j)})
            [G^{(0)}(q^{(j)})]^{-1} , \label{2.39} \\
    G^{(0)}([p^{(i)}]) =i^N \prod_{i=1}^N (p^{(i)2}-m_i^2+i\ve)^{-1} \label{2.40}
\end{gather}
and $e_i$ is the change of $i$-th particle. 

Performing the integration in Eq. \eqref{2.38} according to the definition \eqref{2.29}, passing to the reference frame, where 
\begin{equation}\label{2.41}
    p_+=q_+, \quad \vec{Q}_\bot=0, \quad \Dl^2=(P-Q)^2=-\vec{\Dl}{\,}_\bot^2=
        -(\vec{P}_\bot-\vec{Q}_\bot)^2
\end{equation}
and using the transformation properties of wave functions
\begin{equation}\label{2.42}
    \Phi_{P,\al} ([x^{(i)}, \vec{p}{\,}_\bot^{(i)}]) =
        \Phi_{\vec{P}_\bot=0,\,\al} ([x^{(i)},\vec{p}{\,}_\bot^{(i)}-x^{(i)}\vec{P}_\bot]),
\end{equation}
we obtain \cite{5}
\begin{equation}\label{2.43}
    \la P,\al \mid J_\mu(0) \mid Q,\bt\ra =
        \sum_{k=1}^N \la P,\al \mid J_\mu(0) \mid Q,\bt\ra_k,
\end{equation}
where, for instance,
\begin{multline}\label{2.44}
    \la P,\al \mid J_\mu(0) \mid Q,\bt\ra_k =(P_++Q_+) F_k(-\vec{\Dl}{\,}_\bot^2) \\
    =\frac{-(2\pi)^4 e_k(P_++Q_+)}{(2i)^{N+1}(2\pi i)^{N-1}} \,
        \int_0^1 \prod_{i=1}^N \frac{dx^{(i)}}{x^{(i)}}\, 
        \dl\bigg( 1-\sum_{i=1}^N x^{(i)}\bigg) 
        \int \prod_{i=1}^N d\vec{p}{\,}_\bot^{(i)} 
        \dl^{(2)}\bigg( \sum_{i=1}^N \vec{p}{\,}_\bot^{(i)}\bigg) \\
    \times \Phi_{\vec{P}_\bot=0,\,\al} ([x^{(i)},\vec{p}{\,}_\bot^{(i)}-x^{(i)}\vec{\Dl}_\bot]_{i\neq k},
        x^{(k)}, \vec{p}{\,}_\bot^{(k)}+(1-x^{(k)})\vec{\Dl}_\bot)  \\
    \times \Phi_{\vec{P}_\bot=0,\,\bt} ([x^{(i)},\vec{p}{\,}_\bot^{(i)}]).
\end{multline}

Taking into account the normalization condition for the wave functions 
\begin{multline}\label{2.45}
    i(2\pi)^4 \int_0^{P_+} \prod_{i=1}^N d p_+^{(i)} 
        \dl\bigg(P_+-\sum_{i=1}^N p_+^{(i)}\bigg)
        \int \prod_{i=1}^N d \vec{p}{\,}_\bot^{(i)} 
        \dl^{(2)}\bigg(\vec{P}_\bot-\sum_{i=1}^N \vec{p}{\,}_\bot^{(i)}\bigg) \\
    \times \int_0^{Q_+} \prod_{i=1}^N d q_+^{(i)} 
        \dl\bigg(Q_+-\sum_{i=1}^N q_+^{(i)}\bigg)
        \int \prod_{i=1}^N d \vec{q}{\,}_\bot^{(i)} 
        \dl^{(2)}\bigg(\vec{Q}_\bot-\sum_{i=1}^N \vec{q}{\,}_\bot^{(i)}\bigg) \\
    \times \Psi_{P,\al}^+([p_+^{(i)},\vec{p}{\,}_\bot^{(i)}])\,
        \frac{\pa \wt G^{-1}(P;[p_+^{(i)},\vec{p}{\,}_\bot^{(i)}];[q_+^{(i)},\vec{q}{\,}_\bot^{(i)}])}
            {\pa P^2}\,  
        \Psi_{Q,\bt}([q_+^{(i)},\vec{q}{\,}_\bot^{(i)}])=1,
\end{multline}
for $\vec{\Dl}_\bot=0$ we get 
\begin{equation}\label{2.46}
    F(\Dl^2=0)= \sum_{k=1}^N e_k.
\end{equation}
Thus, the form factor at zero momentum transfer is normalized to the total charge of the system. Note that general problems of normalization of three-dimensional relativistic wave functions have been considered in Ref. \cite{85}. 

Thus Eqs. \eqref{2.43} and \eqref{2.44} define the form factors of a many-body system in terms of the light front relativistic wave functions $\Phi([x^{(i)},\vec{p}{\,}_\bot^{(i)}])$. 

\section{Asymptotic Behaviour of the Pion Form Factor at Large Momentum Transfer}

According to quark counting rules \cite{86}, \cite{87} asymptotic behaviour of the form-factor of composite system is determined by the minimal number $n$ of elementary constituents:
\begin{equation}\label{**}
    F(t) \sim |t|^{1-n}.
\end{equation}

Well-known proof of the relation \eqref{**} consists in using equation for bound state wave function under the assumption of one meson exchange, since this mechanism dominates in the asymptotic region. Asymptotic behaviour of the wave function is then used to calculate the behaviout of form-factor at large momentum transfer.

Let us write the equation for fermion-antifermion system in the one gluon exchange approximation:
\begin{gather*}
    \bigg[ P^2 -\frac{\vec{p}{\,}_\bot^2+m^2}{x(1-x)}\bigg] \Phi_P(\ul{p}) \\=
        \int [d\ul{q}] v(x,\vec{p}_\bot;y,\vec{q}_\bot) 
        (\wh{\ol{p}}_1+m) O^{(1)}\Phi_P(\ul{q}) O^{(2)}(-\wh{\ol{p}}_2+m).
\end{gather*}
Here
\begin{gather}\label{11}
    \Phi_P(\ul{p})\equiv \Phi_P(x,\vec{p}_\bot)=(\wh{\ol{p}}_1+m)
        \psi_P(\ul{p}) (-\wh{\ol{p}}_2+m), \\
    \int [d\ul{q}] =\int_0^1 \frac{dy}{y(1-y)}\int d\vec{q}_\bot. \notag 
\end{gather}
$O^{(i)}$ are matrices in the corresponding fermion-gluon vertices. One gluon exchange quasipotential is given above by Eq. \eqref{2.29n}.

It is known \cite{88} that the most general form of the Bethe--Salpeter amplitude for pion is of the form:
\begin{equation}\label{***}
    \chi_P(p)=\gm_5(\chi_1+\wh p \chi_2+\wh p \chi_3+[\wh P,\wh p]\chi_4) ,
\end{equation}
where $\chi_i$ are scalar functions of invariants $Pp$ and $p^2$. Eq. \eqref{***} can be rewritten as:
$$
    \chi_P(p)=\gm_5(\xi_1+\wh p_1\xi_2+\wh p_2\xi_3+\wh p_1\wh p_2 \xi_4).
$$

Using the equation
$$
    \wh p_i-\wh{\ol{p}}_i=\frac{1}{2}\,(p_--\ol{p}_-)\gm_+,
$$
we obtain
\begin{gather*}
    \chi_P(p)=\gm_5(\xi_1+\xi_2 \wh{\ol{p}}_1 +\xi_3\wh{\ol{p}}_2 +
        \xi_4\wh{\ol{p}}_1 \wh{\ol{p}}_2) 
        +\frac{1}{2}\,\gm_5 \Big[(p_{1,-}-\ol{p}_{1,-})\xi_2 \gm_+\\
    +(p_{2,-}-\ol{p}_{2,-})\xi_3\gm_+ 
        \frac{1}{2} (p_{1,-}-\ol{p}_{1,-}) \xi_4\gm_+\wh{\ol{p}}_2+
        \Big[\frac{1}{2} (p_{2,-}-\ol{p}_{2,-}) \xi_4\wh{\ol{p}}_1\gm_+\Big].
\end{gather*}
Using the relation:
$$
    (\wh{\ol{p}}_i+m) \wh{\ol{p}}_i=m(\wh{\ol{p}}_i+m)
$$
we get the most general form of the quasipotential wave function
\begin{equation}\label{****}
    \Phi_{P}(\ul{p})=(\wh{\ol{p}}_1+m) [\vf_1(\ul{p})+\gm_+\vf_2(\ul{p})]
        \gm_5(-\wh{\ol{p}}_2+m).
\end{equation}
Quasipotential equation \eqref{11} is now rewritten as
\begin{gather*}
    \bigg[ P^2-\frac{\vec{p}{\,}_\bot^2+m^2}{x(1-x)}\bigg] (\wh{\ol{p}}_1+m) 
        [\vf_1(\ul{p})+\gm_+\vf_2(\ul{p})]\gm_5(-\wh{\ol{p}}_2+m) \\
    =\pm \int [d\ul{q}] \mathcal{V}(\ul{p},\ul{q}) (\wh{\ol{p}}_1+m) O^{(1)}(\wh{\ol{q}}_1+m)
        [\vf_1(\ul{q})+\gm_+\vf_2(\ul{q})] \\
    \times (\wh{\ol{q}}_2+m) O^{(2)}(\wh p_2+m) \gm_5.
\end{gather*}
``$-$'' sign is used in the case, if $O^{(2)}\gm_5=-\gm_5 O^{(2)}$. For operators $O^{(i)}$ we use $O^{(i)}=1,\gm_5,\gm_\mu, \gm_\mu\gm_5$. In all four cases we get for the asymptotic behaviour of wave functions:
\begin{equation}\label{delta}
    \vf_1(\ul{p})\sim (\vec{p}{\,}_\bot^2)^{-2}, \quad 
        \vf_2(\ul{p})\sim (\vec{p}{\,}_\bot^2)^{-1}.
\end{equation}

Thus, asymptotic behaviour of wave functions does not depend on the concrete dynamics and is of the above form.

Let us proceed now to the study of the asymptotic behaviour of the form-factor. In the impulse approximation the pion electromagnetic form-factor is of the form:
\begin{gather*}
    \la Q\mid J_\mu \mid P\ra =(P+Q)_\mu F(k^2)=(2P+k)_\mu F(k^2) \\
    =\frac{e_1}{2(2\pi)^3} \int [d\ul{p}] \,[d\ul{q}] \dl(\ul{p}{\,}_2-\ul{q}{\,}_2) 
        Sp\{\Phi_Q(\ul{q}) \gm_\mu^{(1)} \Phi_P(\ul{p})\}+(1\rightleftarrows2).
\end{gather*}

Inserting the wave function in the form \eqref{****} into this equation and integrating over $d\ul{q}$, we obtain
\begin{gather}
    (2P+k)_\mu F(k^2) \notag \\
    \sim \int [d\ul{p}] 
        S_p\{(\wt\vf_1^*+\gm_+\wt\vf_2^*)(\wh{\ol{p}}{\,}^{(1)}+m)\gm_\mu^{(1)}
            (\wh{\ol{q}}{\,}^{(1)}+m)(\vf_1+\gm_+\vf_2).\label{box}
\end{gather}
Here $\wt\vf_i$ depend on $\vec{p}_\bot+(1-x)\vec{\Dl}_\bot$, where
$$
    \vec{\Dl}_\bot=\vec{k}_\bot+\bigg( \frac{P_+}{Q_+}-1\bigg) P_\bot.
$$

In the reference frame, where $P_+=Q_+$, we have: $\vec{\Dl}_\bot=\vec{k}_\bot$ and $k^2=-\vec{\Dl}{\,}_\bot^2$. In the limit $\vec{\Dl}{\,}_\bot^2\to \infty$ asymptotic behaviour of the form-factor is determined by the asymptotics of wave functions. Since asymptotic behaviour of wave functions does not depend on concrete dynamics, the asymptotic behaviour of the form-factor is model independent, as is required by the automodelity \cite{86}, \cite{87}.

Calculating the ``$+$''-component in the expression \eqref{box}, one obtains:
\begin{gather*}
    F_\pi (-\vec{\Dl}{\,}_\bot^2) \sim \int_0^1 \frac{dy_1 dy_2}{y_1y_2} \,
        \dl(1-y_1-y_2) \int d\vec{q}_\bot \bigg\{ \bigg[ \bigg(\frac{x}{y_2}+1\bigg)(\vec{q}{\,}_\bot^2+m^2)\\
    +  (1-x)(\vec{q}_\bot\vec{\Dl}_\bot)\bigg] \wt\vf_1^* \vf_1 
        + 2m P_+^2 [(x+y_2)y_1\wt \vf_1^* \vf_2+ x\wt\vf_2^*\vf_1]+
        6xy_1y_2 P_+^2 \wt\vf_2^*\vf_2\bigg\}.
\end{gather*}

If we take $\vf_2=0$ (such a possibility is also considered sometimes) and use only \eqref{delta} $\vf_1^*\sim (\vec{q}_\bot+(1-x)\Dl_\bot)^{-4}$, in the limit $\vec{\Dl}{\,}_\bot^2\to \infty$ for the form-factor we get $F_\pi \sim (\vec{\Dl}{\,}_\bot^2)^{-2}$, which contradicts the quark counting rules \cite{86}, \cite{87}. At the same time taking into account the second structure, which behaves as $(\vec{q}_\bot+(1-x)\vec{\Dl}_\bot)^{-2}$, we get the correct asymptotic behaviour of the pion form-factor:
$$
    F_\pi(-\vec{\Dl}{\,}_\bot^2)\sim (\vec{\Dl}{\,}_\bot^2)^{-1}, \quad \text{at} \quad 
        \vec{\Dl}{\,}_\bot^2 \to \infty.
$$

It is essential that this behaviour does not depend on the concrete form of dynamics.

\section{Scattering of Relativistic Composite Systems}

Experimental study of high energy processes during the last decades revealed a number of scaling properties of observable quantities. Many of these properties can be understood on the basis of the composite quark-parton structure of elementary particles. In particular, the asymptotic scaling property of differential cross section of hadron-hadron scattering
\begin{equation}\label{2.47}
    \frac{d\sg}{dt} \bigg|_{\substack{s\to \infty \\ |t/s|=const.}} 
        \sim \frac{1}{s^N}\, f(\cos\tht_s),
\end{equation}
where $N$ is integer number, can be explained in the framework of dimensional analysis and assumption on three-quark structure of baryons and quark-anti-quark structure of mesons (quark counting rules) \cite{86}, \cite{87}.

In connection with the development of composite models of elementary particles a problem of the description of their interactions becomes of special interest. Study of intersections of relativistic composite systems is important also in connection with current and future experiments with beams of relativistic nuclei. Here we outline a method for the treatment of problems of that kind \cite{89}.

Below we present a description of the scattering of two composite particles. It will be shown that some simple assumptions on the hadron interactions in the scattering process allow to reproduce the results of quark counting rules. 

Consider the eight-point Green function $G$:
\begin{gather}
    G(x_1,x_2,x_3,x_4;y_1,y_2,y_3,y_4) \notag \\
    =\la 0\mid T(\vf_1(x_1)\vf_2(x_2)\vf_3(x_3)\vf_4(x_4)
        \vf_1^+(y_1)\vf_2^+(y_2)\vf_3^+(y_3)\vf_4^+(y_4))\mid 0 \ra \notag \\
    =\![(2\pi)^4]^{-8} \int \prod_{i=1}^4d^4p_i\,d^4q_i\,
        \exp\bigg[ -i\sum_{i=1}^4 (p_ix_i\!-\!q_iy_i)\bigg] 
        G(p_1,p_2,p_3,p_4;q_1,q_2,q_3,q_4) \notag \\
    =[(2\pi)^4]^{-8} \int d^4P^{(12)}\,d^4p^{(12)}\,d^4P^{(34)}\,d^4p^{(34)}\,
        d^4Q^{(12)}\,d^4q^{(12)}\,d^4Q^{(34)}\,d^4q^{(34)} \notag \\
    \times \exp\big[ -i(P^{(12)}X^{(12)}+P^{(34)}X^{(34)} +
        p^{(12)}x^{(12)}+p^{(34)}x^{(34)} \notag \\
    -Q^{(12)}Y^{(12)}-Q^{(34)}Y^{(34)}-q^{(12)}y^{(12)}-q^{(34)}y^{(34)})\big]
        \notag\\
    \times G(P^{(12)},p^{(12)};P^{(34)},p^{(34)};Q^{(12)},q^{(12)};
        Q^{(34)},q^{(34)}). \label{2.48}
\end{gather}
In \eqref{2.48} the momenta $P^{(12)}$, $p^{(12)}$, $P^{(34)}$, $p^{(34)}$, $Q^{(12)}$, $q^{(12)}$, $Q^{(34)}$, $q^{(34)}$ are introduced according to the following equations
\begin{equation}\label{2.49}
\begin{aligned}
    & P^{(12)}=p_1+p_2, \;\; p^{(12)}=\frac{p_1-p_2}{2}\,, \;\; 
        P^{(34)}=p_3+p_4, \;\; p^{(34)}=\frac{p_3-p_4}{2}\,, \\
    & Q^{(12)}=q_1+q_2, \;\; q^{(12)}=\frac{q_1-q_2}{2}\,, \;\;
        Q^{(34)}=q_3+q_4, \;\; q^{(34)}=\frac{q_3-q_4}{2}\,. 
\end{aligned}
\end{equation}

Passing now to the ``two-time'' description, we introduce the light front variables and define the quantity
\begin{gather}
    \wt G(P^{(12)},p_+^{(12)},\vec{p}{\,}_\bot^{(12)};P^{(34)},p_+^{(34)},
        \vec{p}{\,}_\bot^{(34)};Q^{(12)},q_+^{(12)}, \vec{q}{\,}_\bot^{(12)};
        Q^{(34)},q^{(34)}, \vec{q}{\,}_\bot^{(34)}) \notag \\
    =\!\!\int_{-\infty}^\infty \!\!dp_-^{(12)}  dp_-^{(34)} dq_-^{(12)} dq_-^{(34)}
        G(P^{(12)},p^{(12)};P^{(34)},p^{(34)};Q^{(12)},q^{(12)};
        Q^{(34)},q^{(34)}), \label{2.50}
\end{gather}
where $p_\pm=p_0\pm p_3$, $\vec{p}_\bot=(p_1,p_2)$.

Introduce now the quantity $M$ by the equation 
\begin{gather}
    G(P^{(12)},p^{(12)};P^{(34)},p^{(34)};Q^{(12)},q^{(12)};
        Q^{(34)},q^{(34)}) \notag \\
    =\int d^4 p^{(12)'} d^4p^{(34)'} d^4q^{(12)'} d^4q^{(34)'} \notag \\
    \times G_{12}(P^{(12)},p^{(12)}, p^{(12)'}) 
        G_{34}(P^{(34)},p^{(34)}, p^{(34)'}) \notag \\
    \times M(P^{(12)},p^{(12)'};P^{(34)},p^{(34)'};Q^{(12)},q^{(12)'};
        Q^{(34)},q^{(34)'}) \notag \\
    \times G_{12}(Q^{(12)},q^{(12)'}, q^{(12)}) 
        G_{34}(Q^{(34)},q^{(34)'},q^{(34)}) \equiv (G_{12}G_{34})
        M(G_{12}G_{34}). \label{2.51}
\end{gather}

The quantity $ \wt G$ can be presented in the form (here and in what follows we omit the arguments which are related to the relative momenta and this will not cause any misunderstanding):
\begin{gather}
    \wt G(P^{(12)}, P^{(34)}, Q^{(12)}, Q^{(34)}) \notag \\
    =\!\!\wt G_{12}(P^{(12)}) \wt G_{34}(P^{(34)}) \!\bullet \!
        \wt M(P^{(12)}, Q^{(12)}, P^{(34)},Q^{(34)}) \!\bullet\!
        \wt G_{12}(Q^{(12)}) \wt G_{34}(Q^{(34)}). \label{2.52}
\end{gather}
The symbol $\bullet$ in \eqref{2.52} has to be understood in the following sense
\begin{gather}
    \wt A \bullet \wt B =\int_{-P_+^{(12)}/2}^{P_+^{(12)}/2} dp_+^{(12)}
        \int_{-P_+^{(34)}/2}^{P_+^{(34)}/2} dp_+^{(34)}
        \int d\vec{p}{\,}_\bot^{(12)}\int d\vec{p}{\,}_\bot^{(34)} \notag \\
    \times \wt A(\dots, p_+^{(12)}, \vec{p}{\,}_\bot^{(12)};
            p_+^{(34)}, \vec{p}{\,}_\bot^{(34)})
        \wt B(p_+^{(12)}, \vec{p}{\,}_\bot^{(12)};
        p_+^{(34)}, \vec{p}{\,}_\bot^{(34)},\dots) \label{2.53}
\end{gather}
and dots correspond to the set of other arguments on which the operators $\wt A$ and $\wt B$ can depend.

Knowing the pole singularities of the two-particle Green functions $\wt G_{12}$, $\wt G_{23}$ one can show that in the vicinity of these poles the function $\wt G$ looks as follows: 
\begin{gather}
    \wt G(P^{(12)}, P^{(34)}, Q^{(12)}, Q^{(34)}) \notag \\
    \simeq [2(2\pi)^4]^4 \,
        \frac{\Psi_{12}(P^{(12)}) \Psi_{34}(P^{(34)}) 
                \Psi_{12}^+(Q^{(12)}) \Psi_{34}^+(Q^{(34)})}
            {(P^{(12)2}-M_{12}^2)(P^{(34)2}-M_{34}^2)
                (Q^{(12)2}-M_{12}^{\prime 2})(Q^{(34)2}-M_{34}^{\prime 2})}\notag \\
    \times \Psi_{12}^+(P^{(12)}) \Psi_{34}^+(P^{(34)}) \bullet
        \wt M(P^{(12)},P^{(34)}, Q^{(12)},Q^{(34)}) \notag \\
     \bullet
        \Psi_{12}(Q^{(12)}) \Psi_{34}(Q^{(34)}). \label{2.54}
\end{gather}
Here
\begin{gather}
    \wt M_{1234} =\lim_{\substack{P^{(12)2}\to M_{12}^2, \,P^{(34)2}\to M_{34}^2 \\
                Q^{(12)2}\to M_{12}^{\prime 2},\,Q^{(34)2}\to M_{34}^{\prime 2}}}
        \wt G_{12}^{-1} (P^{(12)}) \wt G_{34}^{-1} (P^{(34)}) \notag \\
    \bullet \wt{G_{12} G_{34}MG_{12}G_{34}}(P^{(12)}, P^{(34)}, Q^{(12)}, Q^{(34)})
        \bullet \wt G_{12}^{-1} (Q^{(12)}) \wt G_{34}^{-1} (Q^{(34)}), \label{2.55}
\end{gather}
$M_{12}^2$, $M_{34}^2$, $M_{12}^{\prime 2}$, $M_{34}^{\prime 2}$ are the masses of corresponding states.

From Eqs. \eqref{2.52} and \eqref{2.54} we get the following expression for the scattering amplitude:
\begin{gather}
    T(P^{(12)}, P^{(34)}, Q^{(12)}, Q^{(34)}) 
    =\Psi_{12}^+(P^{(12)}) \Psi_{34}^+(P^{(34)}) \notag \\
    \bullet
        \wt M(P^{(12)},P^{(34)}, Q^{(12)}, Q^{(34)}) \bullet 
        \Psi_{12}(Q^{(12)}) \Psi_{34}(Q^{(34)}). \label{2.56}
\end{gather}

Eq. \eqref{2.56} gives a general expression for the scattering  amplitude in the case of scattering of composite particles. The detailed form of the scattering amplitude depends on the interaction mechanism in the intermediate  state and on a special form of the wave functions of the scattered objects.

\section{Particle Exchange in the Intermediate State}

\begin{figure}
\begin{center}
\includegraphics*{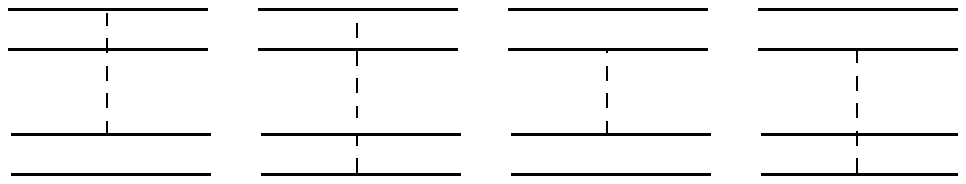}
\end{center}
\caption{}
\label{fig3}
\end{figure}

Below we consider two possible mechanisms: 1) scattering via the exchange of some intermediate  particle (Fig. \ref{fig3}), 
2)~constituent interchange mechanism (Fig. \ref{fig4}). For the sake of simplicity we consider the case of scalar particles. It can be shown that in the first case in the reference frame, where 
$$
    P_+^{(12)}=P_+^{(34)}, \quad Q_+^{(12)}=Q_+^{(34)}, \quad 
        t=(P^{(12)}-Q^{(12)})^2 =-\vec{\Dl}{\,}_\bot^{(t)^2}
$$
\noindent one has the following result for the scattering amplitude
\begin{align}
    T & = \frac{g^2}{4(\vec{\Dl}{\,}_\bot^{(t)^2}+\mu^2)} \,
        \big[ F_{12}^{(1)}(t)F_{34}^{(1)}(t) \notag \\
    & \quad + F_{12}^{(1)}(t)F_{34}^{(2)}(t)+
        F_{12}^{(2)}(t)F_{34}^{(1)}(t)+F_{12}^{(2)}(t)F_{34}^{(2)}(t)], \label{2.57}
\end{align} 
where $F_{12}^{(i)}$, $F_{34}^{(i)}$ are scalar form factors of the scattered particles, which are expressed in terms of the light front wave functions in the following way: 
\begin{equation}\label{2.58}
\begin{aligned}
    F_{12}^{(1)}(t) & =\frac{(2\pi)^3}{2} \int_0^1 \frac{dx}{x^2(1-x)} \,\\
        & \qquad \times \int d\vec{p}_\bot \Phi_{\vec{P}{\,}_\bot^{(12)}=0}^+ 
            (x,\vec{p}_\bot+(1-x)\vec{\Dl}{\,}_\bot^{(t)}) 
            \Phi_{\vec{P}{\,}_\bot^{(12)}=0}(x,\vec{p}_\bot),\\
    F_{34}^{(1)}(t) & =\frac{(2\pi)^3}{2} \int_0^1 \frac{dx}{x^2(1-x)} \, \\
        & \qquad \times \int d\vec{p}_\bot \Phi_{\vec{P}{\,}_\bot^{(34)}=0}^+ 
            (x,\vec{p}_\bot-(1-x)\vec{\Dl}{\,}_\bot^{(t)}) 
            \Phi_{\vec{P}{\,}_\bot^{(34)}=0}(x,\vec{p}_\bot),\\
    F_{12}^{(2)}(t) & =\frac{(2\pi)^3}{2} \int_0^1 \frac{dx}{x(1-x)^2} \, \\
        & \qquad \times \int d\vec{p}_\bot \Phi_{\vec{P}{\,}_\bot^{(12)}=0}^+ 
            (x,\vec{p}_\bot-x\vec{\Dl}{\,}_\bot^{(t)}) 
            \Phi_{\vec{P}{\,}_\bot^{(12)}=0}(x,\vec{p}_\bot),\\
    F_{34}^{(2)}(t) & =\frac{(2\pi)^3}{2} \int_0^1 \frac{dx}{x(1-x)^2} \, \\
        & \qquad \times \int d\vec{p}_\bot \Phi_{\vec{P}{\,}_\bot^{(34)}=0}^+ 
            (x,\vec{p}_\bot+x\vec{\Dl}{\,}_\bot^{(t)}) 
            \Phi_{\vec{P}{\,}_\bot^{(34)}=0}(x,\vec{p}_\bot).
\end{aligned}
\end{equation}

\section{Constituent Interchange Mechanism}

\begin{figure}
\begin{center}
\includegraphics*{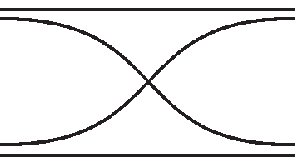}
\end{center}
\caption{}
\label{fig4}
\end{figure}

Considering the constituent interchange mechanism (Fig.\ref{fig4}), one gets the following expression for the scattering amplitude
\allowdisplaybreaks
\begin{align}
    T & = \frac{-1}{2(2\pi)^3} \int_0^1 \frac{dx}{x^2(1-x)^2} \,
        \int d\vec{p}_\bot \Phi_{\vec{P}_\bot=0}^{+{(12)}}
            (x,\vec{p}_\bot-x\vec{\Dl}{\,}_\bot^{(u)} \notag \\
    & \quad +(1-x)\vec{\Dl}{\,}_\bot^{(t)}) 
            \Phi_{\vec{P}_\bot=0}^{+{(12)}}(x,\vec{p}_\bot)
        [M_{12}^2 +M_{34}^2 \notag \\
    & \quad -S(x,\vec{p}_\bot+x\vec{\Dl}{\,}_\bot^{(t)} -
        (1-x)\vec{\Dl}{\,}_\bot^{(u)})-S(x,\vec{p}_\bot)] \notag \\
    & \quad \times \Phi_{\vec{P}_\bot=0}^{(12)}(x,\vec{p}_\bot-x\vec{\Dl}{\,}_\bot^{(u)})
        \Phi_{\vec{P}_\bot=0}^{(34)}(x,\vec{p}_\bot+(1-x)\vec{\Dl}{\,}_\bot^{(t)}), \label{2.59}
\end{align}
where $\vec{\Dl}{\,}_\bot^{(t)}=-t$, $\vec{\Dl}{\,}_\bot^{(u)}=-u$.

Here the notation has been introduced 
\begin{equation}\label{2.60}
    S(x,\vec{p}_\bot) =\frac{m_1^2+\vec{p}{\,}_\bot^2}{1-x} +
        \frac{m_2^2+\vec{p}{\,}_\bot^2}{x}
\end{equation}
and the following properties of wave functions have been used:
\begin{equation}\label{2.61}
\begin{aligned}
    & \Phi(x,\vec{p}_\bot)=\Phi(x,-\vec{p}_\bot), \\
    & \Phi(x,\vec{p}_\bot)=\Phi(1-x,\vec{p}_\bot).
\end{aligned}
\end{equation}

Let us consider now the wave functions of the composite particles in the form 
\begin{equation}\label{2.62}
    \Phi_N(x,\vec{p}_\bot) =\frac{\vf_N(x)}{[S(x,\vec{p}_\bot)]^N}\,,
        \quad N=A,B,C,D,
\end{equation}
$A$, $B$ and $C$, $D$ denote the hadrons before and after the scattering and corresponding powers, respectively.

Inserting the wave functions \eqref{2.62} into Eq. \eqref{2.59} for the scattering amplitude one gets in the asymptotic region
\begin{equation}\label{2.63}
    T_{\substack{s\to \infty \\ |t|\to \infty}} \sim 
        \frac{1}{s^{A+C+D-1}}\,\Big( \frac{1+z}{2}\Big)^{-C}
        \Big( \frac{1-z}{2}\Big)^{-D} f(z),
\end{equation}
where 
\begin{align}
    f(z) & =\int_0^1 \frac{dx\vf_{A}^+(x)\wt{\vf}_B^+(x) \vf_C(x)\vf_D(x)}
                        {\big[(1-x)^2\frac{1-z}{2}+x^2\frac{1+z}{2}\big]^A}
        \,\bigg[(1-x)^2\frac{1+z}{2}+x^2\,\frac{1-z}{2}\bigg] \notag \\
    & \quad \times [x(1-x)^{A+B+C+D-3}x^{-2C} (1-x)^{-2D}, \label{2.64} \\
    \wt\vf_B^+(x) & =\frac{-1}{(2\pi)^3} \int d\vec{p}_\bot 
        \Phi_B^+(x,\vec{p}_\bot)[x(1-x)]^{-B}, \notag 
\end{align}
$z=\cos\vth_s$, where $\vth_s$ is the scattering angle in the c.m.s.
$$
    -t\eqsim \frac{s}{2}\,(1-z), \quad 
        -u\eqsim \frac{s}{2}\,(1+z).
$$

Eq. \eqref{2.63} is in close connection with the results of quark counting rules \cite{86}, \cite{87}. 

%% file: garse3.tex
\section*{\large IV. Deep Inelastic Form Factors of Composite Systems and Multiquark States in Nuclei}

\def\theequation{4.\arabic{equation}}
\setcounter{equation}{0}
\setcounter{section}{0}

The great interest to deep inelastic interaction processes is caused by the possibility of studying the internal structure of hadrons and nuclei experimentally and checking different theoretical models on the assumptions about composite nature of strongly interacting particles. The main part of experimentally observed properties of these processes (in particular, the scale properties of structure functions) have been explained in the framework of composite quark-parton models of hadrons, in which the hadron is considered as bound state of some parallelly moving point-like constituents \cite{90}--\cite{93}. Interaction between constituents and their transverse  motion inside hadron is neglected.

More precise measurements in wider range of kinematic variables have led to the discovery of deviations from exact scale invariance in the behaviour  of structure function \cite{94}--\cite{98}. There were attempts to explain these deviations on the kinematical (search for new scale-invariant variables \cite{99}) and dynamic (taking into account higher chromodynamical corrections \cite{100}--\cite{102}) basis.

Here we incorporate the transverse motion of constituents in the composite system, which leads to the violation of Bjorken scaling of structure functions. In this approach, hadrons are considered as bound states of quarks, described in terms of light front wave functions.

\section{Construction of Tensor $W_{\mu\nu}$}

Consider the quantity $R_{\mu\nu}$, which is defined by the vacuum expectation value of the chronological $T_+$-ordered product of the Heisenberg field operators $\vf_i(x_\mu^{(i)})$ and local currents $J_\mu$ and $J_\nu$:
\begin{gather}
    R_{\mu\nu}([x_\mu^{(i)}]; [y_\mu^{(i)}];z) \notag  \\
    =\la 0 \mid T(\vf_1(x_\mu^{(1)}) \cdots\vf_N(x_\mu^{(N)}) J_\mu(z)J_\nu(0)
        \vf_1^+(y_\mu^{(1)}) \cdots \vf_N^+(y_\mu^{(N)})) \mid 0 \ra \notag \\
    = (2\pi)^{-8N} \int \prod_{i=1}^N d^4 p^{(i)}\, d^4 q^{(i)}\,
        \exp \bigg[ -i\sum_{i=1}^N (p^{(i)}x^{(i)}-q^{(i)}y^{(i)})\bigg] \notag \\
    \times R_{\mu\nu} ([p^{(i)}]; [q^{(i)}];z). \label{3.1}
\end{gather}
Here $[x_\mu^{(i)}]$, $[y_\mu^{(i)}]$, $[p^{(i)}]$, $[q^{(i)}]$ are the sets of corresponding 4-vectors. 

\begin{figure}
\begin{center}
\includegraphics*{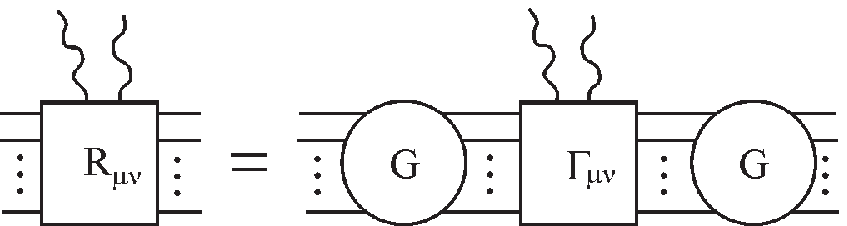}
\end{center}
\caption{}
\label{fig5}
\end{figure}

The quantity $R_{\mu\nu}$ can be presented as (see Fig. \ref{fig5})
$$
    R_{\mu\nu}=G\,\Gm_{\mu\nu}\,G,
$$
where $G$ is the $N$-particle Green function of fields $\vf_i(x_\mu^{(i)})$:
\begin{equation}\label{3.2}
    G([x_\mu^{(i)}]; [y_\mu^{(i)}])=
        \la 0 \mid T(\vf_1(x_\mu^{(1)}) \cdots\vf_N(x_\mu^{(N)}) 
        \vf_1^+(y_\mu^{(1)}) \cdots \vf_N^+(y_\mu^{(N)})) \mid 0 \ra
\end{equation}
and ``two-photon'' vertex $\Gm_{\mu\nu}$ is defined by the sum of irreducible diagrams with $2N+2$ points (legs). 

Let us introduce now the three-dimensional quantity $\wt R_{\mu\nu}$, equating all $x_+^{(i)}=x_+$ and $y_+^{(i)}=y_+$ in \eqref{3.1}
\begin{gather}
    \wt R_{\mu\nu}(x_+,[x_-^{(i)},\vec{x}{\,}_\bot^{(i)}]; y_+, 
        [y_-^{(i)},\vec{y}{\,}_\bot^{(i)}];z) \notag\\
    =\la 0 \mid T(\vf_1(x_+,x_-^{(1)},\vec{x}{\,}_\bot^{(1)}) 
        \cdots\vf_N(x_+,x_-^{(N)},\vec{x}{\,}_\bot^{(N)}) J_\mu(z) J_\nu(0) \notag\\
    \times \vf_1^+(y_+,y_-^{(1)},\vec{y}{\,}_\bot^{(1)}) \cdots 
        \vf_N^+(y_+,y_-^{(N)},\vec{y}{\,}_\bot^{(N)})) \mid 0 \ra\notag \\
    =(2\pi)^{-8N} \int \prod_{i=1}^N \Big(\frac{1}{2}\, d p_+^{(i)}\, d \vec{p}{\,}_\bot^{(i)}\Big)
        \Big(\frac{1}{2}\, d q_+^{(i)}\, d \vec{q}{\,}_\bot^{(i)}\Big)
        \wt R_{\mu\nu}([p_+^{(i)},\vec{p}{\,}_\bot^{(i)}];
            [q_+^{(i)},\vec{q}{\,}_\bot^{(i)}];z) \notag\\
    \times \exp \bigg[ -\frac{1}{2}\,(x_+P_--y_+Q_-) -
        i\sum_{i=1}^N \Big(\frac{1}{2}\,p_+^{(i)}x_-^{(i)}-\vec{p}{\,}_\bot^{(i)}\vec{x}{\,}_\bot^{(i)}\Big)
        \notag \\
    +  i\sum_{i=1}^N \Big(\frac{1}{2}\,q_+^{(i)}y_-^{(i)}-\vec{q}{\,}_\bot^{(i)}\vec{y}{\,}_\bot^{(i)}\Big)
            \bigg]. \label{3.3}
\end{gather}

Fourier transforms of $R_{\mu\nu}$ and $\wt R_{\mu\nu}$ are related to each other in the following way:
\begin{gather}
    \wt R_{\mu\nu}([p_+^{(i)},\vec{p}{\,}_\bot^{(i)}];[q_+^{(i)},\vec{q}{\,}_\bot^{(i)}];z) \notag \\
    =\int_{-\infty}^\infty \prod_{i=1}^N dp_-^{(i)}\,d q_-^{(i)} 
        \dl\bigg( P_--\sum_{i=1}^N p_-^{(i)}\bigg)
        \dl\bigg( Q_--\sum_{i=1}^N q_-^{(i)}\bigg)
        R_{\mu\nu}([p^{(i)}];[q^{(i)}];z). \label{3.4}
\end{gather}

Let us single out now the contribution of $N$-particle bound states in matrix element \eqref{3.3}, expressing the $T$-product via $\tht$-functions and using the completeness of physical states. Using the integral representation of $\tht$-function, we obtain that the quantity $\wt R_{\mu\nu}$ has the double pole singularities in the vicinity of the points corresponding to bound states with masses $M_\al$ and $M_\bt$ and a set of other quantum numbers $\al$ and $\bt$, respectively:
\begin{gather}
    \wt R_{\mu\nu}([p_+^{(i)},\vec{p}{\,}_\bot^{(i)}];[q_+^{(i)},\vec{q}{\,}_\bot^{(i)}];z) \notag \\
    \eqsim \bigg[ \frac{i}{(2\pi)^4}\bigg]^2 \,
        \frac{\Psi_P([p_+^{(i)},\vec{p}{\,}_\bot^{(i)}]) 
            \la P,\al \mid T(J_\mu(z) J_\nu(0)) \mid Q,\bt\ra 
            \Psi_Q^+([q_+^{(i)},\vec{q}{\,}_\bot^{(i)}])}
            {(P^2-M_\al^2)(Q^2-M_\bt^2)}\,. \label{3.5}
\end{gather}
Here $\Psi_P([p_+^{(i)},\vec{p}{\,}_\bot^{(i)}])$ is the three-dimensional wave function of many-body system and is defined in the following way:
\begin{multline}\label{3.6-1}
    2(2\pi)^{-4} \dl\bigg( P_+-\sum_{i=1}^N p_+^{(i)}\bigg)
        \dl^{(2)}\bigg( \vec{P}_\bot-\sum_{i=1}^N \vec{p}{\,}_\bot^{(i)}\bigg)
        \Psi_P([p_+^{(i)},\vec{p}{\,}_\bot^{(i)}]) \\
    =\int \prod_{i=1}^N d x_-^{(i)} \, d\vec{x}{\,}_\bot^{(i)}\,
        \exp \bigg[ i\sum_{i=1}^N 
        \Big(\frac{1}{2}\,p_+^{(i)}x_-^{(i)}-\vec{p}{\,}_\bot^{(i)}\vec{x}{\,}_\bot^{(i)}
            \Big)\bigg]\\
    \times \la 0 \mid \vf_1(0,x_-^{(1)}, \vec{x}{\,}_\bot^{(1)}) \cdots 
        \vf_N(0,x_-^{(N)}, \vec{x}{\,}_\bot^{(N)}) \mid P,\al \ra. 
\end{multline}

Define the three-dimensional vertex function
\begin{gather}
    \wt R_{\mu\nu}([p_+^{(i)},\vec{p}{\,}_\bot^{(i)}];[q_+^{(i)},
        \vec{q}{\,}_\bot^{(i)}];z) \notag \\
    =\int_0^{P_+} \prod_{i=1}^N d p_+^{(i)'} 
        \dl\bigg( P_+-\sum_{i=1}^N p_+^{(i)'}\bigg)
        \int \prod_{i=1}^N d \vec{p}{\,}_\bot^{(i)'} 
        \dl^{(2)}\bigg( \vec{P}_\bot-\sum_{i=1}^N \vec{p}{\,}_\bot^{(i)'}\bigg) \notag \\
    \times \int_0^{Q_+} \prod_{i=1}^N d q_+^{(i)'} 
        \dl\bigg( Q_+-\sum_{i=1}^N q_+^{(i)'}\bigg)
        \int \prod_{i=1}^N d \vec{q}{\,}_\bot^{(i)'} 
        \dl^{(2)}\bigg( \vec{Q}_\bot-\sum_{i=1}^N \vec{q}{\,}_\bot^{(i)'}\bigg) \notag \\
    \times \wt G(P;[p_+^{(i)},\vec{p}{\,}_\bot^{(i)}];[p_+^{(i)'},\vec{p}{\,}_\bot^{(i)'}])
        \wt \Gm_{\mu\nu}([p_+^{(i)'},\vec{p}{\,}_\bot^{(i)'}];
        [q_+^{(i)'},\vec{q}{\,}_\bot^{(i)'}];z) \notag\\
    \times \wt G(Q;[q_+^{(i)'},\vec{q}{\,}_\bot^{(i)'}];[q_+^{(i)},\vec{q}{\,}_\bot^{(i)}]), \label{3.6}
\end{gather}
where $\wt G$ is the three-dimensional Green function of $N$-body system, which has the following pole singularity in the vicinity of bound state $|P,\al \ra$ with mass $M_\al$ and set of other quantum numbers $\al$:
\begin{equation}\label{3.7}
    \wt G([p_+^{(i)},\vec{p}{\,}_\bot^{(i)}];[q_+^{(i)},\vec{q}{\,}_\bot^{(i)}])\big|_{P^2\to M_\al^2}
        \simeq \frac{1}{(2\pi)^4}\, \frac{\Psi_P([p_+^{(i)},\vec{p}{\,}_\bot^{(i)}])
            \Psi_Q^+([q_+^{(i)},\vec{q}{\,}_\bot^{(i)}])}{P^2-M_\al^2}\,.
\end{equation}

Taking into account the pole singularities of three-dimensional Green functions at $P^2 \to M_\al^2$ and $Q^2\to M_\bt^2$ we can present the quantity $\wt R_{\mu\nu}$ as follows:
\begin{gather}
    \wt R_{\mu\nu}([p_+^{(i)},\vec{p}{\,}_\bot^{(i)}];[q_+^{(i)},\vec{q}{\,}_\bot^{(i)}];z)
        \Big|_{\substack{P^2\to M_\al^2 \\ Q^2\to M_\bt^2}} \notag \\
    \simeq \bigg[ \frac{i}{(2\pi)^2}\bigg]^2 \,
        \frac{\Psi_P([p_+^{(i)},\vec{p}{\,}_\bot^{(i)}]) 
            \Psi_Q^+([q_+^{(i)},\vec{q}{\,}_\bot^{(i)}])}
            {(P^2-M_\al^2)(Q^2-M_\bt^2)} \notag \\
    \times \int_0^{P_+} \prod_{i=1}^N d p_+^{(i)'} 
        \dl\bigg( P_+-\sum_{i=1}^N p_+^{(i)'}\bigg)
        \int \prod_{i=1}^N d \vec{p}{\,}_\bot^{(i)'} 
        \dl^{(2)}\bigg( \vec{P}_\bot-\sum_{i=1}^N \vec{p}{\,}_\bot^{(i)'}\bigg) \notag \\
    \times \int_0^{Q_+} \prod_{i=1}^N d q_+^{(i)'} 
        \dl\bigg( Q_+-\sum_{i=1}^N q_+^{(i)'}\bigg)
        \int \prod_{i=1}^N d \vec{q}{\,}_\bot^{(i)'} 
        \dl^{(2)}\bigg( \vec{Q}_\bot-\sum_{i=1}^N \vec{q}{\,}_\bot^{(i)'}\bigg) \notag \\
    \times \Psi_P^+([p_+^{(i)'},\vec{p}{\,}_\bot^{(i)'}]) 
        \wt\Gm_{\mu\nu}([p_+^{(i)'},\vec{p}{\,}_\bot^{(i)'}];
            [q_+^{(i)'},\vec{q}{\,}_\bot^{(i)'}];z) 
            \Psi_Q([q_+^{(i)'},\vec{q}{\,}_\bot^{(i)'}]). \label{3.8}
\end{gather} 
Comparing this expression with \eqref{3.5}, we get the following expression for the matrix element of $T$-product of currents:
\begin{gather}
    \la P,\al \mid T(J_\mu(z) J_\nu(0)) \mid Q,\bt\ra \notag \\
    =\int_0^{P_+} \prod_{i=1}^N d p_+^{(i)} 
        \dl\bigg( P_+-\sum_{i=1}^N p_+^{(i)}\bigg)
        \int \prod_{i=1}^N d \vec{p}{\,}_\bot^{(i)} 
        \dl^{(2)}\bigg( \vec{P}_\bot-\sum_{i=1}^N \vec{p}{\,}_\bot^{(i)}\bigg) \notag \\
    \times \int_0^{Q_+} \prod_{i=1}^N d q_+^{(i)} 
        \dl\bigg( Q_+-\sum_{i=1}^N q_+^{(i)}\bigg)
        \int \prod_{i=1}^N d \vec{q}{\,}_\bot^{(i)} 
        \dl^{(2)}\bigg( \vec{Q}_\bot-\sum_{i=1}^N \vec{q}{\,}_\bot^{(i)}\bigg) \notag \\
    \times \Psi_P^+([p_+^{(i)},\vec{p}{\,}_\bot^{(i)}]) 
        \wt\Gm_{\mu\nu}([p_+^{(i)},\vec{p}{\,}_\bot^{(i)}];[q_+^{(i)},\vec{q}{\,}_\bot^{(i)}];z) 
            \Psi_Q([q_+^{(i)},\vec{q}{\,}_\bot^{(i)}]). \label{3.9}
\end{gather}

Fourier transform of this matrix element defines the amplitude of virtual Comptin scattering of photon with space-like momentum $q_\mu$ on the hadron with momentum $P_\mu$:
\begin{gather}
    T_{\mu\nu}(P,q)=i\int d^4 z\, e^{iqz}
        \la P,\al \mid T(J_\mu(z) J_\nu(0)) \mid Q,\bt\ra \notag \\
    =i \int_0^{P_+} \prod_{i=1}^N d p_+^{(i)} 
        \dl\bigg( P_+-\sum_{i=1}^N p_+^{(i)}\bigg)
        \int \prod_{i=1}^N d \vec{p}{\,}_\bot^{(i)} 
        \dl^{(2)}\bigg( \vec{P}_\bot-\sum_{i=1}^N \vec{p}{\,}_\bot^{(i)}\bigg) \notag \\
    \times \int_0^{Q_+} \prod_{i=1}^N d q_+^{(i)} 
        \dl\bigg( Q_+-\sum_{i=1}^N q_+^{(i)}\bigg)
        \int \prod_{i=1}^N d \vec{q}{\,}_\bot^{(i)} 
        \dl^{(2)}\bigg( \vec{Q}_\bot-\sum_{i=1}^N \vec{q}{\,}_\bot^{(i)}\bigg) \notag \\
    \times \Psi_P^+([p_+^{(i)},\vec{p}{\,}_\bot^{(i)}]) 
        \int d^4 z \, e^{iqz}\,
        \wt\Gm_{\mu\nu}([p_+^{(i)},\vec{p}{\,}_\bot^{(i)}];[q_+^{(i)},\vec{q}{\,}_\bot^{(i)}];z) 
            \Psi_Q([q_+^{(i)},\vec{q}{\,}_\bot^{(i)}]). \label{3.10}
\end{gather}

According to optical theorem the tensor $W_{\mu\nu}$ which defines the hadronic part of deep inelastic lepton-hadron scattering cross section is related to the imaginary part of the amplitude of the zero angle virtual Compton scattering in the following way:
\begin{equation}\label{3.11}
    W_{\mu\nu}(P,q)\! =\!\sum_\al \int d^4 z\, e^{iqz}
        \la P,\al \mid J_\mu(z) J_\nu(0) \mid Q,\bt\ra \!=\!
        \frac{1}{2\pi}\, \IIm T_{\mu\nu}(P,q),
\end{equation}

Taking into account the current conservation, the tensor $W_{\mu\nu}$ can be expressed via two invariant structure functions $W_1$ and $W_2$:
\begin{align}
    W_{\mu\nu} (P,q) & =\bigg( -g_{\mu\nu}+\frac{q_\mu q_\nu}{q^2}\bigg) W_1(q^2,\nu) \notag \\
    & \quad + \frac{1}{M^2} 
        \bigg( P_\mu-\frac{Pq}{q^2}\,q_\mu\bigg) 
        \bigg( P_\nu-\frac{Pq}{q^2}\,q_\nu\bigg)W_2(q^2,\nu), \label{3.12}
\end{align}
where $M\nu=Pq$, $M$ is the hadron mass. 

Thus, using Eqs. \eqref{3.10}--\eqref{3.12} one can express the structure functions of deep inelastic lepton-hadron scattering in terms of relativistic many-body wave functions, describing the internal motion of partons inside hadron, and the ``two-photon'' vertex function:
\begin{gather}
    \wt \Gm_{\mu\nu}([p_+^{(i)},\vec{p}{\,}_\bot^{(i)}];[q_+^{(i)},\vec{q}{\,}_\bot^{(i)}];q)
        =\int d^4 z\, e^{iqz} 
        \wt \Gm_{\mu\nu}([p_+^{(i)},\vec{p}{\,}_\bot^{(i)}];[q_+^{(i)},\vec{q}{\,}_\bot^{(i)}];z) \notag \\
    = \int d^4 z\, e^{iqz} 
        \int_0^{P_+} \prod_{i=1}^N d p_+^{(i)'} 
        \dl\bigg( P_+-\sum_{i=1}^N p_+^{(i)'}\bigg)
        \int \prod_{i=1}^N d \vec{p}{\,}_\bot^{(i)'} 
        \dl^{(2)}\bigg( \vec{P}_\bot-\sum_{i=1}^N \vec{p}{\,}_\bot^{(i)'}\bigg) \notag \\
    \times \int_0^{Q_+} \prod_{i=1}^N d q_+^{(i)'} 
        \dl\bigg( Q_+-\sum_{i=1}^N q_+^{(i)'}\bigg)
        \int \prod_{i=1}^N d \vec{q}{\,}_\bot^{(i)'} 
        \dl^{(2)}\bigg( \vec{Q}_\bot-\sum_{i=1}^N \vec{q}{\,}_\bot^{(i)'}\bigg) \notag \\
    \times {\wt G}^{-1}(P;[p_+^{(i)},\vec{p}{\,}_\bot^{(i)}];[p_+^{(i)'},\vec{p}{\,}_\bot^{(i)'}])
        [\wt{G \Gm_{\mu\nu}G}] ([p_+^{(i)'},\vec{p}{\,}_\bot^{(i)'}];[q_+^{(i)'},\vec{q}{\,}_\bot^{(i)'}];z)
        \notag \\
    \times {\wt G}^{-1} (Q;[q_+^{(i)'},\vec{q}{\,}_\bot^{(i)'}];[q_+^{(i)},\vec{q}{\,}_\bot^{(i)}]). \label{3.13}
\end{gather}
Here wave over the $G\Gm_{\mu\nu}G$ denotes the integration over the ``$-$'' components of 4-momenta $p^{(i)'}$ and $q^{(i)'}$. 

\section{Lowest Order in the Electromagnetic Interaction}

The ``two-photon'' vertex operator $\wt \Gm_{\mu\nu}$ can be constructed using methods of perturbation theory and expanding the functions $\wt G^{-1}$ and $\wt R_{\mu\nu}=\wt{G\Gm_{\mu\nu}G}$ in the series in coupling constant. In lowest order we obtain:
\begin{equation}\label{3.15}
    \wt G_{\mu\nu}^{(0)}=[\wt G^{(0)}]^{-1} [\wt{G^{(0)}\Gm_{\mu\nu}^{(0)} G^{(0)}}]
        \, [\wt G^{(0)}]^{-1}\,,
\end{equation}
where multiplication is understood as integration over all ``$+$'' and ``${\,}_\bot$'' components of 4-momenta $p^{(i)'}$ and $q^{(i)'}$, $\wt G^{(0)}$ is the three-dimensional Green function of $N$ free particles.

\begin{figure}
\begin{center}
\includegraphics*{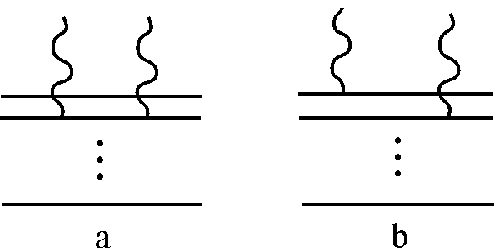}
\end{center}
\caption{}
\label{fig6}
\end{figure}

In the lowest order two types of diagrams, shown in Fig. \ref{fig6} will contribute in $\Gm_{\mu\nu}^{(0)}$. The contribution of $i$-th diagram a) is equal to 
\begin{gather}
    \wt \Gm_{\mu\nu,i}^{(0)}([p_+^{(i)},\vec{p}{\,}_\bot^{(i)}];
        [q_+^{(i)},\vec{q}{\,}_\bot^{(i)}];q) \notag\\
    =\frac{e_j^2}{2i(4\pi)^{N-1}} \,\bigg( \prod_{i=1}^N 
        \dl(p_+^{(i)}-q_+^{(i)}) \dl^{(2)}(\vec{p}{\,}_\bot^{(i)}-\vec{q}{\,}_\bot^{(i)})\bigg)
        \bigg( \prod_{i=1}^N q_+^{(i)}\bigg) \,
        \frac{1}{p_+^{(i)}(p_+^{(i)}+q_+^{(i)})} \notag \\
    \times \frac{(2\ol{p}{\,}^{(i)}+q)_\mu (2\ol{p}{\,}^{(i)}+q)_\nu}
        {P_--\sum\limits_{j=1}^N \frac{\vec{p}{\,}_\bot^{(j)2}+m_j^2}{p_+^{(j)}}+
            q_-+\frac{\vec{p}{\,}_\bot^{(i)2}+m_i^2}{p_+^{(i)}} +
            \frac{(\vec{p}{\,}_\bot^{(i)}+\vec{q}{\,}_\bot^{(i)})^2+m_i^2}{p_+^{(i)}+q_+^{(i)}}
            +i\ve}\;. \label{3.16}
\end{gather}

The contribution of diagrams of type b) is equal to: 
\begin{gather}
    \wt \Gm_{\mu\nu,b}^{(0)}([p_+^{(i)},\vec{p}{\,}_\bot^{(i)}];
        [q_+^{(i)},\vec{q}{\,}_\bot^{(i)}];q) 
        =\frac{e_i e_j}{2i(4\pi)^{N-1}} \,
        \bigg( \prod_{i=1}^N q_+^{(i)}\bigg) \,
        \frac{1}{q_+^{(i)}q_+^{(j)}} \notag \\
    \times \dl(p_+^{(i)}-q_+-q_+^{(i)}) 
        \dl^{(2)}(\vec{p}{\,}_\bot^{(i)}-\vec{q}_\bot-\vec{q}{\,}_\bot^{(i)})
        \dl(p_+^{(j)}-q_+-q_+^{(j)}) 
        \dl^{(2)}(\vec{p}{\,}_\bot^{(j)}-\vec{q}_\bot-\vec{q}{\,}_\bot^{(j)}) \notag \\
    \times \prod_{k=1}^N
        {}_{i,j}\; \dl(p_+^{(k)}-q_+^{(k)}) 
        \dl^{(2)}(\vec{p}{\,}_\bot^{(k)}-\vec{q}{\,}_\bot^{(k)})\,
        \frac{(2\ol{p}{\,}^{(i)}+q)_\mu (2\ol{p}{\,}^{(i)}+q)_\nu}
        {q_--\frac{\vec{p}{\,}_\bot^{(i)2}+m_i^2}{p_+^{(i)}} -
            \frac{\vec{q}{\,}_\bot^{(i)2}+m_i^2}{q_+^{(i)}}+i\ve}  \notag \\
    \times \Bigg[ 
        \frac{Q_--\sum\limits_{j=1}^N \frac{\vec{q}{\,}_\bot^{(j)2}+m_j^2}{q_+^{(j)}}+i\ve}
            {Q_--\sum\limits_{j=1}^N \frac{\vec{q}{\,}_\bot^{(j)2}+m_j^2}{q_+^{(j)}}-q_--
            \frac{\vec{p}{\,}_\bot^{(i)2}+m_i^2}{p_+^{(i)}}+
            \frac{\vec{q}{\,}_\bot^{(i)2}+m_i^2}{q_+^{(i)}}+i\ve} \notag \\
    - \frac{P_--\sum\limits_{j=1}^N \frac{\vec{p}{\,}_\bot^{(j)2}+m_j^2}{p_+^{(j)}}+i\ve}
            {P_--\sum\limits_{j=1}^N \frac{\vec{p}{\,}_\bot^{(j)2}+m_j^2}{p_+^{(j)}}+q_--
            \frac{\vec{p}{\,}_\bot^{(i)2}+m_i^2}{p_+^{(i)}}-
            \frac{\vec{q}{\,}_\bot^{(i)2}+m_i^2}{q_+^{(i)}}+i\ve} 
            \Bigg]. \label{3.17}
\end{gather}

Assuming that the partons constituting hadron are on the mass shell and neglecting small terms of order $P_--\sum\limits_{i=1}^N (\vec{p}{\,}_\bot^{(i)2}+m_i^2)/p_+^{(i)}$, we obtain that diagrams of type b) do not give contribution to $\wt\Gm_{\mu\nu}^{(0)}$ and contribution of $i$-th diagram of type a) is equal to 
\begin{gather}
    \wt \Gm_{\mu\nu,i}^{(0)}([p_+^{(i)},\vec{p}{\,}_\bot^{(i)}];
        [q_+^{(i)},\vec{q}{\,}_\bot^{(i)}];q) 
    =\frac{e_i^2}{2i(4\pi)^{N-1}} \, 
        \frac{(2\ol{p}{\,}^{(i)}+q)_\mu (2\ol{p}{\,}^{(i)}+q)_\nu}
            {p_+^{(i)}[(\ol{p}{\,}^{(i)}+q)^2-m_i^2+i\ve]} \,\notag \\
    \times \prod_{i=1}^N p_+^{(i)} \dl(p_+^{(i)}-q_+^{(i)})
            \dl^{(2)} (\vec{p}{\,}_\bot^{(i)}-\vec{q}{\,}_\bot^{(i)})\,, \label{3.18}
\end{gather}
where $e_i$ os the electric charge of $i$-th parton, and $\ol{p}{\,}^{(i)}$ is the momentum of parton on the mass shell:
$$
    \ol{p}{\,}^{(i)}=\bigg( \frac{\vec{p}{\,}_\bot^{(i)2}+m_i^2}{p_+^{(i)}}\,, 
        p_+^{(i)}, \vec{p}{\,}_\bot^{(i)}\bigg).
$$

Summing the contributions of all diagrams of type a), inserting into the expression \eqref{3.10} for virtual Compton scattering amplitude and extracting the imaginary part of the expression obtained, we get for tensor $W_{\mu\nu}$:
\begin{align}
    W_{\mu\nu}(P,q) & = \frac{1}{4(4\pi)^{N-1}} \int_0^{P_+} 
        \prod_{i=1}^N d p_+^{(i)} 
        \dl\bigg( P_+-\sum_{i=1}^N p_+^{(i)}\bigg) \notag \\
    & \quad \times \int \prod_{i=1}^N d \vec{p}{\,}_\bot^{(i)} 
        \dl^{(2)}\bigg( \vec{P}_\bot-\sum_{i=1}^N \vec{p}{\,}_\bot^{(i)}\bigg)
        \bigg(\prod_{i=1}^N d p_+^{(i)} \bigg)
        \big| \Psi_P([p_+^{(i)},\vec{p}{\,}_\bot^{(i)}])\big|^2 \notag \\
    & \quad \times \sum_{i=1}^N 
        \bigg\{ \frac{e_i^2}{p_+^{(i)}}\, (2\ol{p}{\,}^{(i)}+q)_\mu (2\ol{p}{\,}^{(i)}+q)_\nu
        \dl [(\ol{p}{\,}^{(i)}+q)^2-m_i^2]\bigg\}. \label{3.19}
\end{align}

Introducing the wave function $\Phi_P([x^{(i)}, \vec{p}{\,}_\bot^{(i)}])$ (see Eq. \eqref{1.41}), we can rewrite the tensor $W_{\mu\nu}$ in the form:
\begin{align}
    W_{\mu\nu}(P,q) & = \frac{1}{4(4\pi)^{N-1}} \int_0^1 
        \prod_{i=1}^N d  \frac{dx^{(i)}}{x^{(i)}} \,
        \dl\bigg( 1-\sum_{i=1}^N x^{(i)}\bigg) \notag \\
    & \quad \times \int \prod_{i=1}^N d \vec{p}{\,}_\bot^{(i)} 
        \dl^{(2)}\bigg( \vec{P}_\bot-\sum_{i=1}^N \vec{p}{\,}_\bot^{(i)}\bigg)
        \big| \Phi_P([x^{(i)},\vec{p}{\,}_\bot^{(i)}])\big|^2 \notag \\
    & \quad \times \sum_{i=1}^N 
        \bigg\{ \frac{e_i^2}{x^{(i)}}\, (2\ol{p}{\,}^{(i)}+q)_\mu (2\ol{p}{\,}^{(i)}+q)_\nu
        \dl [(\ol{p}{\,}^{(i)}+q)^2-m_i^2]\bigg\}. \label{3.20}
\end{align} 

Expressions for the structure functions $W_1$ and $W_2$ can be obtained multiplicating the tensor $W_{\mu\nu}$ by the ``projection'' operators $L_{\mu\nu}^{(i)}$:
\begin{equation}\label{3.21}
    L_{\mu\nu}^{(i)} Q_{\mu\nu}=W_i, \quad i=1,2.
\end{equation}

Using the following expressions for these ``projection'' operators
\addtocounter{equation}{1}
\begin{align}
    L_{\mu\nu}^{(1)} & = \frac{1}{2}\,\bigg[ -g_{\mu\nu} +
        \frac{P_\mu P_\nu}{M^2(1-\nu^2/q^2)} \bigg], \tag{\theequation a} \label{3.22a} \\
    L_{\mu\nu}^{(2)} & = \frac{1}{2(2-\nu^2/q^2)}\,\bigg[ -g_{\mu\nu} +
        \frac{3P_\mu P_\nu}{M^2(1-\nu^2/q^2)} \bigg], \tag{\theequation b} \label{3.22b} 
\end{align}
for the structure functions $W_1$ and $\nu W_2$ we obtain:
\addtocounter{equation}{1}
\begin{align}
    W_1& (q^2,\nu)  =\frac{1}{8(4\pi)^{N-1}} \int_0^1 \prod_{i=1}^N 
        \frac{dx^{(i)}}{x^{(i)}} \,
        \dl\bigg( 1-\sum_{i=1}^N x^{(i)}\bigg) \notag \\
    & \times \int \prod_{i=1}^N d \vec{p}{\,}_\bot^{(i)} 
        \dl^{(2)}\bigg( \vec{P}_\bot-\sum_{i=1}^N \vec{p}{\,}_\bot^{(i)}\bigg)
        \big| \Phi_P([x^{(i)},\vec{p}{\,}_\bot^{(i)}])\big|^2 \notag \\
    & \times \sum_{i=1}^N 
        \bigg\{ \frac{e_i^2}{x^{(i)}}\, \bigg[ \frac{(M\nu+2\ol{p}{\,}^{(i)}P)^2}{M^2(1-\nu^2/q^2)}
        -(4m_i^2-q^2)\bigg] \dl (q^2+2\ol{p}{\,}^{(i)}q)\bigg\}, 
            \tag{\theequation a} \label{3.23a}  \\
    \nu W_2&(q^2,\nu)  =\frac{1}{8(4\pi)^{N-1}} \,\frac{\nu}{(1-\nu^2/q^2)} 
        \int_0^1 \prod_{i=1}^N 
        \frac{dx^{(i)}}{x^{(i)}} \,
        \dl\bigg( 1-\sum_{i=1}^N x^{(i)}\bigg) \notag \\
    & \times \int \prod_{i=1}^N d \vec{p}{\,}_\bot^{(i)} 
        \dl^{(2)}\bigg( \vec{P}_\bot-\sum_{i=1}^N \vec{p}{\,}_\bot^{(i)}\bigg)
        \big| \Phi_P([x^{(i)},\vec{p}{\,}_\bot^{(i)}])\big|^2 \notag \\
    & \times \sum_{i=1}^N 
        \bigg\{ \frac{e_i^2}{x^{(i)}}\, \bigg[ \frac{3(M\nu+2\ol{p}{\,}^{(i)}P)^2}{M^2(1-\nu^2/q^2)}
        -(4m_i^2-q^2)\bigg] \dl (q^2+2\ol{p}{\,}^{(i)}q)\bigg\}.
            \tag{\theequation b} \label{3.23b} 
\end{align}

For further consideration we proceed to the frame, where the virtual photon and hadron are moving along the $z$ axis:
$$
    P=(P_-,P_+,\vec{0}_\bot), \quad q=(q_-, q_+, \vec{0}_\bot).
$$
In this frame 
$$
    2\ol{p}{\,}^{(i)}P =x^{(i)} M^2 +\frac{\vec{p}{\,}_\bot^{(i)2}+m_i^2}{x^{(i)}}
$$
and $\dl$-function in \eqref{3.23a}, \eqref{3.23b} can be rewritten in the form:
\begin{equation}\label{3.24}
    \dl(q^2+2\ol{p}{\,}^{(i)}P) =\frac{1}{\xi}\, 
        \dl\bigg[ \frac{\vec{p}{\,}_\bot^{(i)2}+m_i^2}{x^{(i)}} +
            \frac{Q^2(\xi-x^{(i)})}{\xi^2}\bigg].
\end{equation}
Here we have introduced the variables $Q^2=-q^2$ and 
\begin{equation}\label{3.25}
    \xi=-\frac{q_+}{P_+} =\frac{Q^2}{M(\nu+\sqrt{\nu^2+Q^2}\,)}\,. 
\end{equation}

Scale properties of structure functions with respect of variables $\xi$ are discussed in several papers \cite{99}, \cite{101}--\cite{104}. The Nachtman variables $\xi$ is the generalization of usual Bjorken variable $x_B$ taking into account the hadron mass and is related to $x_B$ by the following relation:
\begin{equation}\label{3.26}
    \xi=\frac{2x_B}{1+\sqrt{1+\frac{4Mx_B}{Q^2}}}\;.
\end{equation}

Expressing the kinematic variable $\nu$ via the variables $Q^2$ and $\xi$ we obtain the following expressions for structure functions:
\addtocounter{equation}{+1}
\begin{gather}
    W_1(Q^2,\xi) =\frac{1}{2(4\pi)^{N-1}\xi} \int_0^1 \prod_{i=1}^N 
        \frac{dx^{(i)}}{x^{(i)}} \,
        \dl\bigg( 1-\sum_{i=1}^N x^{(i)}\bigg) \notag \\
    \times \int \prod_{i=1}^N d \vec{p}{\,}_\bot^{(i)} 
        \dl^{(2)}\bigg( \vec{P}_\bot-\sum_{i=1}^N \vec{p}{\,}_\bot^{(i)}\bigg)
        \big| \Phi_P([x^{(i)},\vec{p}{\,}_\bot^{(i)}])\big|^2 \notag \\
    \times \sum_{i=1}^N 
        \bigg\{ e_i^2 \bigg[ \frac{Q^2x^{(i)}(x^{(i)}-\xi)}{\xi^2}-m_i^2\bigg] 
        \dl \bigg[ \vec{p}{\,}_\bot^{(i)2} +m_i^2 -
            \frac{Q^2x^{(i)}(x^{(i)}-\xi)}{\xi^2} \bigg] \bigg\},
            \tag{\theequation a} \label{3.27a}  \\
    \nu W_2(Q^2,\xi) =\frac{MQ^2}{2(4\pi)^{N-1}\xi^2} \frac{(Q^2/\xi^2-M^2)}{(Q^2/\xi^2+M^2)^2}
        \int_0^1 \prod_{i=1}^N 
        \frac{dx^{(i)}}{x^{(i)}} \,
        \dl\bigg( 1-\sum_{i=1}^N x^{(i)}\bigg) \notag \\
    \times \int \prod_{i=1}^N d \vec{p}{\,}_\bot^{(i)} 
        \dl^{(2)}\bigg( \vec{P}_\bot-\sum_{i=1}^N \vec{p}{\,}_\bot^{(i)}\bigg)
        \big| \Phi_P([x^{(i)},\vec{p}{\,}_\bot^{(i)}])\big|^2 \notag \\
    \times \sum_{i=1}^N 
        \bigg\{ e_i^2 \bigg[ \frac{6Q^2x^{(i)}(x^{(i)}-\xi)}{\xi^2}+Q^2-2m_i^2\bigg] \notag \\
    \times \,\dl \bigg[ \vec{p}{\,}_\bot^{(i)2} +m_i^2 -
            \frac{Q^2x^{(i)}(x^{(i)}-\xi)}{\xi^2} \bigg] \bigg\}.
            \tag{\theequation b} \label{3.27b} 
\end{gather}

If one neglects the masses and transverse momenta of partons $(m_i^2 \ll Q^2$, $\vec{p}{\,}_\bot^{(i)2} \ll Q^2)$, the $\dl$-function takes the form
$$
    \dl(2\ol{p}{\,}^{(i)}q-Q^2)=\frac{\xi}{Q^2}\, \dl(x^{(i)}-\xi).
$$
Then the structure function $W_1$ vanishes and for the structure function $W_2$ we obtain
\begin{gather}
    \nu W_2(Q^2,\xi)=\frac{MQ^2}{2(4\pi)^{N-1}\xi} \frac{(Q^2/\xi^2-M^2)}{(Q^2/\xi^2+M^2)^2}
        \int_0^1 \prod_{i=1}^N 
        \frac{dx^{(i)}}{x^{(i)}} \,
        \dl\bigg( 1-\sum_{i=1}^N x^{(i)}\bigg) \notag \\
    \times \int \prod_{i=1}^N d \vec{p}{\,}_\bot^{(i)} 
        \dl^{(2)}\bigg(\sum_{i=1}^N \vec{p}{\,}_\bot^{(i)}\bigg)
        \big| \Phi_P([x^{(i)},\vec{p}{\,}_\bot^{(i)}])\big|^2 
        \sum_{i=1}^N e_i^2 \dl (x^{(i)}-\xi). \label{3.28}
\end{gather} 

In the asymptotic limit $(\nu,Q^2\gg M^2$, $x_B$ is fixed) the variable $\xi$ coincides with the Bjorken variable $x_B$ and we obtain that in this limit the structure function $\nu W_2$ is scale-invariant with respect to the variable $x_B$:
\begin{gather}
    \nu W_2(x_B)=\frac{Mx_B}{2(4\pi)^{N-1}} 
        \int_0^1 \prod_{i=1}^N 
        \frac{dx^{(i)}}{x^{(i)}} \,
        \dl\bigg( 1-\sum_{i=1}^N x^{(i)}\bigg) \notag \\
    \times \int \prod_{i=1}^N d \vec{p}{\,}_\bot^{(i)} 
        \dl^{(2)}\bigg(\sum_{i=1}^N \vec{p}{\,}_\bot^{(i)}\bigg)
        \big| \Phi_P([x^{(i)},\vec{p}{\,}_\bot^{(i)}])\big|^2 
        \sum_{i=1}^N e_i^2 \dl (x^{(i)}-x_B). \label{3.29}
\end{gather} 

Normalization condition for wave function $\Phi([x^{(i)},\vec{p}{\,}_\bot^{(i)}])$ is of the form 
\begin{gather}
    iP_+\int_0^1 \prod_{i=1}^N 
        \frac{dx^{(i)}}{x^{(i)}} \,
        \dl\bigg( 1-\sum_{i=1}^N x^{(i)}\bigg) 
        \int \prod_{i=1}^N d \vec{p}{\,}_\bot^{(i)} 
        \dl^{(2)}\bigg(\vec{P}{\,}_\bot^{(i)}-\sum_{i=1}^N \vec{p}{\,}_\bot^{(i)}\bigg) \notag \\
    \times \int_0^1 \prod_{i=1}^N 
        \frac{dx^{(i)'}}{x^{(i)'}} \,
        \dl\bigg( 1-\sum_{i=1}^N x^{(i)'}\bigg) 
        \int \prod_{i=1}^N d \vec{p}{\,}_\bot^{(i)'} 
        \dl^{(2)}\bigg(\vec{P}{\,}_\bot^{(i)}-\sum_{i=1}^N \vec{p}{\,}_\bot^{(i)'}\bigg) \notag \\
    \times \Phi_P([x^{(i)},\vec{p}{\,}_\bot^{(i)}])\, 
        \frac{\pa \wt G^{-1}(P;[p_+^{(i)}, \vec{p}{\,}_\bot^{(i)}]; [p_+^{(i)'},\vec{p}{\,}_\bot^{(i)'}])}
            {\pa P^2}\,\Phi_P([x^{(i)'},\vec{p}{\,}_\bot^{(i)'}])=1. \label{3.30}
\end{gather}

Assuming that the interaction kernel does not depend on the total energy and using explicit expression for the Green function of $N$ free particles we obtain the following sum rule:
\begin{equation}\label{3.31}
    \int_0^1 \frac{\nu W_2(x_B)}{Mx_B}\,d x_B =\sum_{i=1}^N e_i^2.
\end{equation}

\section{Model Parametrizations of Wave Functions}

Let us consider now the case, when the hadron consists of two constituents. This case corresponds to meson, which consists of quark and antiquark. We will neglect contributions of gluons and quark-antiquark sea. 

Expressions for structure functions in the case $N=2$ have the form:
\addtocounter{equation}{+1}
\begin{align}
    W_1(Q^2,& \xi)  =\frac{e_1^2+e_2^2}{8\pi \xi} \int_a^1 \frac{dx}{x(1-x)}\,
        \int d\vec{p}_\bot |\Phi_P(x,\vec{p}_\bot)|^2 \notag \\
    &  \times \bigg[ \frac{Q^2 x(x-\xi)}{\xi^2}-m^2 \bigg]\,
        \dl \bigg[ \vec{p}{\,}_\bot^2+m^2-\frac{Q^2 x(x-\xi)}{\xi^2} \bigg],
        \tag{\theequation a} \label{3.32a} \\
    \nu W_w(Q^2,& \xi)  =\frac{e_1^2+e_2^2}{8\pi \xi} \, MQ^2\, 
        \frac{Q^2/\xi^2-M^2}{(Q^2/\xi^2+M^2)^2}  \notag \\
    &  \times  \int_a^1 \frac{dx}{x(1-x)}\,
        \int d\vec{p}_\bot |\Phi_P(x,\vec{p}_\bot)|^2 \notag \\
    &  \times \bigg[ \frac{6Q^2 x(x-\xi)}{\xi^2}+Q^2-2m^2 \bigg]\,
        \dl \bigg[ \vec{p}{\,}_\bot^2+m^2-\frac{Q^2 x(x-\xi)}{\xi^2} \bigg].
        \tag{\theequation b} \label{3.32b}
\end{align}
Here we assume that masses of constituents are equal to each other: $m_1=m_2=m$.

In \eqref{3.32a}, \eqref{3.32b} the limit of integration over $x$ is defined from the $\dl$-function and equals to:
\begin{equation}\label{3.33}
    a=\frac{\xi}{2}\,\bigg( 1+\sqrt{1+\frac{4m^2}{Q^2}}\,\bigg).
\end{equation}

Neglecting masses and transverse momenta of quarks we obtain that the structure function $W_1$ vanishes and the structure function $\nu W_2$ takes the following form:
\begin{equation}\label{3.34}
    \nu W_2(Q^2,\xi)=\frac{(e_1^2+e_2^2)MQ^2}{8\pi \xi(1-\xi)} \,
        \frac{Q^2/\xi^2-M^2}{(Q^2/\xi^2+M^2)^2}  
        \int d\vec{p}_\bot |\Phi_P(x,\vec{p}_\bot)|^2 . 
\end{equation}

If we choose the following parametrization for wave function $\Phi_P$
\begin{equation}\label{3.35}
    \Phi_P(x,\vec{p}_\bot) =C\bigg[ \frac{\vec{p}{\,}_\bot^2+m^2}{x(1-x)}-\al\bigg]^{-n}\,, 
\end{equation}
then for the structure function $\nu W_2$ we get 
\begin{gather}
    \nu W_2(Q^2,\xi)=\frac{(e_1^2+e_2^2)MQ^2}{8} \,
        \frac{Q^2/\xi^2-M^2}{(Q^2/\xi^2+M^2)^2}  \notag \\
    \times \frac{|C|^2 \xi^{2n-1}(1-\xi)^{2n-1}}{(2n-1) [m^2-\al\xi(1-\xi)]^{2n-1}}\,.\label{3.36}
\end{gather}

In the Bjorken limit $(Q^2\gg m^2$, $\xi\to x_B)$ for $n=1$ we obtain \cite{105}:
\begin{equation}\label{3.37}
    \nu W_2(x_B)=\frac{e_1^2+e_2^2}{8} \,M|C|^2\,
        \frac{x_B(1-x_B)}{m^2-\al x_B(1-x_B)}\,.
\end{equation}

Let us consider now another parametrization of the wave function 
\begin{equation}\label{3.38}
    \Phi_P(x,\vec{p}_\bot) =C\exp \bigg[ -\bt\,\frac{\vec{p}{\,}_\bot^2+m^2}{x(1-x)}\bigg].
\end{equation}

Inserting this wave function into \eqref{3.32a} and \eqref{3.32b} and neglecting quark masses $(m^2\ll Q^2)$, for the structure functions we get:
\addtocounter{equation}{+1}
\begin{align}
    W_1(Q^2,\xi) & =\frac{(e_1^2+e_2^2)Q^2(1-\xi)}{8 \xi^3}\,|C|^2 \notag \\
    & \times \bigg\{ -1-\bigg( 1+2\bt\,\frac{Q^2}{\xi^2}\bigg) 
        \exp \bigg( 2\bt\frac{Q^2}{\xi^2}\bigg) 
        Ei\bigg( -2\bt\frac{Q^2}{\xi^2}\bigg) \bigg\},
        \tag{\theequation a} \label{3.39a} \\
    \nu W_2(Q^2,\xi) & =\frac{6MQ^2}{\xi}\, 
        \frac{Q^2/\xi^2-M^2}{(Q^2/\xi^2+M^2)^2} \, W_1(Q^2,\xi) \notag \\
    & +\frac{(e_1^2+e_2^2) MQ^4}{4 \xi^2} \, 
        \frac{Q^2/\xi^2-M^2}{(Q^2/\xi^2+M^2)^2}\,|C|^2 \notag \\
    & \times \bigg\{ -\exp \bigg( 2\bt\frac{Q^2}{\xi^2}\bigg) 
        Ei\bigg( -2\bt\frac{Q^2}{\xi^2}\bigg) \bigg\}.
        \tag{\theequation b} \label{3.39b}
\end{align}
Here $Ei$ is the integral exponent function.

Note that more realistic consideration needs inclusion of spin and color degrees of freedom of constituents, which can be done directly by means of corresponding formalism, developed in the previous sections.

\section{Quark Degrees of Freedom in Nuclei and the EMC-Effect}

As has been mentioned in the Introduction, recent investigations have led to the conclusion that the consideration of nuclei as the systems of quasi-independent nonrelativistic nucleons is incomplete and they require a relativistic description of the internal motion of nucleons in nuclei and taking into account quark degrees of freedom. These are first of all the prediction \cite{106} and observation \cite{107}--\cite{109} of the cumulative production of particles in hadron-nucleus and nucleus-nucleus collisions and change of an exponential fall-off of form-factors of light nuclei at small momentum transfers to a power-law fall-off at large momentum transfers \cite{110}, \cite{111} according to the quark counting rules \cite{86}, \cite{87}.

In this sense the most impressive are the results of deep inelastic lepton-nucleus scattering experiments \cite{112}--\cite{116} (the so-called EMC-effect). Different models are suggested for the explanation of this phenomenon (see, e.g., Refs. \cite{117}--\cite{ 131}). It seems that the above mentioned regularities are of the same nature and are determined by the possibility of formation of multiquark configurations (multiquark bags) in nuclei. Here we analyze the EMC-effect and show that the effect can be explained by taking into account the scattering on colorless multiquark configurations which are contained in the medium of spectator nucleons. 

Let us consider deep inelastic scattering of charged lepton on a nucleus $A$. We shall assume that in the nucleus, together with nucleons (three-quark bags), the configurations with six, nine, etc. quarks  are formed with definite probabilities and leptons interact with the nucleus by the exchange of virtual photons with quarks from these bags. Then the nucleus structure function can be represented by the sum:
\begin{equation}\label{3.40}
    F_2(x,Q^2) =\sum_{K=1}^A N(A,K) F_2^K(x,Q^2),
\end{equation}
where $F_2^K$ is the structure function of the nucleus $A$ which contains a $3K$-quark bag and $(A-K)$ nucleons. The coefficients $N(A,K)$ of these structure function have the meaning of the effective number of $3K$-quark bags in the nucleus $A$ and obey the normalization condition:
\begin{equation}\label{3.41}
    \sum_{K=1}^A KN(A,K)=A.
\end{equation}

We use the parametrization of $N(A,K)$ in the form of the Bernoulli distribution:
\begin{equation}\label{3.42}
    N(A,K)=\frac{A!}{K!(A-K)!}\, p(A)^{K-1} 
        [1-p(A)]^{A-K}.
\end{equation}

For the parameter $p(A)$ determining the probability of a three-quark nucleon to get a $3K$-quark bag we consider two possibilities:

1) $p(A)$ is determined by the ratio of the bag and nucleus volumes:
\begin{equation}\label{3.43}
    p(A)=\frac{V_K}{V_A}=\frac{r_K^3}{R_A^3} =
        \frac{r_K^3}{(R_0A^{1/3})^3} \sim A^{-1}.
\end{equation}

Taking the bag radius $r_K$ close to the nucleon radius $r_K\approx 0,8$~fm and $R_0=1,4$~fm \cite{132} we obtain $p(A)=1,1867\,A^{-1}$.

2) $p(A)$ is determined by the ratio of the bag and nucleus cross sections:
\begin{equation}\label{3.44}
    p(A)\sim \frac{r_K^2}{R_A^2} \sim A^{-2/3}.
\end{equation}

The coefficient of proportionality in \eqref{3.44} for this case has been obtained in Ref. \cite{133} by fitting the data on production of $\pi$-mesons with large transverse momenta in photon-nucleus scattering, and we shall also use this parametrization $p(A)=0,085\,A^{-0,57}$.

In both cases $N(A,K)$ are rapidly decreasing functions of $K$ and the main contribution to the structure function is given by first few terms of the sum \eqref{3.40}.

We proceed now to calculation of structure functions $F_2^K$. As it is well-known, the structure function $F_2$ appears in the decomposition of the deep inelastic tensor $W_{\mu\nu}$ into gauge-invariant structures. The tensor $W_{\mu\nu}$ itself is proportional to the imaginary part of the forward virtual Compton scattering amplitude:
\begin{equation}\label{3.45}
    T_{\mu\nu}(P_A,q)=i\int d^4 z\, \exp(iqz) 
        \la P_A \mid T\{J_\mu(z)J_\nu(0)\} \mid P_A\ra
\end{equation}
that can be expressed through the two-photon vertex function and relativistic wave functions of composite systems:
\begin{gather}
    T_{\mu\nu}(P_A,q) =i\int_0^{P_{A,+}} 
        \prod_{i=K+1}^A d p_+^{(i)} d P_{K,+}
        \int \prod_{i=K+1}^A d \vec{p}{\,}_\bot^{(i)} d \vec{P}_{K,\bot} \notag \\
    \times \dl \bigg( P_{A,+} -P_{K,+} -\sum_{i=K+1}^A p_+^{(i)}\bigg)
        \dl^{(2)} \bigg( \vec{P}_{A,\bot} -\vec{P}_{K,\bot} -
            \sum_{i=K+1}^A \vec{p}{\,}_\bot^{(i)}\bigg)\notag\\
    \times \int_0^{Q_{A,+}} 
        \prod_{i=K+1}^A d q_+^{(i)} d Q_{K,+}
        \int \prod_{i=K+1}^A d \vec{q}{\,}_\bot^{(i)} d \vec{Q}_{K,\bot} \notag \\
    \times \dl \bigg( Q_{A,+} -Q_{K,+} -\sum_{i=K+1}^A q_+^{(i)}\bigg)
        \dl^{(2)} \bigg( \vec{Q}_{A,\bot} -\vec{Q}_{K,\bot} -
            \sum_{i=K+1}^A \vec{q}{\,}_\bot^{(i)}\bigg)\notag\\
    \times \Psi_{P_A}^+(P_{K,+} \vec{P}_{K,\bot},[p_+^{(i)}, \vec{p}{\,}_\bot^{(i)}])
        \Gm_{\mu\nu}(P_{K,+} \vec{P}_{K,\bot},[p_+^{(i)}, \vec{p}{\,}_\bot^{(i)}]; q; 
            [q_+^{(i)}, \vec{q}{\,}_\bot^{(i)}],Q_{K,+}, \vec{Q}_{K,\bot}) \notag \\
    \times \Psi_{Q_A}(Q_{K,+} \vec{Q}_{K,\bot},[q_+^{(i)}, \vec{q}{\,}_\bot^{(i)}]). \label{3.46}
\end{gather}
Here $\Gm_{\mu\nu}$ is the two=particle vertex functions, $\Psi_{P_A}$ is the relativistic wave function of the composite system consisting of a $3K$-quark bag and $(A-K)$ nucleons, $P_{A,\mu}$ $(\mu=0,1,2,3)$ is the 4-momentum of nucleus $A$, $P_{K,\mu}$ and $Q_{K,\mu}$ are the 4-momenta of the $3K$-quark bag, $p_\mu^{(i)}$, $q_\mu^{(i)}$ are the 4-momenta of nucleons. The square brackets in the arguments of $\Psi_{P_A}$ and $\Gm_{\mu\nu}$ denote the sets of corresponding variables:
\begin{equation}\label{3.47}
    [p_+^{(i)}, \vec{p}{\,}_\bot^{(i)}]=p_+^{(K+1)}, \vec{p}{\,}_\bot^{(K+1)};
        p_+^{(K+2)}, \vec{p}{\,}_\bot^{(K+2)};\dots; p_+^{(A)}, \vec{p}{\,}_\bot^{(A)}.
\end{equation}

Representing now the relativistic wave function $\Psi_{P_A}(P_{K,+},\vec{P}_{K,\bot},[p_+^{(i)}, \vec{p}{\,}_\bot^{(i)}])$ in the form 
\begin{gather}
    \Psi_{P_A}(P_{K,+},\vec{P}_{K,\bot},[p_+^{(i)}, \vec{p}{\,}_\bot^{(i)}]) =
        \Psi_{A-K}([p_+^{(i)}, \vec{p}{\,}_\bot^{(i)}]) \notag \\
    \times \int \prod_{j=1}^{3K} dk_{j,+} 
        \dl\bigg( P_{K,+}-\sum_{j=1}^{3K} k_{j,+}\bigg) \notag \\
    \times \int \prod_{j=1}^{3K} d\vec{k}_{j,\bot} 
        \dl^{(2)}\bigg( \vec{P}_{K,\bot}-\sum_{j=1}^{3K} \vec{k}_{j,\bot}\bigg)
        \Psi_{3K}([k_{j,+},\vec{k}_{j,\bot}]), \label{3.48}
\end{gather}
where $k_{j,+}$ and $\vec{k}_{j,1}$ are components of the quark momentum in the $3K$-quark bag, and calculating the function $\Gm_{\mu\nu}$ in the lowest order in electromagnetic interaction, we get the following expression for the structure function $F_2^K$:
\begin{gather}
    F_2^K(x,Q^2) =\frac{\pi}{(4\pi)^{A+2K}} \,
        \frac{M_AQ^2(Q^2/\xi_A^2-M_A^2)}{\xi_A^2(Q^2/\xi_A^2-M_A^2)} \notag \\
    \times \int_0^1 \prod_{i=K+1}^A \frac{x_i}{x_i}\, Z_k\, dZ_k
        \dl\bigg( 1-Z_k-\sum_{i=K+1}^A x_i\bigg) \notag \\
    \times \int \prod_{i=K+1}^A d\vec{p}_{i,\bot} d\vec{P}_{K,\bot}
        \dl^{(2)} \bigg( \vec{P}_{K,\bot}+\sum_{i=K+1}^A \vec{p}_{i,\bot}\bigg)
        |\Phi_{A-K}([x_i,\vec{p}_{i,\bot}])|^2 \notag \\
    \times \sum_{j=1}^{3K} e_j^2 \int_0^1 \prod_{l=1}^{3K} \frac{dz_l}{z_l}\,
        \dl\bigg( 1-\sum_{l=1}^{3K} z_l\bigg) \notag \\
    \times \int \prod_{l=1}^{3K} d\vec{k}_{l,\bot} 
        \dl^{(2)}\bigg( \vec{P}_{K,\bot}-\sum_{l=1}^{3K} \vec{k}_{l,\bot}\bigg)
        |\Phi_{3K}([z_l,\vec{k}_{l,\bot}])|^2 \notag \\
    \times \bigg[ Q^2+2m_j^2+\frac{6Q^2}{(\xi_A/Z_K)^2}\, z_j(z_j-\xi_A/Z_K)\bigg] \notag \\
    \times \dl^{(2)}\bigg[ \vec{k}{\,}_{l,\bot}^2 +m_j^2 -
        \frac{Q^2z_j(z_j-\xi_A/Z_k)}{(\xi_A/Z_K)^2}\bigg]. \label{3.49}
\end{gather}
Here $q^2=-Q^2$ is the 4-momentum transfer squared. The variable $\xi_A$ is defined as 
\begin{equation}\label{3.50}
    \xi_A=\frac{2x_A}{1+(1+4M_A^2x_A^2/Q^2)^{1/2}}=z_jZ_K=\frac{k_{j,+}}{P_{A,+}}\,,
\end{equation}
where $x_A=Q^2/2M_A\nu$, $\nu=(P_Aq)/M_A$, $M_A$ is the nucleus mass. The variable $x_A$ varies in the interval $0<x_A<1$ and is related to the Bjorken variable $x=Q^2/2M\nu$ by $x_A=(M/M_A)x$ ($M$ is the nucleon mass). It is evident that $0<x<M_A/M\approx A$ and 
\begin{equation}\label{3.51}
    \xi_A=\frac{M}{M_A}\,\xi\approx \frac{\xi}{A}\,, \quad 
        \xi=\frac{2x}{1+(1+4M^2x^2/Q^2)^{1/2}}\,.
\end{equation}

In \eqref{3.49} $e_j$ and $m_j$ are the electric charge and the mass of an $j$-th quark, respectively. 

The wave functions $\Phi_{A-K}([x_i,\vec{p}_{i,\bot}])$ and $\Phi_{3K}([z_j,\vec{k}_{j,\bot}])$ are related to the functions $\Psi_{A-K}([\ol{p}{\,}_i])$ and $\Psi_{3K}([\ol{k}{\,}_j])$ by the following formulas:
\addtocounter{equation}{+1}
\begin{align}
    \Phi_{A-K}([x_i,\vec{p}_{i,\bot}]) & =(P_{A,+})^{A-K} 
        \bigg( \prod_{i=K+1}^A x_i\bigg) \Psi_{A-K} ([\ol{p}{\,}_i])
        \tag{\theequation a} \label{3.52a} \\
    \Phi_{3K}([z_j,\vec{k}_{j,\bot}]) & =(P_{K,+})^{3K-1} 
        \bigg( \prod_{j=1}^{3K} z_j\bigg) \Psi_{3K} ([\ol{k}{\,}_j]).
        \tag{\theequation b} \label{3.52b}
\end{align}

The variables $x_i$ and $z_j$ are defined as:
\begin{align*}
    x_i & =\frac{p_{i,+}}{P_{A,+}}\,, \quad 0<x_i<1, \quad 
        \sum_{i=K+1}^A x_i=1-Z_K, \\
    z_j & =\frac{k_{j,+}}{P_{K,+}}\,, \quad 0<z_j<1, \quad 
        \sum_{j=1}^{3K} z_j=1.
\end{align*}

The variable $Z_K$ is the ratio of ``$+$''-components of the 4-momenta of the $3K$-quark bag and nucleus $A$, $\vec{p}{\,}_{i,\bot}$ and $\vec{k}_{j,\bot}$ are transverse momenta of nucleons in the nuclei and of a quarks in the bag, respectively.

Let us choose now the wave functions $\Phi_{A-K}([x_i,\vec{p}_{i,\bot}])$ and $\Phi_{3K}([z_j,\vec{k}_{j,\bot}])$ in the following form:
\begin{align}
    \Phi_{A-K}([x_i,\vec{p}_{i,\bot}]) & \sim \exp \bigg[ -\al_A 
        \sum_{i=K+1}^A \frac{\vec{p}{\,}_{i,\bot}^2+M_i^2}{x_i} \bigg], \label{3.53} \\
    \Phi_{3K}([z_j,\vec{k}_{j,\bot}]) & \sim \exp \bigg[ -\bt_K 
        \sum_{j=1}^{3K} \frac{\vec{k}{\,}_{j,\bot}^2+m_j^2}{z_j} \bigg], \label{3.54}
\end{align}
$M_i$ are the nucleon masses, $m_j$ are the quark masses. 

In the deep inelastic limit ($Q^2\gg M_i^2, m_j^2$; $\xi_A\to x_A$; $\xi\to x$) the $Q^2$-dependence in the structure functions $F_2^K$ disappears, and after the corresponding calculations from \eqref{3.49} for the quark distribution function $f^K(x)$, which is defined by the following relation
\begin{equation}\label{3.55}
    F_2^K(x)=\la e_q^2\ra \times f^K(x),
\end{equation}
($\la e_q^2\ra$ is the average value of quark charges squared), we obtain \cite{129}:
\begin{equation}\label{3.56}
    f^K(x)=\frac{3K}{A} \,\frac{I_K(x_A)}{\int_0^1 dx_A I_K(x_A)}\,;\quad 
    x_A=\frac{x}{A}\,, \quad K=1,2,\dots,A-1,
\end{equation}
where 
\begin{gather}
    I_K(x_A) = \int_{x_A}^1 d Z_K \,\frac{(1-Z_K)^{A-K-1}}{1+(\bt_K/\al_A)(1-Z_K)}\,
        \bigg(1-\frac{x_A}{Z_K}\bigg)^{3K-2} \notag \\
    =\frac{x_A(1-x_A)^{(A-K)+(3K-2)}}{1+(\bt_K/\al_A)(1-x_A)}\, B(A-K,3K-1) \notag \\
    \times {\,}_2 F_1\bigg(3K\!-\!1,A\!-\!K,1,A\!+\!2K\!-\!1;1\!-\!x_A;
        \frac{(1+\bt_K/\al_A)(1-x_A)}{1+(\bt_K/\al_A)(1-x_A)}\bigg). \label{3.57}
\end{gather}
Here $B$ is the Euler beta-function, ${\,}_2F_1$ is the hypergeometric function of two variables (the Appel function). 

A characteristic feature of the structure function $F_2^K$ is possible existence of a superfast quark in a nucleus which in expreme situation takes all the momentum of nucleus.

The structure functions \eqref{3.55}--\eqref{3.57} are normalized to the number of quarks in the corresponding bag
\begin{equation}\label{3.58}
    \int_0^A dx\, f^K(x)=3K.
\end{equation}

One can relate the parameters $\al_A$ and $\bt_K$ of the relativistic wave functions $\Phi_{A-K}([x_i,\vec{p}_{i,\bot}])$ and $\Phi_{3K}([z_j,\vec{k}_{j,\bot}])$ to the radii of the nucleus $A$ and $3K$-quark bag, respectively,
\addtocounter{equation}{+1}
\begin{align}
    & R_A^2 \sim 8A\al_A, \tag{\theequation a} \label{3.59a} \\
    & r_K^2 \sim 24K\bt_K. \tag{\theequation b} \label{3.59b} 
\end{align}

As can be seen from \eqref{3.56} and \eqref{3.57} the structure functions $F_2^K$ depend on the ratio
\begin{equation}\label{3.60}
    \frac{\bt_K}{\al_A} =\frac{A}{3K}\,\frac{r_K^2}{R_A^2} =\frac{r_K^2 A^{1/3}}{3R_0^2 K}\,.
\end{equation}
Choosing $r_K=0.8$~fm, $R_0=1,2$~fm we get $\bt_K/\al_A=0.109\,A^{1/3}/K$.

In the experiments \cite{112}--\cite{116} the ratio of structure functions $F_2(Fe)/F_2(D)$ is measured in the region $x<1$ that corresponds to the region of small $x_A<1/A$. It is evident from \eqref{3.55}--\eqref{3.56} and it can be checked by straightforward numerical calculations that the structure functions weakly depend on the parameter $\bt_K/\al_A$ in this region. That is why in the numerical calculations we have supposed that the ratio $\bt_K/\al_A$ is the same for all bags. For the iron nucleus we have taken $(\bt_K/\al_A)_{Fe}=0.1$. But for the deuteron (considering that the average distance between nucleons in the deuteron is somewhat larger than in other nuclei) we have chosen the value $(\bt_K/\al_A)_D=0.05$.

\begin{figure}
\begin{center}
\includegraphics*{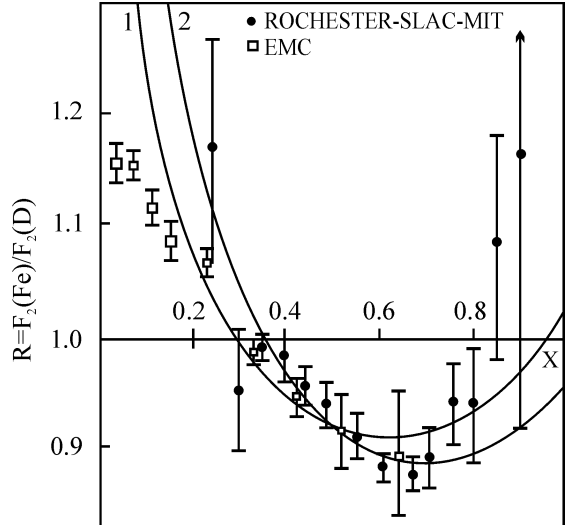}
\end{center}
\caption{The ratio of the structure functions of iron and deuteron. Curves 1 and 2 -- calculations with parametrizations $p(A) \sim A^{-1}$ and $p(A)\sim A^{-2/3}$, respectively.}
\label{fig7}
\end{figure}

Experimental data on the ratio $F_2(Fe)/F_2(D)$ and the curves calculated by the formulas \eqref{3.40}, \eqref{3.42}, \eqref{3.55}--\eqref{3.57} with two parametrizations for $P(A)$ are shown in Fig. \ref{fig7}. The curves reproduce rather well the experimental data in the region $x>0.2$ and somewhat differ from the data in the region $x<0.2$. This difference is probably due to the fact that we restrict our consideration by the valence quarks and do not take into account contributions of the sea quarks and gluons.

The analysis performed shows that deviation of the ratio $F_1(Fe)/F_2(D)$ from unity is due to a larger contribution of the multiquark configurations in the iron nucleus as compared to the deuteron. Note that according to \eqref{3.42} the probabilities of formation of six-, nine-, etc., quark configurations in heavy nuclei are larger than in light nuclei. In the deuteron the contribution of a six-quark state is only a small admixture to the contribution of a two-nucleon state. The deviation from unity of the ratio of the sum of two-nucleon and six-quark contributions to the contribution of a pure two-nucleon state does not exceed 5\% for the deuteron in the whole region $0<x<2$ (see in this connection also \cite{134}, \cite{135}).

\begin{figure}
\begin{center}
\includegraphics*{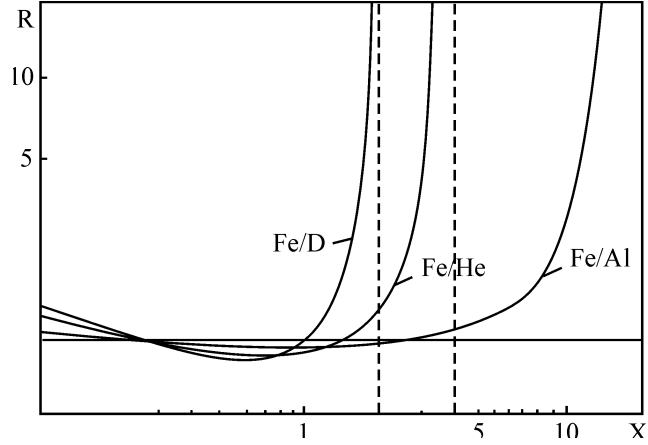}
\end{center}
\caption{The ratios of structure functions of different nuclei in the whole region of variable $x$ $(0<x<A)$}
\label{fig8}
\end{figure}

In Refs. \cite{24}, \cite{25} the possibility was established for extracting information on the quark parton functions of nuclei from the data on cumulative pion production, and in Ref \cite{136} similarity in the $x$-behaviour of the ratio of pion production cross sections on different nuclei and the ratio of the deep inelastic structure functions of the same nuclei was pointed out. It should be noticed that the data on cumulative production allow us to investigate the region $x>1$ (which is not yet reached in the experiments on deep inelastic lepton-nucleus scattering). We have calculated the ratios $F_2(Fe)/F_2(D)$, $F_2(Fe)/F_2(He)$ and $F_2(Fe)/F_2(Al)$ in the whole region of variable $x$ $(0<x<A)$. The curves of these calculations in the parametrization $p(A)\sim A^{-1}$ in the double logarithmic presentation are given in Fig. \ref{fig8}. The similarity in the behaviour of these curves and the ratios of cross sections of cumulative pion production is observed.

The analysis performed shows that the explanation of the observed deviation from unity of the ration $F_1(Fe)/F_2(D)$ requires the consideration of multiquark configurations in nuclei. It seems very interesting to analyze the $A$-dependence of the ratio $F_2(A)/D_2(D)$ contained in \eqref{3.40}, \eqref{3.42}, \eqref{3.54}--\eqref{3.57} (there are the data on this dependence \cite{115}) and carry out experimental investigations of the deep inelastic lepton-nucleus scattering in the region, where the variable $x>1$. It seems that until now in the region $1<x<1,4$ only the structure function of ${\,}^{12}C$ nucleus has been measured \cite{137}.

\begin{figure}
\begin{center}
\includegraphics*{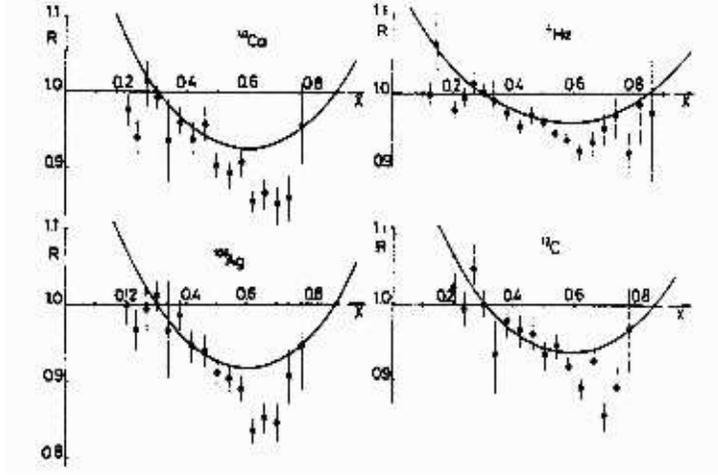}
\end{center}
\caption{The ratios of structure functions of ${\,}^4He$, ${\,}^{12}C$, ${\,}^{40}Ca$, ${\,}^{106}Ag$ nuclei to structure function of deuteron versus variable $x$. Data from Ref. \cite{115}.}
\label{fig9}
\end{figure}

\begin{figure}
\begin{center}
\includegraphics*{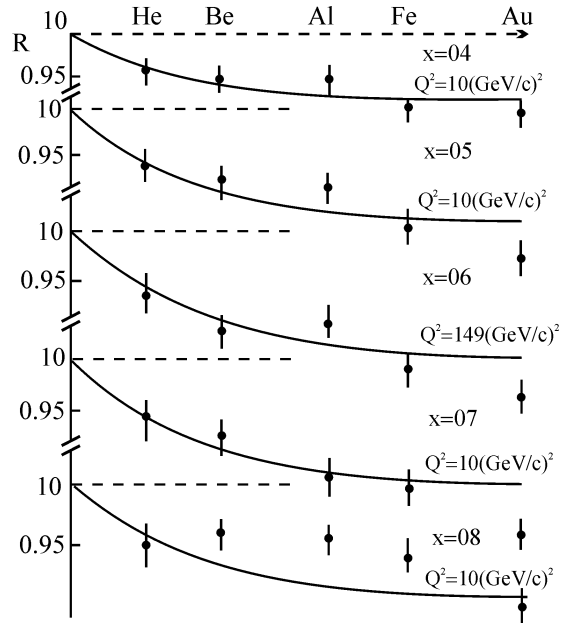}
\end{center}
\caption{The $A$-dependence of structure function ratios at various values of variable $Q^2$. Data from Ref. \cite{115}.}
\label{fig10}
\end{figure}

\begin{figure}
\begin{center}
\includegraphics*{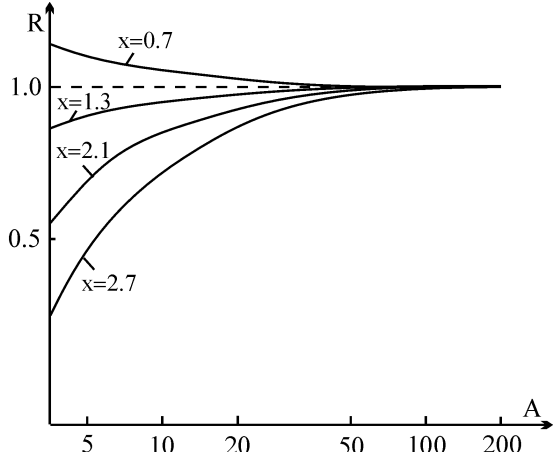}
\end{center}
\caption{The $A$-dependence of ratio $F_2(A)/F_2(Pb)$ at different values of variables $x$.}
\label{fig11}
\end{figure}

On Fig. \ref{fig9} the theoretical curves on the ratio $F_1(A)/F_2(D)$ for several different nuclei and experimental data on deep inelastic scattering cross section ration $\sg_A/\sg_D$ are presented and good agreement is observed. More correct comparison of these quantities needs taking into account the contributions from structure function $F_1$ to the cross section. On Fig. \ref{fig10} the $A$-dependence of these ratios is presented at various values of variable $Q^2$. The Fig. \ref{fig11} shows the $A$-dependence of ration $F_2(A)/F_2(Pb)$ at different values of $x$. Quantitevely these curves at $x>1$ have the same behaviour as the data on cumulative pion production.

Note that in the region of very small $x$ the experimental situation has been changed significantly \cite{138}, \cite{139} and the problem of the theoretical explanation of this phenomenon is of special interest (see in this connection Ref. \cite{140}). 

%% file: garse4.tex
\section*{\large V. Processes Involving High Energy Nuclei and Problem of Relativisation of Nuclear Wave Functions}

\def\theequation{5.\arabic{equation}}
\setcounter{equation}{0}
\setcounter{section}{0}

\section{Scale-Invariant Parametrization of the Deuteron Relativistic \;\; Wave Function}

In this Chapter we consider the relativization of nuclear wave functions. Let us start our consideration with simplest case of a two-nucleon nucleus, i.e. deuteron. To construct a relativistic wave function of deuteron, we use the fact that in the framework of light front formalism the Lorenz-invariant combination $(\vec{p}{\,}_\bot^2+m^2)/x(1-x)$ (where $\vec{p}_\bot$ is the relative transverse momentum of the constituent, $x=1/2+(p_0+p_z)/(P_0+P_z)$, $P_\mu$ and $P_\nu$ are the relative 4-momentum of the internal motion of the constituents and the total 4-momentum of the composite system, respectively, $m$ is the constituent mass) plays the role analogous to that of $\vec{p}{\,}^2$ -- the square of three-dimensional relative momentum. The relativistic wave functions can be obtained from the corresponding nonrelativistic expressions by substitution \cite{6}, \cite{141}, \cite{142}
$$
    \vec{p}{\,}^2 \to \frac{\vec{p}{\,}_\bot^2+m^2}{x(1-x)}
$$
and changing the nonrelativistic numerical parameters by the relativistic ones.

In this way one can obtain, for example, the relativistic analog of the Hulten wave function:
\begin{equation}\label{4.1}
    \Phi_R(x,\vec{p}_\bot)=C_R \bigg[ \frac{\vec{p}{\,}_\bot^2+m^2}{x(1-x)}-\al_R\bigg]^{-1}
        \bigg[ \frac{\vec{p}{\,}_\bot^2+m^2}{x(1-x)}-\bt_R\bigg]^{-1}
\end{equation}
which is written in an arbitrary reference frame for the arbitrary momenta of the deuteron as a whole and arbitrary intrinsic momenta of its constituent nucleons. Note that relativization of other, more accurate, deuteron wave functions does not meet any principal difficulties. In particular, in what follows we shall use the relativistic analog of the Gartenhaus--Moravchik wave function. We shall not describe here other possible ways of relativization of the wave functions, which can be found in Refs. \cite{143}--\cite{149}. 

In the deuteron rest frame, when the momenta of internal motion of its constituent obey the condition $|\vec{p}|/m_N \ll 1$, the wave function \eqref{4.1} turns to the well-known nonrelativistic Hulten wave function
\begin{equation}\label{4.2}
    \Phi_{NR}(\vec{p}) =C_{NR} (\vec{p}{\,}^2+\al_{NR}^2)^{-1}
        (\vec{p}{\,}^2+\bt_{NR}^2)^{-1}.
\end{equation}
In \eqref{4.1} and \eqref{4.2} $\al_R$, $\bt_R$ and $\al_{NR}$, $\bt_{NR}$ are adjustable parameters of the relativistic and nonrelativistic wave functions, respectively, $C_R$ and $C_{NR}$ are normalization coefficients. 

Neglecting the dependence of variable $x$ on binding energy from the condition that in the nonrelativistic limit the wave function $\Phi_R$ goes over into the wave function $\Phi_{NR}$ we obtain the following relation between the parameters
\begin{equation}\label{4.3}
    \al_R=4(m^2-\al_{NR}^2); \qquad 
    \bt_R=4(m^2-\bt_{NR}^2).
\end{equation}

From the normalization condition for wave function \eqref{4.2}
\begin{equation}\label{4.4}
    \int d\vec{p} |\Phi_{NR}(\vec{p})|^2=1
\end{equation}
we get the following expression for normalization coefficient $C_{NR}$
\begin{equation}\label{4.5}
    C_{NR}=\frac{1}{\pi} (\al_{NR}+\bt_{NR})^{3/2} \al_{NR}^{1/2} \bt_{NR}^{1/2}.
\end{equation}

For the normalization of the wave function \eqref{4.1} it is necessary to know the form of all interaction within the two-particle bound state. Assuming that the total interaction kernel does not depend on the total 4-momentum of the deuteron, we obtain the following normalization condition:
\begin{equation}\label{4.6}
    \int_0^1 \frac{dx}{x(1-x)}\, \int d\vec{p}_\bot |\Phi_R(x,\vec{p}_\bot)|^2=8\pi. 
\end{equation}

Inserting into this normalization condition the wave function \eqref{4.1} we get the following  expression for normalization coefficient $C_R$
\begin{equation}\label{4.7}
    C_R=2^{3/2}(\al_R-\bt_R) \big[ f(\al_R,\bt_R)+f(\bt_R,\al_R)\big]^{-1/2},
\end{equation}
where 
\begin{align}
    f(\al_R,\bt_R) & = \frac{4[m^2(\al_R-\bt_R)+\al_R(4m^2-\al_R)]}
            {(\al_R-\bt_R)\al_R^{3/2} (4m^2-\al_R)^{1/2}} \notag \\
    & \quad \times \arctg \bigg( \frac{\al_R}{4m^2-\al_R}\bigg)^{1/2} -\frac{1}{\al_R}\,.
\end{align}

The relation between the wave functions $\Phi_R$ and $\Phi_{NR}$ in the nonrelativistic limit with account of \eqref{4.3} takes the form
$$
    \Phi_R(x,\vec{p}_\bot) \to 2\pi^{1/2} m^{1/2} \Phi_{NR}(\vec{p}).
$$
Here, the relativistic wave function $\Phi_R$ is normalized by the condition \eqref{4.6} and the nonrelativistic wave function $\Phi_{NR}$ is normalized by the condition \eqref{4.4}. 

\section{Break-up of the Relativistic Deuteron and the Verification of Scaling Properties of Its Wave Function}

Let us consider the interaction process of relativistic deuteron with target in which incoming deuteron breaks-up and a system of hadrons $X_N$ is created. In the impulse approximation this process is described by two diagrams in which only the one nucleon of the deuteron interacts with a target. Another nucleon, the so-called spectator-nucleon, does not interact with the target and observation of its distributions allows obtain information on the character of internal motion of constituent in deuteron to be obtained. In the range of small transverse momenta this approximation seems very reasonable. 

For the study of the dynamics of the relativistic deuteron experiments with hydrogen targets are the most convenient, since in this case there are no effects associated with disintegration of the target, and the selection of the spectator nucleons makes it possible to obtain direct information about the deuteron wave function. 

We consider now the process of interaction of the relativistic deuteron with hydrogen target in which a spectator-nucleon and the system $X_N$ of hadrons are produced $D+p \to p_{sp}(n_{sp})+X_N$. One of the simplest processes of this type is the process of direct break-up of the deuteron $Dp\to p\,p\,n$. 

Assuming that the spectator-nucleon does not interact with the target we can obtain the following invariant distribution for spectator-nucleon \cite{6}, \cite{142}, \cite{150}:
\begin{equation}\label{4.9}
    E_{sp}\,\frac{d\sg}{\vec{P}_{sp}} \sim 
        \frac{\lb^{1/2}(s_{NN}, m^2,m^2)}{\lb^{1/2}(s,m^2,m_D^2)}\,
        \sg_{in}(s_{NN}) \,\bigg| \frac{\Phi_R(x,\vec{p}_\bot)}{1-x}\bigg|^2\,.
\end{equation}
Here $s$ is the usual Mandelstam variable for the deuteron--proton system, $s_{NN}$ is the similar variable for the subsystem consisting of the interacting nucleon of the deuteron and the target proton. Energy-momentum conservation leads to the following relation between these variables:
\begin{equation}\label{4.10}
    s_{NN} =s(1-X_{sp})+m^2 -\frac{\vec{P}{\,}_{sp,\bot}^2+m^2}{X_{sp}}\,,
\end{equation}
$\sg_{in}(s_{NN})$ is the total inelastic cross section of the nucleon-nucleon interaction in the given channel $NN\to X_N$, $m$ is the nucleon mass, $m_D$ is the deuteron mass. This flux factor is defined in the following way: $\lb(x,y,z)=(x-y-z)^2-4yz$. The variable $X_{sp}$ is defined as follows:
\begin{equation}\label{4.11}
    X_{sp} =\frac{(E_{sp}+P_{sp,z})}{(E_d+P_{d,z})+(E_p+P_{p,z})}\,.
\end{equation}
Here $E_{sp}$, $E_d$, $E_p$ and $P_{sp,z}$, $P_{d,z}$, $P_{p,z}$ are the energies and the $z$-components of momenta of the spectator nucleon, colliding deuteron and proton, respectively. Note that the variable $X_{sp}$ is scale-invariant and Lorenz-invariant under transformations of the reference frames along the collision axes (the $z$-axes). 

The arguments of the wave function $\Phi_R(x,\vec{p}_\bot)$ are related to the variables $X_{sp}$ and $\vec{P}_{sp,\bot}$ in the following way:
\begin{gather}
    x=1-\bigg( 1+\frac{E_p+P_{p,z}}{E_d+P_{d,z}}\bigg) X_{sp}, \label{4.12} \\
    \vec{p}_\bot=-\vec{P}_{sp,\bot} \;\; \text{in the reference frame, in which $\vec{P}_{d,\bot}=0$}\,. \notag
\end{gather}

It follows from these expressions that the experimental observation of the spectator-nucleon distributions allows information on the internal motion of nucleons inside the relativistic deuteron to be obtained. 

In the high energy limit the distribution of the spectator, summed over all possible hadron systems $X_N$, takes the form
\begin{equation}\label{4.13}
    \frac{1}{\sg_{tot}(\infty)}\,\frac{d\sg}{d\vec{P}_{sp}/E_{sp}}\bigg|_{s\to \infty} 
        \sim \frac{|\Phi_R(X_{sp},\vec{P}_{sp,\bot})|^2}{1-X_{sp}}\,,
\end{equation}
which is very close to the predictions for inclusive distributions obtained in the framework of the hypothesis of limiting fragmentation \cite{151} and those of the parton model \cite{90}--\cite{93} and of the automodelity principle for strong interaction \cite{152}, \cite{153}. Deviations from the automodel behaviour may occur because of a possible weak dependence of wave function parameters on the energy. So, the experimental study of processes with beams of deuterons of different energies seems very interesting. We shall discuss this point later on when comparing theoretical calculations with experimental data. 

A characteristic feature of the spectator-nucleon distribution \eqref{4.9} in the reference frame, in which the target proton is at rest, is the prediction of a maximum at the value of variable $X_{sp}$
\begin{equation}\label{4.14}
    \wt X_{sp} =\frac{1}{2\big(1+\frac{m}{E_d+P_{d,z}}\big)}\,,
\end{equation}
which approaches its limiting value $\wt X_{sp}=0,5$ as energy increases. As it will be seen from what follows the positions of the maxima of the experimental $X_{sp}$-distributions agree with the values predicted by \eqref{4.14}.

To compare the relativistic parametrization \eqref{4.1} with the nonrelativistic Hulten wave function \eqref{4.2}, we consider the momentum distribution of the spectator nucleons in the deuteron rest frame \cite{154}. The momentum distribution of the spectators is related to the invariant differential cross section \eqref{4.9} in the following way
\begin{equation}\label{4.15}
    \frac{d\sg}{d P_{sp}}=\frac{2\pi P_{sp}^2}{(P_{sp}^2+m^2)^{1/2}}
        \int_{-1}^1 d\,\cos \vth_{sp} \bigg( E_{sp}\,\frac{d\sg}{d\vec{P}_{sp}}\bigg)\,. 
\end{equation}

\begin{figure}
\begin{center}
\includegraphics*{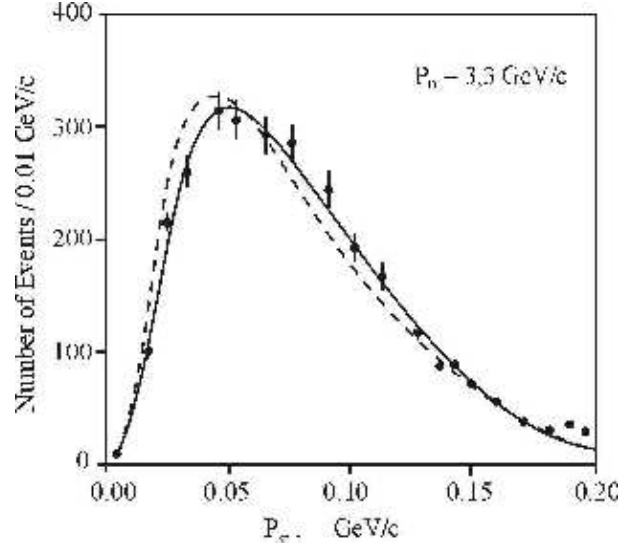}
\end{center}
\caption{Momentum distribution of the spectator neutrons in the deuteron rest frame in the reaction $Dp\to ppn_{sp}$. The solid and dashed curves are the results of calculation with the relativistic and the nonrelativistic wave functions, respectively.}
\label{fig12}
\end{figure}

In Fig. \ref{fig12} the results of theoretical calculations with the relativistic wave function \eqref{4.1} and with the nonrelativistic Hulten wave function \eqref{4.2} are compared with the experimental distribution of spectator neutrons in the process of direct break-up of the deuteron $D+p \to p+p+n_{sp}$ in the deuteron bubble chamber bombarded by deuterons with 3,3 GeV/c momentum and converted to the antilaboratory frame in which the momentum of the incident protons is equal to 1,66 GeV/c \cite{155}--\cite{157}.

\noindent Experimentally, the spectators were chosen as particles having the smallest momentum in the deuteron rest frame. The small admixture of other nucleons cannot affect significantly our considerations. Numerical values of the parameters $\al_R$ and $\bt_R$ of the relativistic wave function of deuteron were calculated by Eq. \eqref{4.3} from the values $\al_{NR}=0,0456$~GeV/c and $\bt_{NR}=0,26$~GeV/c \cite{158} of the nonrelativistic Hulten wave function. The cross section of elastic nucleon-nucleon scattering at these energies is almost constant and equal to 24~mb \cite{159}, \cite{160}. The solid curve on Fig. \ref{fig12} corresponds to the theoretical calculation with the relativistic wave function \eqref{4.1} and the values of the parameters $\al_R=3,521$~(GeV/c)$^2$ and $\bt_R=3,390$~(GeV/c)$^2$ ($\chi^2/N_p=27/20)$. The dashed curve corresponds to the calculation with the nonrelativistic wave function \eqref{4.2} and the parameters $\al_{NR}=0,0456$~GeV/c and $\bt_{NR}=0.26$~GeV/c $(\chi^2/N_p=65/20)$. Comparison with experimental data shows that in the range of spectator momenta $P_{sp}<0,2$~GeV/c the relativistic wave function gives somewhat better description of the experimental data than the nonrelativistic one. 

\begin{figure}
\begin{center}
\includegraphics*{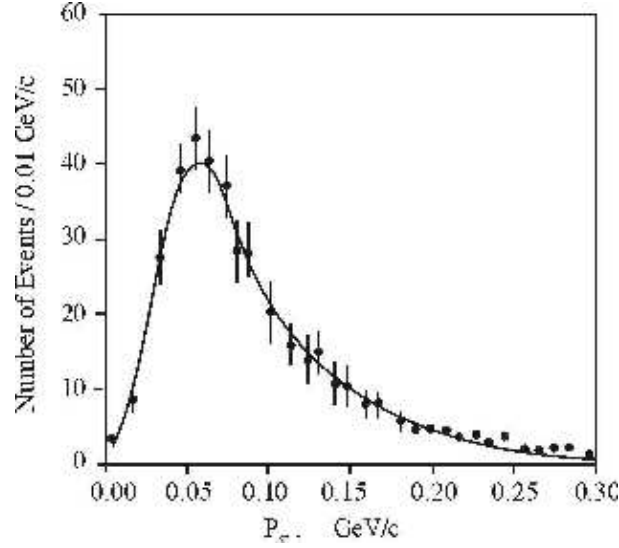}
\end{center}
\caption{Momentum distribution of the spectator protons in the deuteron rest frame in the reaction $Dp\to p(p\pi^-) p_{sp}$.}
\label{fig13}
\end{figure} 

\begin{figure}
\begin{center}
\includegraphics*{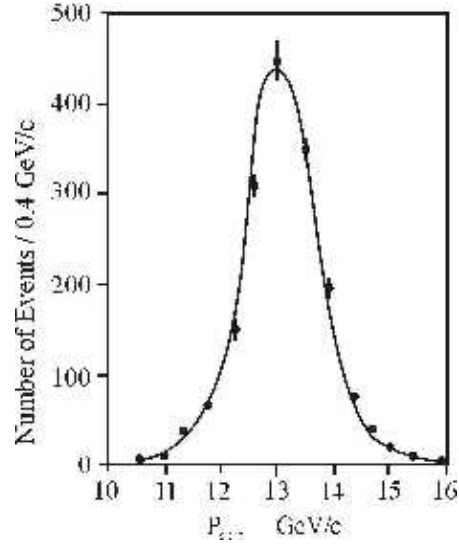}
\end{center}
\caption{Momentum distribution of the spectator protons in the frame of the colliding beams in the reaction $Dp\to p(p\pi^-) p_{sp}$.}
\label{fig14}
\end{figure}

In Fig. \ref{fig13} the momentum distribution of the spectator protons in the reaction $D+p\to p+(p\pi^-)+P_{sp}$ at the energy $\sqrt{s}=52$~GeV is presented in the deuteron rest frame. The data were obtained in an experiment at the CERN ISR with the colliding deuteron-proton beams \cite{36}, \cite{37}. In Fig. \ref{fig14} the momentum distribution of the spectator protons in the same reaction and at the same energy is presented in the frame of the colliding beams. The curves in Figs. \ref{fig13} and \ref{fig14} correspond to the calculations with the wave function \eqref{4.1}. The cross section $\sg_{in}(s_{NN})$ of neutron diffractive dissosiation $n+p\to (p\pi^-)+p$ at the considered energies was taken to be constant and equal to 185~mb \cite{36}. The numerical values of the parameters $\al_R$ and $\bt_R$ obtained by fitting the experimental data and the corresponding values of $\chi^2/N_p$ are given in Table 1. The agreement of the results of theoretical calculations with the experimental data confirms the validity of using the relativistic parametrization \eqref{4.1}. 

\begin{table}
\begin{center}
Table 1. Parameters of the Relativistic Hulthen Wave Function \\[2mm]
\begin{tabular}{lccc} \hline
\qquad Reaction & $\al_{{}_{R}}$, (GeV/c)$^2$ & $\bt_{{}_{R}}$, (GeV/c)$^2$ & $\chi^2/N_p$ \\[1mm] \hline
Calculation by Eq. \eqref{4.3} & 3,521 & 3,390 \\ [1mm] \hline
$D+p\to p+(p\pi^-)+P_{sp}$ \\
\qquad $\sqrt{s}=52$ GeV & $3,522\pm 0,006$ & $3,383\pm 0,033$ & 11/32 \\
Deuteron rest frame \\[1mm] \hline
$D+p\to p+(p\pi^-)+P_{sp}$ \\
\qquad $\sqrt{s}=52$ GeV & $3,513\pm 0,005$ & $3,478\pm 0,019$ & 28/14 \\
Center-of-mass frame \\[1mm] \hline
$D+Al\to n_{sp}+X_N$ \\
\quad $P_D=3,46$ GeV/c & 3,5156-fixed & $3,5240\pm 0,0004$ & 1,9 \\
\quad $P_D=4,46$ GeV/c & 3,5156-fixed & $3,5222\pm 0,0006$ & 1,1 \\
\quad $P_D=7,66$ GeV/c & 3,5156-fixed & $3,5203\pm 0,0010$ & 2,9 \\
\quad $P_D=10$ GeV/c & 3,5156-fixed & $3,5170\pm 0,0021$ & 1,1 \\[1mm] \hline
$D+p\to p+p+n_{sp}$ \\
\quad $P_D=3,3$ GeV/c & 3,5156-fixed & $3,4572\pm 0,0005$ & 232/150 \\
\quad $d\sg/d X_{sp}$ \\
in three intervals of $P_{sp,\bot}$ \\[1mm] \hline
$D+p\to p+p+n_{sp}$ \\
\quad $P_D=3,3$ GeV/c & 3,5156-fixed & $3,4579\pm 0,0005$ & 311/150 \\
\quad $d\sg/d P_{sp,\bot}$ \\
in three intervals of $X_{sp}$ \\[1mm] \hline
\end{tabular}
\end{center}
\label{t1}
\end{table}

In order to verify the scaling properties of the relativistic wave functions, we have compared the theoretical calculations with the experimental distributions of the spectator nucleons at various momenta of incident deuterons \cite{161}. The distributions of the spectator neutron calculated using the wave function \eqref{4.1} were compared with the experimental distributions of neutron from deuteron stripping, which were obtained in the following way. A beam of deuteron accelerated at the JINR Synchrophasotron to momentum $P_D$ collided with an internal Alluminum target. Neutrons scattered from the target under the angles $\tht<0,001$~rad were directed in the 100-cm bubble chamber. This set-up gives with a good accuracy the neutron with zero transverse momentum $P_{sp,\bot}$. Momentum spectra of neutrons were measured in the reaction $n\,p\to p\,p\,\pi^-$, which can be identified clearly in the bubble chamber. Events selected by this method were used for obtaining the momentum distributions of neutrons detected in the 100-cm bubble chamber. Detailed description of the experimental procedure can be found in \cite{162}--\cite{164}. 

Assuming that deuteron break-up process occurs on the separate nucleons of target nuclei one can suppose that distribution of spectator nucleons does not depend on the target. Independence of the spectator fragment distributions on the target for this type of processes seems to be general property and allows one to use theoretical distribution \eqref{4.9} in this case. 

\begin{figure}
\begin{center}
\includegraphics*{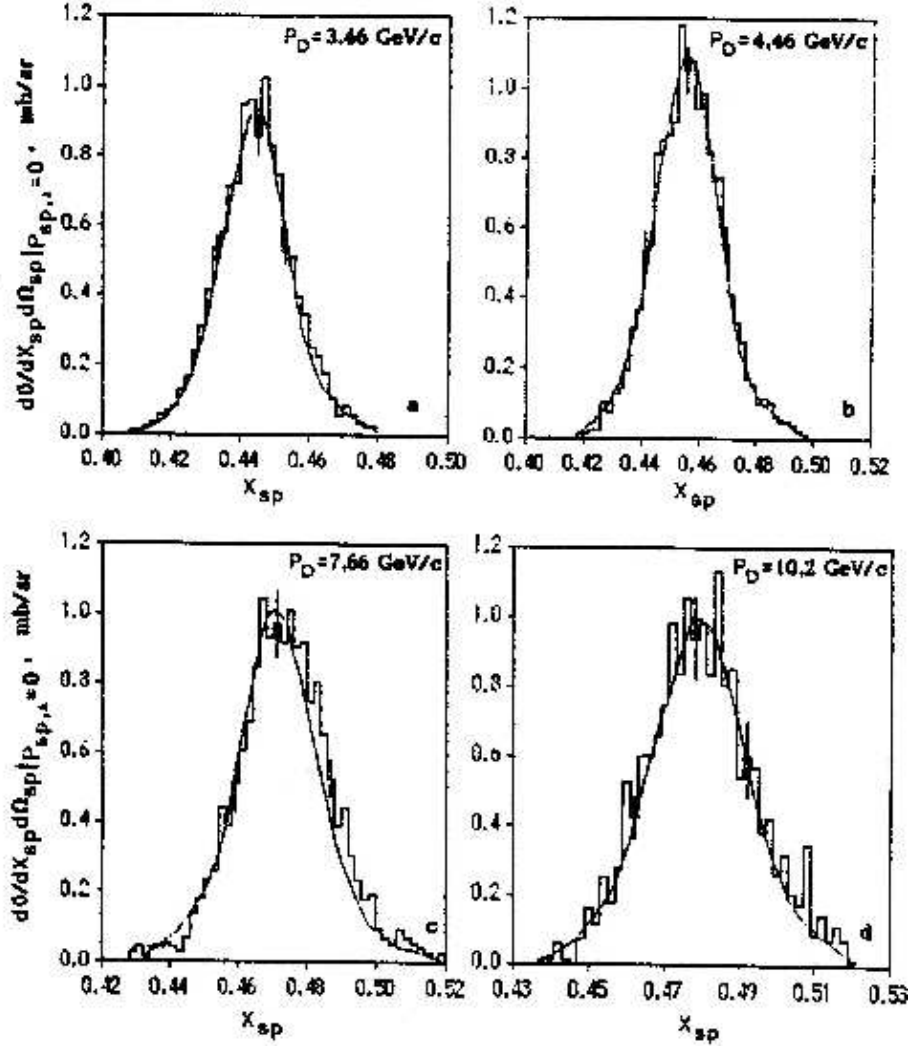}
\end{center}
\caption{The distributions $\frac{d\sg}{d X_{sp} d\Omega_{sp}}\Big|_{P_{sp,\bot}=0}$ of spectator neutrons at different momenta of the incident deuterons.}
\label{fig15}
\end{figure}

In Figs. \ref{fig15} a,b,c,d the experimental  distributions
$$
    \frac{d\sg}{d X_{sp} d\Omega_{sp}} \bigg|_{\vec{P}_{sp,\bot}=0}
$$
of spectator neutrons for four different values of the incident deuteron energy and the corresponding theoretical distributions normalized to the unity in the points of their maxima are presented. The relation between this distribution and the invariant distribution \eqref{4.9} at zero value of transverse momentum of the spectator has the form:
$$
    \frac{d\sg}{d X_{sp} d\Omega_{sp}} \bigg|_{\vec{P}_{sp,\bot}=0}=
        \frac{[(m+P_D+E_D)X_{sp}]^2-m^2}{4(m+P_D+E_D)^2X_{sp}^3} \,
        \bigg( E_{sp}\,\frac{d\sg}{d\vec{P}_{sp}} \bigg) \bigg|_{\vec{P}_{sp,\bot}=0}.
$$

Performed analysis shows that in the considered region of momenta of the incident deuteron the parameters of the relativistic wave function $\al_R$ and $\bt_R$ depend weakly on the energy of the incident beam (see Table \ref{t1}). This fact indicates that in wide range of energy the relativistic wave function of deuteron $\Phi_R$ contains no other dependence on the energy apart from the dependence on the variable $x$ and allows to say that wave function with the scale-invariant parametrization of the ``longitudinal motion'' of constituents in terms of light front variables gives a good description of the relativistic deuteron. 

The positions of maxima of the spectator neutron distributions in Figs. \ref{fig15} a,b,c,d coincide rather well with the values of $\wt X_{sp}$ predicted by Eq. \eqref{4.14}. 

In order to study the transverse momentum distributions of the spectators \cite{150} we have used the experimental data on the interaction of deuteron with momentum 3,3~GeV/c in hydrogen bubble chamber. From the experimental point of view, quite comprehensive data are available for the direct break-up channel $dp\to p\,p\,n$. In this case $\sg_{in}(s_{NN})$ in Eq. \eqref{4.9} must be replaced by the total elastic cross section $\sg_{el}(s_{NN})$ of the nucleon-nucleon interaction. We restricted ourselves to the case of the neutron-spectator, since there are much more experimental data on $\sg_{el}^{pp}$ than on $\sg_{el}^{np}$ (see, e.g., \cite{159}). It should be noted that in the experiment with a beam of 3,3~GeV/c deuterons, $s_{pp}$ varies in an interval, in which $\sg_{el}(s_{pp})$ is almost constant and equal to 24~mb.

\begin{figure}
\begin{center}
\includegraphics*{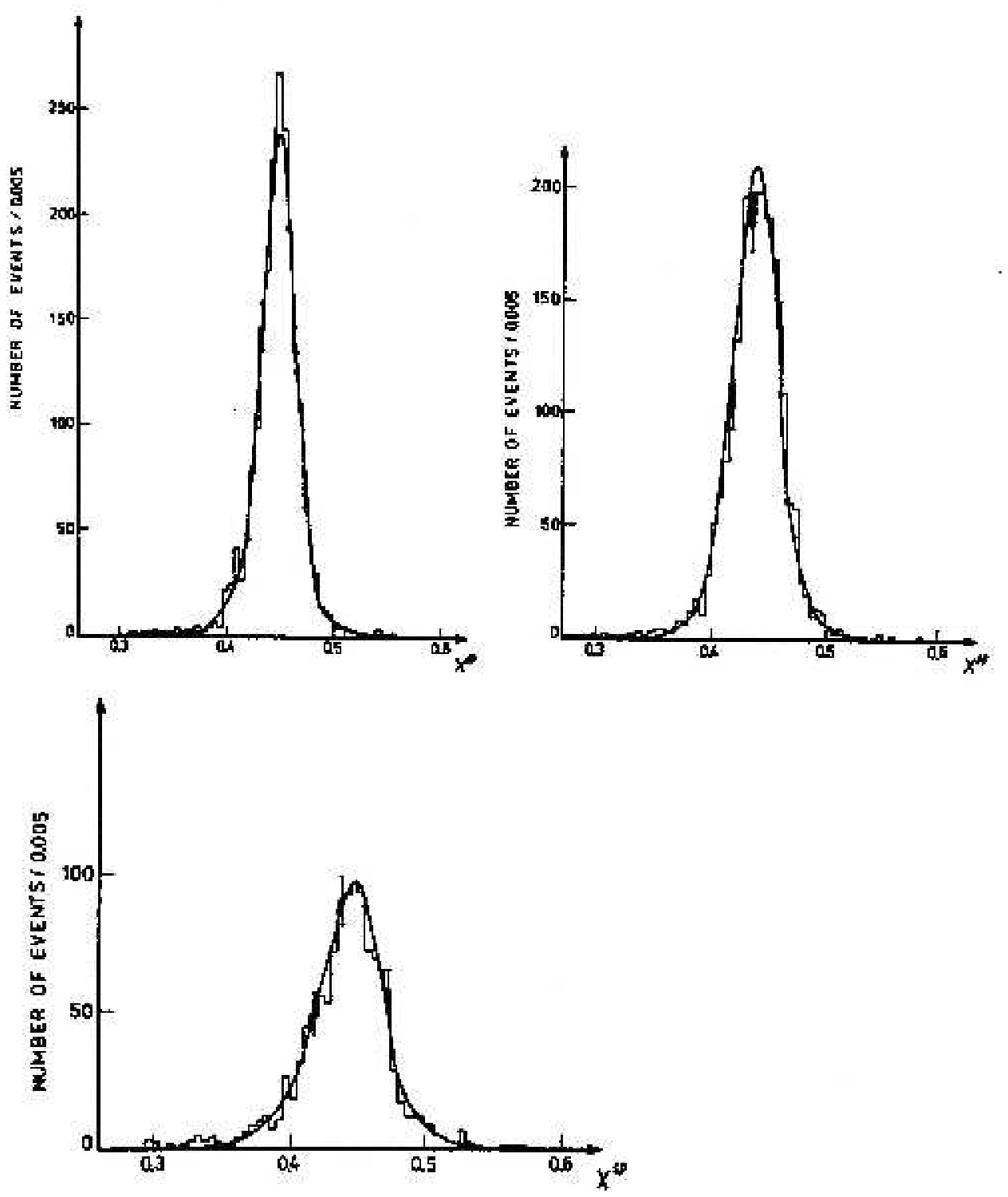}
\end{center}
\caption{The $d\sg/dX_{sp}$ distributions of spectator neutron in the reaction $Dp\to ppn_{sp}$ in the intervals a)~$0.01<P_{sp,\bot}<0.04$~GeV/c, b)~$0.04<P_{sp,\bot}<0,07$~GeV/c, c)~$0.07<P_{sp,\bot}<0.1$~GeV/c.}
\label{fig16}
\end{figure}

\begin{figure}
\begin{center}
\includegraphics*{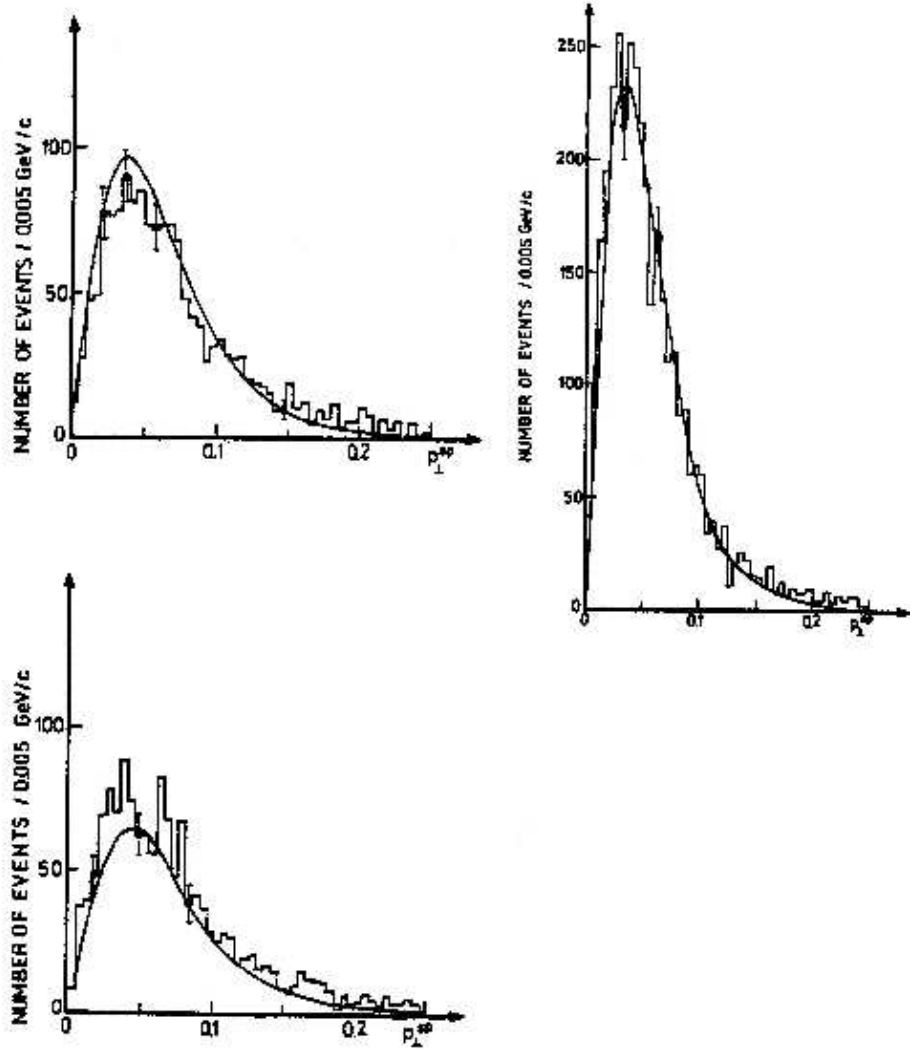}
\end{center}
\caption{The $d\sg/dP_{sp,\bot}$ distributions of spectator neutron in the reaction $Dp\to ppn_{sp}$ in the intervals a)~$0.40<X_{sp}<0.43$, b)~$0.43<X_{sp}<0,46$, c)~$0.45<X_{sp,\bot}<0.49$.}
\label{fig17}
\end{figure} 

The experimental distributions $d\sg/d X_{sp}$ and $d\sg/d P_{sp,\bot}$ in the rest frame of the target proton were analyzed. They are related to the invariant differential cross section as follows:
\addtocounter{equation}{+1}
\begin{gather}
    \frac{d\sg}{d X_{sp}} =\int_{P_{sp,\bot_{\min}}}^{P_{sp,\bot_{\max}} }
        \frac{d\sg}{d X_{sp}\,d P_{sp,\bot}}\, dP_{sp,\bot}, 
        \tag{\theequation a} \label{4.16a} \\
    \frac{d\sg}{d P_{sp,\bot}} =\int_{X_{sp,{}_{\min}}}^{X_{sp,{}_{\max}}} 
        \frac{d\sg}{d X_{sp}\,d P_{sp,\bot}}\, dX_{sp}, 
        \tag{\theequation b} \label{4.16b} \\
    \frac{d\sg}{d X_{sp} dP_{sp,\bot}} =2\pi \bigg( \frac{P_{sp,\bot}}{X_{sp}} \bigg)
        \frac{d\sg}{d\vec{P}_{sp}/E_{sp}} \,.
        \tag{\theequation c} \label{4.16c} 
\end{gather}

The experimental distributions $d\sg/d X_{sp}$ $(d\sg/d P_{sp,\bot})$ integrated in three different intervals of $P_{sp,\bot}(X_{sp})$ were compared with the results of the theoretical scheme described above. The results of the analysis are given in Figs. \ref{fig16} a,b,c and Figs. \ref{fig17} a,b,c. The theoretical curves in these Figures correspond to the values of parameters of the relativistic wave function \eqref{4.1} given in Table 1. Values of $\chi^2/N_P$ (where $N_p$ is the number of experimental points) allow to consider the agreement of the model with the experimental data to be satisfactory. In accordance with the prediction \eqref{4.14}, the maximum in the $X_{sp}$ distribution  is at $\wt X_{sp}\approx 0.44$. 

\section{The Break-up of More Complicated Nuclei}

It is interesting to attempt to find relativistic analogues of the wave functions of more complicated nuclei and to verify the universality of scale-invariant properties of these relativistic wave functions in the framework of light front formalism \cite{5}, \cite{9}, \cite{165}--\cite{167} by comparison of corresponding results with experimental data.

In this section the process of knocking out the nucleon from the relativistic nucleus in the collision with hydrogen target (proton) is considered. The relativistic nuclear wave functions having scale-invariant properties and turning in the nonrelativistic limit to the well-known Gaussian nuclear wave functions are introduced. The invariant differential cross section of the nucleon knock-out reaction from nucleus is expressed via the overlap integral of these wave functions. This overlap integral is calculated in an analytical form using the Gaussian parametrization of the light front wave functions. The results of theoretical calculation are compared with experimental data on the partial break-up of ${}^4He$ nuclei with momentum 8,56~GeV/c and ${}^3He$ nuclei with momentum 13,5~GeV/c in collision with proton target. The data obtained on the 100-cm bubble chamber of the JINR \cite{168}, \cite{169} and the data on the process ${}^4He+p\to {}^3He+X$ at different value of momentum of incident nuclei obtained at SATURNE \cite{170} have been used. 

Let us consider the process of the knocking out one nucleon from the relativistic nucleus $A$ on the hydrogen target. Assuming that only knocked out nucleon interacts with the target and that the remaining $(A-1)$ nucleons of the initial nucleus continue to exist in the form of a fragment nucleus (the so-called spectator fragment), one can calculate the distributions of these fragments.

The nucleus consisting of $A$ nucleons with total 4-momentum $P_A$ is described by means of the relativistic wave function $\Phi_{P_A}^{(A)}([x_i^{(A)},\vec{p}_{i,\bot}])$ in which the ``longitudinal motion'' of the constituents is parametrized in terms of scale-invariant variables
$$
    x_i^{(A)}=\frac{p_0^{(i)}+p_3^{(i)}}{P_{A,0}+P_{A,3}}\,,
$$
where $P_\mu^{(i)}$ $(\mu=0,1,2,3$ is the Lorentz index) and $P_{A,\mu}$ are the individual 4-momentum of the $i$-th nucleon in the composite system and the total 4-momentum of the composite system as a whole. 
Square brackets in the argument of the wave function $\Phi_{P_A}^{(A)}$ denote the set of the corresponding variables $x_i^{(A)}$ and $\vec{p}_{i,\bot}$, which satisfy the following conditions:
\begin{equation}\label{4.17}
    \sum_{i=1}^A x_i^{(A)}=1; \quad 0<x_i^{(A)}<1; \quad 
        \sum_{i=1}^A \vec{p}_{i,\bot}=\vec{P}_{A,\bot}.
\end{equation}
The superscript of the variables $x_i^{(A)}$ means that this variable is defined in a system of particles whose number is equal to this index.

The distribution of the spectator fragments in the process of nucleon knock out from the nucleus in the laboratory frame (incoming nucleus $A$ moves along the $z$-axis and the target proton is at rest) looks as follows
\begin{equation}\label{4.18}
    E_{sp}\,\frac{d\sg}{d\vec{P}_{sp}} \sim 
        \frac{\lb^{1/2}(s_{NN},m^2,m^2)}{\lb^{1/2}(s,M_A^2,m^2)}\,
        \sg_{NN}^{el}(s_{NN}) \bigg| \frac{I(X_{sp},\vec{P}_{sp,\bot})}{1-\al X_{sp}}\bigg|^2.
\end{equation}
Here $s$ is the usual Mandelstam variable for the system consisting of the incident nucleus $A$ and the target proton, $s_{NN}$ is the analogous variable for subsystem consisting of the interacting nucleon of nucleus $A$ and the target proton. Energy-momentum conservation leads to the following relation between these variables
\begin{equation}\label{4.19}
    s_{NN}=s(1-X_{sp})+M_{sp}^2 -\frac{\vec{P}{}_{sp,\bot}^2+M_{sp}^2}{X_{sp}}\,,
\end{equation}
$\sg_{NN}^{el}(s_{NN})$ is the total elastic cross section for interaction of the nucleon from nucleus $A$ with target. $\lb(x,y,z)$ is the flux factor, which is defined as 
$$
    \lb(x,y,z)=(x-y-z)^2-4yz,
$$
$m$ is the nucleon mass. $M_A$ is the mass of the incident nucleus, $M_{sp}$ is the mass of the spectator fragment. $\al$ is given by 
\begin{equation}\label{4.20}
    \al=1+\frac{m}{E_A+P_{A,3}}\,.
\end{equation}
The variable $X_{sp}$ is defined in the following way
\begin{equation}\label{4.21}
    X_{sp}=\frac{E_{sp}+P_{sp,3}}{M+E_A+P_{A,3}}\,,
\end{equation}
$P_{A,3}$, $E_A$ and $P_{sp,3}$, $E_{sp}$ are the $z$-components of the momenta and the energies of the incident nucleus $A$ and of the spectator-fragment $(A-1)$, respectively.

The overlap integral $I(X_{sp},\vec{P}_{sp,\bot})$ of the relativistic wave functions of the incident nucleus and of the spectator fragment is given by:
\begin{gather}
    I(X_{sp},\vec{P}_{sp,\bot})=\int_0^1 \prod_{i=1}^{A-1} 
        \frac{dx_i^{(A-1)'}}{x_i^{(A-1)'}}\,
        \dl\bigg(1-\sum_{i=1}^{A-1} x_i^{(A-1)'}\bigg) \notag \\
    \times \int \prod_{i=1}^{A-1} d\vec{p}{\,}_\bot^{(i)'} 
        \dl^{(2)}\bigg(\vec{P}_{sp,\bot}-\sum_{i=1}^{A-1} \vec{p}{\,}_\bot^{(i)'}\bigg) \notag \\
    \times \Phi_f^{+(A-1)'}([x_i^{(A-1)'},\vec{p}{\,}_\bot^{(i)'}-x_i^{(A-1)'} \vec{P}_{sp,\bot}])
        \Phi_{in}^{(A)}([x_i^{(A)},\vec{p}{\,}_\bot^{(i)}]). \label{4.22}
\end{gather}
The overlap integral is a direct analogue of the corresponding notion which appears in the nonrelativistic theory of nuclear reactions (see, e.g., Ref. \cite{39} and the references therein).

The arguments of the wave function $\Phi_{in}^{(A)}$ of the incident nucleus are related to the integration variables and the observable quantities $X_{sp}$ and $\vec{P}_{sp,\bot}$ in the following way:
\begin{equation}\label{4.23}
\begin{aligned}
    & x_i^{(A)}=\al X_{sp} x_i^{(A-1)'}; \quad 
        \vec{p}{\,}_\bot^{(i)}=\vec{p}{\,}_\bot^{(i)'}; \quad i=1,2,\dots,A-1, \\
    & x_A^{(A)}=1-\al X_{sp}; \quad 
        \vec{p}{\,}_\bot^{(A)}=-\vec{P}_{sp,\bot}.
\end{aligned}
\end{equation}

Thus observation of the spectator fragment makes it possible to obtain information about the character of the ``longitudinal'' and transverse momentum distributions of the nucleons in the incident nucleus $A$. 

Note that the variable $X_{sp}$ is defined in an arbitrary Lorentz frame, in which the nucleus $A$ and the target collide along the $z$-axis as follows
\begin{equation}\label{4.24}
    X_{sp} =\frac{(E_{sp}+P_{sp,3})}{(E_N+P_{N,3})+(E_A+P_{A,3})}\,,
\end{equation}
where $P_3$'s and $E$'s are the longitudinal momenta and the energies of the corresponding particles. As is easily seen, $X_{sp}$ is a Lorentz-invariant and scale-invariant variable. In the proton rest frame it turns to the form given by Eq. \eqref{4.21}.

In the case $A=2$, which corresponds to the process of the direct break-up of the deuteron, the overlap integral in Eq. \eqref{4.18} is replaced by the deuteron relativistic wave function. 

If together with the knock-out of the nucleon from the nucleus the particle production takes place, the differential cross section of this process looks like Eq. \eqref{4.18}, but the cross section $\sg_{NN}^{el}$ of elastic scattering of the active nucleon from incident nucleus on the target nucleon is replaced by the cross section $\sg_{NN}^{tot}$ of particle production in the nucleon-nucleon scattering. 

In Eq. \eqref{4.22} the wave functions $\Phi_{in}^{(A)}$ and $\Phi_f^{(A-1)}$ of the incoming and outgoing nuclei are defined in the reference frame, in which their total transverse momenta are zero. They are related to the wave functions with arbitrary total 4-momentum $P$ as follows \cite{52}
\begin{equation}\label{4.25}
    \Phi_P^{(A)}([x_i^{(A)},\vec{p}{\,}_\bot^{(i)}])=
        \Phi_{P_\bot=0}^{(A)}([x_i^{(A)},\vec{p}{\,}_\bot^{(i)}-x_i^{(A)}\vec{P}_\bot]).
\end{equation}

As an example, we consider the following simplest parametrization of the relativistic wave functions of the incident and fragment nuclei:
\begin{equation}\label{4.26}
    \Phi^{(A)}([x_i^{(A)},\vec{p}{\,}_\bot^{(i)}]) =C_A
        \exp \bigg( -a_A^R \sum_{i=1}^A \frac{\vec{p}{}_\bot^{(i)2}+m_i^2}{x_i^{(A)}}\bigg)
\end{equation}
and similarly for $\Phi^{(A-1)}$ with the substitution $A\to (A-1)$.

In Eq. \eqref{4.26} $a_A^R$ is an adjustable parameter, $C_A$ is a normalization factor. If the scale-invariant parametrization of ``longitudinal motion'' of the constituent put into the wave function $\Phi^{(A)}([x_i^{(A)},\vec{p}{\,}_\bot^{(i)}])$ is really valid, then the parameters $a_A^R$ would be the same at different energies of incoming nuclei. 

Since we do not distinguish protons from neutrons in our consideration, the wave function $\Phi^{(A)}([x_i^{(A)},\vec{p}{\,}_\bot^{(i)}])$ is a symmetric function of its arguments. Solving the problem of a conditional extremum under the conditions \eqref{4.17}, we find the the wave function \eqref{4.26} has a maximum at zero values of the transverse momenta of the constituent nucleons and at values of the variables $x_i^{(A)}$ equal to 
\begin{equation}\label{4.27}
    \wt x_i^{(A)}=\frac{m_i}{\sum\limits_{i=1}^A m_i}=\frac{1}{A}\,.
\end{equation}

Taking into account the relation between the variables $x_i^{(A)}$ and $X_{Sp}$ (see Eqs. \eqref{4.23}), we find that the $X_{sp}$ distributions of the spectator fragments must have a maximum at 
\begin{equation}\label{4.28}
    \wt X_{sp} =\frac{A-1}{A\big(1+\frac{m}{E_A+P_{A,3}}\big)}\,.
\end{equation}

Note that properties such as the scale invariance of the relativistic wave functions and the position of the maximum in the $X_{sp}$ distribution of the spectator fragments, do not depend on the concrete form of \eqref{4.26} of the wave functions $\Phi^{(A)}([x_i^{(A)}, \vec{p}{\,}_\bot^{(i)}])$ and remain valid for their arbitrary parametrization. 

In order to normalize the relativistic wave functions correctly, one has to know in general the form of all the interactions inside the relativistic system. Assuming, however, that the total interaction kernel does not depend on the total 4-momentum of the composite system, one gets the following normalization condition:
\begin{gather}
    \int_0^1 \prod_{i=1}^A \frac{dx_i^{(A)}}{x_i^{(A)}}\,
        \dl\bigg( 1-\sum_{i=1}^A x_i^{(A)}\bigg)
        \int \prod_{i=1}^A d\vec{p}{\,}_\bot^{(i)}
        \dl^{(2)}\bigg( \vec{P}{}_{A,\bot}-\sum_{i=1}^A \vec{p}{\,}_\bot^{(i)}\bigg) \notag \\
    \times |\Phi_{P_A}^{(A)}([x_i^{(A)},\vec{p}{\,}_\bot^{(i)}])|^2=2(4\pi)^{A-1}. \label{4.29}
\end{gather}

Substituting the wave function $\Phi^{(A)}([x_i^{(A)},\vec{p}{\,}_\bot^{(i)}])$ in the form of \eqref{4.26} into the normalization condition \eqref{4.29}, we obtain the following approximate  expression for the normalization factor $C_A$:
\begin{gather}
    C_A =\sqrt{2} (4\pi)^{\frac{A-1}{2}} 
        \bigg( \sum_{i=1}^A m_i\bigg)^{\frac{3(A-1)+1}{4}}
        \bigg( \prod_{i=1}^A m_i\bigg)^{-\frac{1}{4}} \notag \\
    \times \bigg(\frac{2a_A^R}{\pi}\bigg)^{\frac{3(A-1)}{4}} 
        \exp\bigg[a_A^R \bigg( \sum_{i=1}^A m_i\bigg)^2\bigg]. \label{4.30}
\end{gather}

One of the guiding points in the choice of relativistic wave functions is their correct nonrelativistic limit. The nonrelativistic wave function $\Phi_{NR}^{(A)}([\vec{p}{\,}^{(i)}])$, the relativistic analogue of which is given by Eq. \eqref{4.26} has the Gaussian form
\begin{equation}\label{4.31}
    \Phi_{NR}^{(A)}([\vec{p}{\,}^{(i)}])=
        \bigg(\frac{4a_A^{NR}}{\pi}\bigg)^{\frac{3(A-1)}{4}} 
        \exp\bigg(a_A^R \sum_{i=1}^A \vec{p}{\,}^{(i)2}\bigg)
\end{equation}
and is normalized by the condition 
\begin{equation}\label{4.32}
    \int \prod_{i=1}^A d\vec{p}{\,}^{(i)} 
        \dl^{(3)}\bigg(\sum_{i=1}^A \vec{p}{\,}^{(i)}\bigg) 
        |\Phi_{NR}^{(A)}([\vec{p}{\,}^{(i)}])|^2=1.
\end{equation}

Let us expand the combination $(\vec{p}{\,}_\bot^{(i)2}+m_i^2)/x_i^{(A)}$ in the power series of the parameters $p_3^{(i)}/m_i$, $p_\bot^{(i)}/m_i$, which are small in the nonrelativistic limit. Restricting ourselves to the quadratic terms and writing the result in the nucleus rest frame $\vec{P}_A=0$, we get
\begin{gather}
    \frac{\vec{p}{\,}_\bot^{(i)2}+m_i^2}{x_i^{(A)}} \simeq (m_i-p_3^{(i)})
        \bigg(\sum_{i=1}^A m_i\bigg) +\frac{\vec{p}{\,}^{(i)2}}{2m_i}
        \bigg(\sum_{i=1}^A m_i\bigg)
    +m_i\bigg(\sum_{i=1}^A \frac{\vec{p}{\,}^{(i)2}}{m_i}\bigg). \label{4.33}
\end{gather}
Substituting this expression into the wave function \eqref{4.26} we obtain in the nonrelativistic limit (without taking into account the normalization factor)
\begin{gather}
    \Phi^{(A)}([x_i^{(A)},\vec{p}{\,}_\bot^{(i)}])\sim 
        \exp\bigg( -a_A^R \sum_{i=1}^A \frac{\vec{p}{\,}_\bot^{(i)2}+m_i^2}{x_i^{(A)}}\bigg)\notag\\
    \to \exp \bigg[ -a_A^R\bigg(\sum_{i=1}^A m_i\bigg)^2\bigg] 
        \exp \bigg[ -a_A^R\bigg(\sum_{i=1}^A m_i\bigg)
        \bigg(\sum_{i=1}^A \frac{\vec{p}{\,}^{(i)2}}{m_i}\bigg)\bigg]. \label{4.34}
\end{gather}

The condition of the correct nonrelativistic limit gives the following relation between the parameters of the relativistic and nonrelativistic wave functions
\begin{equation}\label{4.35}
    a_A^R=\frac{m_i}{\sum\limits_{i=1}^A m_i} =\frac{1}{A}\,a_A^{NR}.
\end{equation}

For the normalized wave functions, the nonrelativistic limit looks as follows
\begin{equation}\label{4.36}
    \Phi_R^{(A)}([x_i^{(A)},\vec{p}{\,}_\bot^{(i)}]) \to \sqrt{2}(2\pi^2m^2)^{\frac{A-1}{4}}
        A^{\frac{1}{4}} \Phi_{NR}^{(A)}([\vec{p}{\,}^{(i)}]).
\end{equation}
Here $\Phi_R^{(A)}$ is the relativistic wave function \eqref{4.26} normalized by the condition \eqref{4.29} and $\Phi_{NR}^{(A)}$ is the nonrelativistic wave function \eqref{4.31} normalized by the condition \eqref{4.32}. Masses of all nucleons are assumes to be equal $m_i=m$.

Substituting now the wave functions $\Phi^{(A)}$ and $\Phi^{(A-1)}$ into the overlap integral \eqref{4.22}, taking into account the relation \eqref{4.23} and integrating over the transverse momenta, we obtain
\begin{gather}
    I(X_{sp},\vec{P}_{sp,\bot})=C_AC_{A-1} \bigg( \frac{\pi}{a_{A-1}^R+a_A^R/(\al X_{sp})}\bigg)^{A-2}
        \notag \\
    \times \exp \bigg( -\frac{a_A^R m_A^2}{1-\al X_{sp}}\bigg) 
        \exp \bigg[ -\frac{a_A^R \vec{P}{}_{sp,\bot}^2}{\al X_{sp}(1-\al X_{sp})}\bigg]
        J(X_{sp}), \label{4.37}
\end{gather}
where
\begin{gather}
    J(X_{sp})=\int_0^1 \prod_{i=1}^{A-1}dx_i^{(A-1)}
        \dl\bigg(1-\sum_{i=1}^{A-1} x_i^{(A-1)}\bigg) \notag \\
    \times \exp \bigg[ -\bigg(a_{A-1}^R +\frac{a_A^R}{\al X_{sp}}\bigg) 
        \sum_{i=1}^{A-1} \frac{m_i^2}{x_i^{(A-1)}}\bigg]. \label{4.38}
\end{gather}

Thus, the $\vec{P}_{sp,\bot}$-distribution is obtained in the analytic form. The integral over the variables $x_i^{(A-1)}$ can be calculated approximately by means of the multidimensional saddle point method \cite{171} in the form of an asymptotic expansion in inverse powers of the large parameter $a_{A-1}^R+a_A^R/\al X_{sp}$. The leading term of this expansion has the form:
\begin{gather}
    J_0(X_{sp})\simeq \bigg(\prod_{i=1}^{A-1} m_i\bigg)^{1/4} 
        \bigg(\sum_{i=1}^{A-1}m_i\bigg)^{-\frac{3(A-2)+1}{2}}
        \bigg(\frac{\pi}{a_{A-1}^R+a_A^R/(\al X_{sp})}\bigg)^{\frac{A-2}{2}} \notag \\
    \times \exp \bigg[ -\bigg(a_{A-1}^R +\frac{a_A^R}{\al X_{sp}}\bigg) 
        \bigg(\sum_{i=1}^{A-1} m_i^2\bigg)^2\bigg]. \label{4.39}
\end{gather}
One can estimate the first order correction to the value of integral \eqref{4.38}. With account of this correction the integral over the variables $x_i^{(A-1)}$ takes the following form
\begin{equation}\label{4.40}
    J(X_{sp})=J_0(X_{sp}) K(X_{sp}),
\end{equation}
where 
\begin{gather}
    K(X_{sp})=1-\frac{3}{16}\,\bigg(\sum_{i=1}^{A-1} m_i\bigg)^{-2} 
        \bigg(a_{A-1}^R +\frac{a_A^R}{\al X_{sp}}\bigg)^{-1} \notag \\
    \times \bigg[ 9\bigg(\sum_{i=1}^{A-1} m_i\bigg)
        \bigg(\sum_{i=1}^{A-1} \frac{1}{m_i}\bigg)-
        3(A-1)^2-14(A-1)+8\bigg]. \label{4.41}
\end{gather}

Substituting into Eq. \eqref{4.37} the expressions for the normalization factors $C_A$ and $C_{A-1}$ from \eqref{4.30} and the expression for the integral $J(X_{sp})$, we obtain finally for the overlap integral
\begin{gather}
    I(X_{sp},\vec{P}_{sp,\bot})=2(4\pi)^{\frac{2A-3}{2}} 
        \bigg(\frac{2a_A^R}{\pi}\bigg)^{\frac{3(A-1)}{2}}
        \bigg(\frac{2a_{A-1}^R}{\pi}\bigg)^{\frac{3(A-2)}{2}} \notag \\
    \times \bigg(\sum_{i=1}^{A} m_i\bigg)^{\frac{3(A-1)+1}{4}}
        \bigg(\sum_{i=1}^{A-1} m_i\bigg)^{-\frac{3(A-2)+1}{4}} 
        \bigg(\prod_{i=1}^{A} m_i\bigg)^{-\frac{1}{4}}
        \bigg(\prod_{i=1}^{A-1} m_i\bigg)^{\frac{1}{4}} \notag \\
    \times \bigg(\frac{\pi}{a_{A-1}^R+a_A^R/(\al X_{sp})}\bigg)^{\frac{3(A-2)}{2}}
        \exp \bigg[ -\frac{a_A^R \vec{P}{}_{sp,\bot}^2}{\al X_{sp}(1-\al X_{sp})}\bigg] \notag \\
    \times \exp \bigg\{ -a_A^R \frac{\Big[\Big(\sum\limits_{i=1}^{A-1} m_i\Big) \al X_{sp}-
        \Big(\sum\limits_{i=1}^{A-1} m_i\Big)\Big]^2}{\al X_{sp}(1-\al X_{sp})}\bigg\}K(X_{sp}). 
        \label{4.42}
\end{gather}

It can be seen from the overlap integral \eqref{4.42} that the distribution with respect to the transverse momentum $\vec{P}_{sp,\bot}$ of the spectator fragment must have a Gaussian form and $X_{sp}$-distribution will have a maximum at the point predicted by Eq. \eqref{4.28}.

Now Eq. \eqref{4.18} for the differential cross section of the nucleon knock-out from the relativistic nucleus with the overlap integral \eqref{4.42} can be used for a comparison with experimental data. 

\section{Scaling Properties of the Relativistic Wave Functions of ${}^4He$ and ${}^3He$ Nuclei}

To verify the scaling properties of the relativistic wave functions and to extract the information about the values of their parameters, we used the experimental data on ${}^4He\,p$ interaction at 8,56~GeV/c and 13,5~GeV/c momenta of incident ${}^4He$ nucleus and ${}^3He\,p$ interaction at 13,5~GEV/c momentum of incident ${}^3He$ nucleus. These data were obtained in the 100-cm hydrogen bubble chamber \cite{168}, \cite{169} of the JINR at Dubna. We also used the data on the momentum distribution of the ${}^3He$ nuclei emitted at angle $0^\circ 65'$ in the ${}^4He\,p \to {}^3 He\,X$ reaction with an incident ${}^4He$ nucleus having momentum 6,85~GeV/c \cite{170}. 

The experimental $d\sg/d X_{sp}$ and $d\sg/d P_{sp,\bot}$ distributions of the ${}^3H$ and ${}^3He$ spectator fragments in the ${}^4He\,p \to {}^3 H\,p\,p$ and ${}^4He\,p \to {}^3He\,p\,n$ reactions were analyzed in the rest frame of the target proton. (The spectator nuclei ${}^3H$ and ${}^3He$ were defined as the fragments having the smallest momentum among the reaction products in the ${}^4He$ rest frame.) They are related to the invariant differential cross section \eqref{4.18} as follows
\addtocounter{equation}{+1}
\begin{align}
    \frac{d\sg}{d X_{sp}}  & = \int_0^{P_{sp,\bot_{\max}}} 
        \frac{d\sg}{d X_{sp}\,d P_{sp,\bot}}\,d P_{sp,\bot}, \tag{\theequation a} \label{4.43a} \\
    \frac{d\sg}{d P_{sp,\bot}}  & = \int_{X_{sp_{\min}}}^{X_{sp_{\max}}} 
        \frac{d\sg}{d X_{sp}\,d P_{sp,\bot}}\,d X_{sp}, \tag{\theequation b} \label{4.43b} \\
    \frac{d\sg}{d X_{sp}\, dP_{sp,\bot}}  & = 2\pi\bigg(\frac{P_{sp,\bot}}{X_{sp}}\bigg)\, 
        \frac{d\sg}{d \vec{P}_{sp}/E_{sp}}\,. \tag{\theequation c} \label{4.43c} 
\end{align}

The limits of integration $X_{sp_{\min}}$ and $X_{sp_{\max}}$ were taken from the corresponding experimental $X_{sp}$-distributions of the spectator fragments. The upper limit $P_{sp,\bot_{\max}}$ is the kinematical limit determined by the condition of the positivity of the factor $\lb(s_{NN},m^2,m^2)$ in Eq. \eqref{4.18}:
\begin{equation}\label{4.44}
    (P_{sp,\bot_{\max}})^2=(sX_{sp}-M_{sp}^2)^2(1-X_{sp})-4m^2X_{sp}\,.
\end{equation}

The total elastic cross section $\sg_{NN}^{el}$ can be assumed to be effectively constant in the considered range of energies and approximately equal to 24~mb \cite{159}.

The overlap integral \eqref{4.42} depends weakly on the values of the parameter of the spectator nucleus. Therefore, when the parameters of the relativistic wave functions of the final nuclei ${}^3H$ and ${}^3He$ are determined by fitting the experimental data, there are large errors. For this reason, the fitting was done for a fixed value of the parameter of the wave function of the three-nucleon nucleus. 

For the determination of the parameter $a_3^R$ we used relation \eqref{4.35} between the parameters of relativistic and nonrelativistic wave functions. It should be noted that the parameters of the nonrelativistic Gaussian wave functions of ${}^4He$, ${}^3He$ and ${}^3H$ nuclei are not determined well enough and we do not have reliable information about them. The values of the parameters $a_4^{NR}$ and $a_3^{NR}$, given in the literature, vary in rather wide range (see, g.e., Refs. \cite{172}--\cite{175}). The greatest number of data are available for the parameters of the nonrelativistic ${}^4He$ wave function, but the values of the parameter $a_4^{NR}$ of the Gaussian parametrization \eqref{4.31} vary in the range $a_4^{NR}=20-28({\rm GeV/c})^{-2}$. 

\begin{figure}
\begin{center}
\includegraphics*{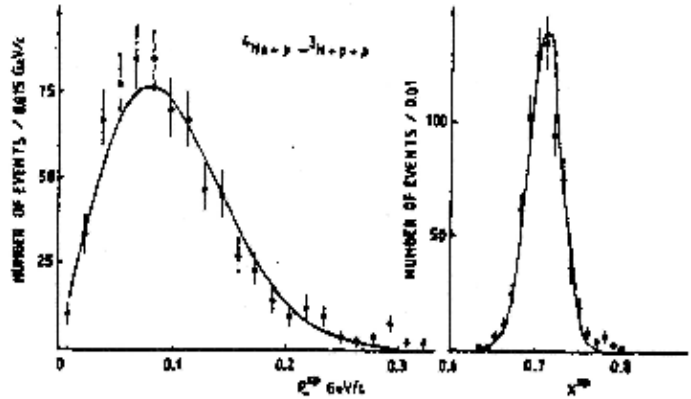}
\end{center}
\caption{The $P_{sp,\bot}$ and $X_{sp}$ distributions of the spectator fragment ${}^3H$ in the ${}^4He\,p\to {}^3H\,p\,p$ reaction at ${}^4He$ momentum 8,56~GeV/c.}
\label{fig18}
\end{figure}

\begin{figure}
\begin{center}
\includegraphics*{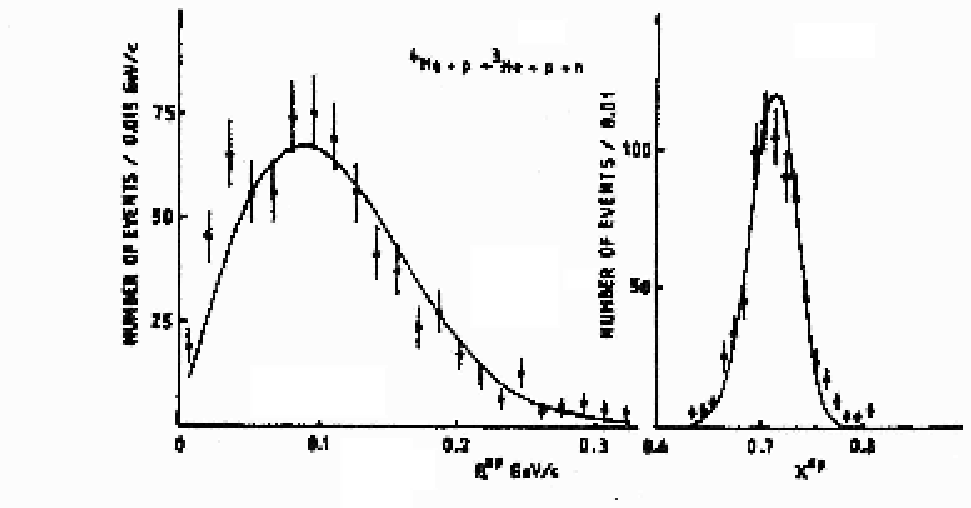}
\end{center}
\caption{The $P_{sp,\bot}$ and $X_{sp}$ distributions of the spectator fragment ${}^3He$ in the ${}^4He\,p\to {}^3He\,p\,p$ reaction at ${}^4He$ momentum 8,56~GeV/c.}
\label{fig19}
\end{figure}

\begin{figure}
\begin{center}
\includegraphics*{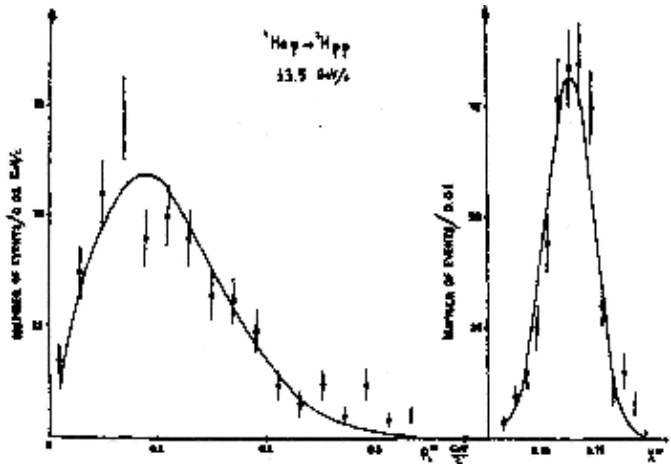}
\end{center}
\caption{The $P_{sp,\bot}$ and $X_{sp}$ distributions of the spectator fragment ${}^3H$ in the ${}^4He\,p\to {}^3H\,p\,p$ reaction at ${}^4He$ momentum 13,5~GeV/c.}
\label{fig20}
\end{figure}

\begin{figure}
\begin{center}
\includegraphics*{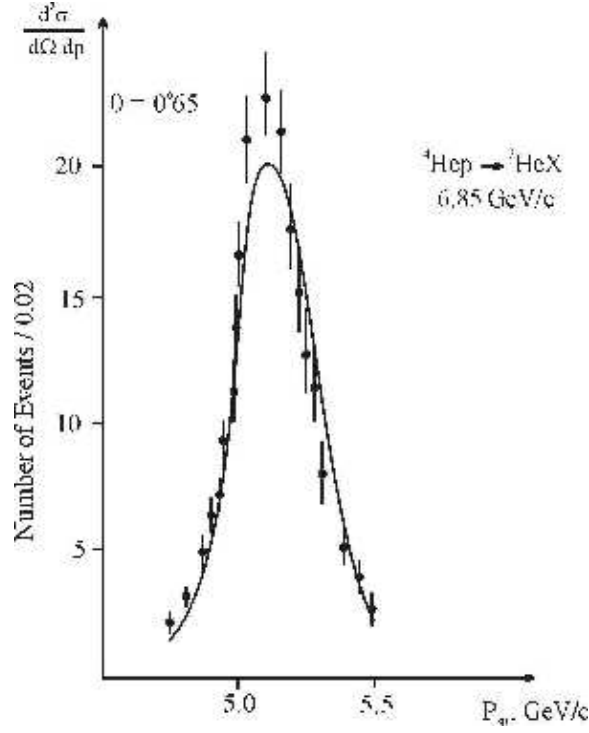}
\end{center}
\caption{Momentum Spectrum of ${}^3H$ fragments emitted at angle $0^\circ65'$ in the ${}^4He\,p\to {}^3He\,X$ reaction at ${}^4He$ momentum 6,85~GeV/c.}
\label{fig21}
\end{figure}

The experimental and theoretical $P_{sp,\bot}$ and $X_{sp}$-distributions of the spectator nucleus ${}^3H$ in the ${}^4He\,p \to {}^3H\,p\,p$ reaction at 8,56~GeV/c momentum of the incident ${}^4He$ are given in Fig. \ref{fig18}. The same distributions of the ${}^3He$ spectator fragments in the ${}^4He\,p\to {}^3He\,p\,n$ reaction at the same momentum are presented in Fig. \ref{fig19}. The $P_{sp,\bot}$ and $X_{sp}$ distributions of ${}^3H$ fragments in ${}^4He\,p \to {}^3H\,p\,p$ reaction at 13,5GeV/c momentum of incident ${}^4He$ are given in Fig. \ref{fig20} and momentum spectrum of ${}^3He$ fragments scattered on $0^\circ 65'$ in the laboratory frame in the ${}^4He\,p \to {}^3 He\,X$ reaction at 6,85~GeV/c momentum of incident ${}^4He$ nucleus is shown in Fig. \ref{fig21}. The solid curves in Figs. \ref{fig18}--\ref{fig21} correspond to the theoretical calculations with values of the parameters obtained by fitting the experimental data and presented in Table \ref{t2}. As can be seen from this Table the values of parameter $a_4^R$ of the relativistic wave function obtained by fitting the data at different momenta of the incident nucleus are close to each other and agree satisfactorily with the values $a_4^R \approx 5-7$~(GeV/c)$^2$. predicted by Eq. \eqref{4.35}.

The positions of the maxima in the $X_{sp}$ of spectator fragments agree well with the values of $\wt X_{sp}$ predicted by Eq. \eqref{4.28}.

The approximate energy-independence of the values of the parameters of the relativistic wave function in the considered region of energies of the incident ${}^4He$ nucleus allows to assume that in the wave function \eqref{4.26} there is no other dependence on energy apart from the dependence on the scale-invariant variables $x_i^{(A)}$.

\begin{table}
\begin{center}
Table 2. The Parameters of Relativistic Wave Functions of Three and Four-Nucleon Nuclei \\[2mm]
\begin{tabular}{lccc} \hline
Reaction, & $\al_3^R$, (GeV/c)$^{-2}$ & $a_4^R$, (GeV/c)$^{-2}$ & $\chi^2/N_p$\\
momentum  \\[1mm] \hline
&&& $d\sg/d X_{sp}$ \quad $d\sg/d P_{sp,\bot}$ \\
${}^4He\,p \to {}^3H\,p\,p$  \\
$P_{He}=8,56$ GeV/c & 8 - fixed & $7,39\pm 0,26$ & 31,13/18 \quad 27,65/22 \\[1mm]\hline
${}^4He\,p \to {}^3He\,p\,p$  \\
$P_{He}=8,56$ GeV/c & 8 - fixed & $5,86\pm 0,21$ & 54,63/18 \quad 43,15/22 \\[1mm]\hline
${}^4He\,p \to {}^3H\,p\,p$  \\
$P_{He}=13,5$ GeV/c & 8 - fixed & $6,23\pm 0,25$ & 29,01/14 \quad 37,14/17 \\[1mm]\hline
${}^4He\,p \to {}^3He\,X$  \\
$P_{He}=6,85$ GeV/c & 8 - fixed & $7,21\pm 0,34$ & 34,93/20 \\[1mm]\hline
\end{tabular}
\end{center}
\label{t2}
\end{table}

For the experimental determination of the parameter $a_3^R$ of the relativistic wave function of the three-nucleon nucleus, we consider the ${}^3He\,p \to D\,p\,p$ reaction. In this case, the overlap integral has the following form
\begin{gather}
    I(X_{sp},\vec{P}_{sp,\bot}) =4\pi c_D 3^{7/4} m(a_3^R)^{3/2} \notag \\
    \times \exp \bigg[ -\frac{a_3^R \vec{P}{}_{sp,\bot}^2}{\al X_{sp}(1-\al X_{sp})}\bigg]
        \exp \bigg[ -a_3^R\frac{m^2(3\al X_{sp}-2)^2}{\al X_{sp}(1-\al X_{sp})}\bigg]\notag \\
    \times \sum_{i=1}^5 A_i^R \exp (-4m^2\al_i^R)
        \bigg( \al_i^R+\frac{a_3^R}{\al X_{sp}}\bigg)^{-3/2}. \label{4.45}
\end{gather}
Here $C_D$ is a normalization factor
\begin{gather*}
    C_D=2^{5/2} \pi^{-1/4} m^{1/2} 
        \bigg\{ \sum_{i=1}^5 \sum_{j=1}^5 \frac{A_i^R A_j^R}{(\al_i^R+\al_j^R)^{3/2}} 
    \,\exp [-4m^2(\al_i^R+\al_j^R)]\bigg\}^{-1/2} 
\end{gather*}
and $A_i^R$ and $\al_i^R$ are the parameters of the relativistic wave function of deuteron:
\begin{equation}\label{4.46}
    \Phi_D^R(x,\vec{p}_\bot)=C_D\sum_{i=1}^5 A_i^R 
        \exp \bigg[ -\al_i^R \,\frac{\vec{p}{\,}_\bot^2+m^2}{x(1-x)}\bigg]
\end{equation}
which we chose in the form of the relativistic analog of the Garthenhaus--Moravcsik wave function
\begin{equation}\label{4.47}
    \Phi_D^{NR}(\vec{p}) =C_D^{NR} \sum_{i=1}^5 A_i^{NR} \exp (-\al_i^{NR}\vec{p}{\,}^2).
\end{equation}

The parameters $A_i^R$ and $\al_i^R$ are related to the parameters of the nonrelativistic wave function \eqref{4.47} in the following way
\begin{equation}\label{4.48}
    A_i^R=A_i^{NR} \exp(m^2 \al_i^{NR}), \quad \al_I^R=\frac{1}{4}\, \al_i^{NR}.
\end{equation}

When comparing the theoretical calculations with the experimental data on the distribution of the spectator deuterons in the reaction ${}^3He\,p\to D\,p\,p$, the parameters of the deuteron wave function \eqref{4.46} were fixed by the values obtained in accordance with the expression \eqref{4.48} from the values of the parameters of the nonrelativistic wave function \eqref{4.47} given in Ref. \cite{176}. 

\begin{figure}
\begin{center}
\includegraphics*{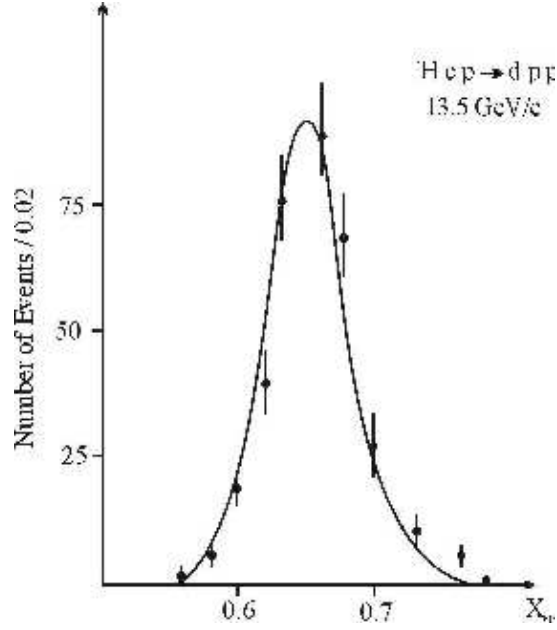}
\end{center}
\caption{The $X_{sp}$ distribution of spectator deuterons on the reaction ${}^3He\,p\to D\,p\,p$ at ${}^3He$ momentum 13,5~GeV/c.}
\label{fig22}
\end{figure}

In Fig. \ref{fig22} the $X_{sp}$-distribution of spectator deuterons in the ${}^3He\,p \to D\,p\,p$ reaction at 13,5~GeV/c momentum of incident ${}^3He$ is presented. 

\noindent The theoretical curve corresponds to the calculation with the overlap integral \eqref{4.45} with the value of parameter $a_3^R=8,28 \pm 0,71$~(GeV/c)$^{-2}$, obtained by fitting the experimental data. This value of the parameter $a_3^R$ agrees well with the value $a_3^R=6$~(GeV/c)$^{-2}$, predicted by Eq. \eqref{4.35} and used by us in previous calculations. 

It is interesting to compare the theoretical calculations on the distributions of the parameter spectator deuterons with experimental data at different energies of incident ${}^3He$ nucleus in order to check more precisely the scaling properties of the relativistic wave function of ${}^3He$ nucleus. It is also of interest to consider wave functions of a more complicated form. However, our simplest parametrizations of the relativistic nuclear wave functions make it possible to establish a number of features (such as scale-invariant character of ``longitudinal'' motion of nucleons inside the nucleus) that do not depend on their specific parametrization. 

%% file: garse.bbl
\begin{thebibliography}{9999}

\bibitem{1}  H. A. Bethe and E. E. Salpeter,  Phys. Rev. \textbf{84}, 1232  (1951).
\bibitem{2}  A. A. Logunov and  A. N. Tavkhelidze, Nuovo Cimento \textbf{29}, 380  (1963). 
\bibitem{3}  V. G. Kadyshevsky, Nucl. Phys. B \textbf{6}, 125 (1968).
\bibitem{4}  V. R. Garsevanishvili,  A. N. Kvinikhidze,  V. A. Matveev, A. N. Tavkhelidze, and R. N. Faustov,  Sov. J. Theor. Mat. Phys. \textbf{23}, 310  (1975).
\bibitem{5}  V. R. Garsevanishvili and V. A. Matveev, Sov. J. Theor. Mat. Phys. \textbf{24}, 3  (1975). 
\bibitem{6}  A. A. Khelashvili, Preprint JINR P2-8750, Dubna (1975).
\bibitem{7}  V. R. Garsevanishvili, Lectures at the 13th Intern. School of Theoretical Physics, Karpacz, Poland (1976). 
\bibitem{8}  M. I. Strikman and L. L. Frankfurt, Phys. Rep. C \textbf{76}, 215 (1981).
\bibitem{9}  V. R. Garsevanishvili, Z. R. Menteshashvili, D. G. Mirianashvili and M. S. Nioradze, Sov. J. Part. Nucl. \textbf{15}, 497 (1984).
\bibitem{10}  V. A. Karmanov, Sov. J. Part. Nucl. \textbf{19}, 525 (1988).
\bibitem{11}  A. M. Baldin, S. B. Gerasimov et al. In: Lectures at the School on High Energy Physics, Sukhumi, USSR, 1972. R2-6867, JINR, Dubna 374 (1972). 
\bibitem{12}  V. P. Alekseev, A. M. Baldin et al. Preprint JINR 9-7148, Dubna (1973).
\bibitem{13}  H. Steiner, Talk presented at the Intern. Conf. on High Energy Collisions Involving Nuclei, Trieste, Italy, 1974. Prepring LBL-3613, Berkeley (1974).
\bibitem{14}  A. M. Baldin, Talk presented at the 6th Intern. Conf. on High Energy Physics and Nuclear Structure, Fanta-Fe, USA, 1975. Preprint JINR E2-9138, Dubna (1975).
\bibitem{15}  G. A. Leksin, Lectures at the School on Some Problems of Experimental Investigations in High Energy Physics, Moscow Engineering Physics Institute (1975).
\bibitem{16}  P. Zielinski, In: Proc. of the 18th Intern. Conf. on High Energy Physics, Tbilisi, USSR, 1976. JINR D1,2-10400, Dubna (1976).
\bibitem{17}  A. M. Baldin, Sov. J. Part. Nucl. \textbf{8}, 175 (1977).
\bibitem{18}  H. Steiner, In: Proc. of the 7th Intern. Conf. on High Energy Physics and Nuclear Structure, Zurich, Switzerland, 1977. Ed. M. Locher (1977).
\bibitem{19}  A. S. Goldhaber and H. H. Heckman, Ann. Rev. Nucl. Part. Sci. \textbf{28}, 161 (1978).
\bibitem{20}  V. A. Karmanov and I. S. Shapiro, Sov. J. Part. Nucl. \textbf{9}, 134 (1978).
\bibitem{21}  A. M. Baldin, In: Fundamental Problems of Theoretical and Mathematical Physics. JINR D-12831, Dubna (1979).
\bibitem{22}  V. K. Luk'yanov and A. I. Titov, Sov. J. Part. Nucl. \textbf{10}, 321 (1979).
\bibitem{23}  V. S. Stavinskii, Sov. J. Part. Nucl. \textbf{10}, 373 (1979).
\bibitem{24}  A. M. Baldin, Talk at the Intern. Conf. on Extreme States in Nuclear Systems, Dresden, GDR, 1980. Preprint JINR E1-80-174, Dubna (1980).
\bibitem{25}  A. M. Baldin, In: Proc. of the 1981 CERN-JINR School of Physics, Hanko, Finland, 1981. CERN 82-04, Geneva (1982).
\bibitem{26}  E. M. Friedlander and H. H. Heckman, Preprint LBL-13864, Berkeley (1982).
\bibitem{27}  A. V. Efremov, Sov. J. Part. Nucl. \textbf{13}, 254 (1982).
\bibitem{28}  J. Cleymans, M. Dechantreiter and F. Halzen, Univ. of Wisconsin preprint MAD/TH/50, Madison (1982).
\bibitem{29}  V. K. Luk'yanov and A. I. Titov, In: Proc. of the 15th Intern. School on High Energy Physics, Dubna, USSR, 1982. JINR D2,4-83-179, Dubna (1983).
\bibitem{30}  A. P. Kobushkin and V. P. Shelest, Sov. J. Part. Nucl. \textbf{14}, 483 (1983).
\bibitem{31}  L. Van Hove, Lecture at the Intern. Conf. on Nuclear Physics, Florence, Italy, 1983. Preprint CERN TH-3725, Geneva (1983).
\bibitem{32}  V. R. Garsevanishvili, In: Proc. of the 9th European Conf. on Few Body Problems in Physics, Tbilisi, USSR, 1984. Eds. L.D. Faddeev and T.I. Kopaleishvili, World Scientific, Singapore (1984).
\bibitem{33}  M. A. Faessler, Phys. Rep. \textbf{115}, 1 (1984).
\bibitem{34}  M. I. Strikman and L. L. Frankfurt, Phys. Rep. \textbf{160}, 235 (1988).
\bibitem{35}  L. Kluberg, In: Proc. of the 24th Intern. Conf. on High Energy Physics, Munich, Germany, 1988. Eds. R. Kotthaus and J.~H. Kuhn, Springer-Verlag, 250 (1988).
\bibitem{36}  G.~C. Mantovani et al. Phys. Lett. \textbf{64B}, 471 (1976).
\bibitem{37}  G.~C. Mantovani et al. Phys. Lett. \textbf{65B}, 401 (1976).
\bibitem{38}  G. Alberi and G. Goggi, Phys. Rep. \textbf{74}, 1 (1981).
\bibitem{39}  G. Jakob and Th. Maris, Rev. Mod. Phys. \textbf{45}, 6 (1973).
\bibitem{40}  C. Ciofi degli Atti, In: Photonuclear Reactions. Eds. S. Costa and C.Schaerf, Springer-Verlag, Berlin, 521 (1977). 
\bibitem{41}  H. Leutwyler and J. Stern, Ann. Phys. \textbf{112}, 94 (1978).
\bibitem{42}  S. J. Brodsky, Preprint SLAC-PUB-4342, Stanford (1987).
\bibitem{43}  V. R. Garsevanishvili and Z. R. Menteshashvili, ``Relativistic Nuclear Physics in the Ligh Front Formalism'', Nova Science Publishers, New York (1993).
\bibitem{44}  S. J. Brodsky, In: Proc. of the ELFE Summer School and Workshop on Confinement Physics, Cambridge, England 1995. Preprint SLAC-PUB-95-7056.
\bibitem{45}  J. Carbonel, B. Desplanques, V. A. Karmanov and J.-F. Mathiot, Phys. Rep. \textbf{300}, 215 (1998).
\bibitem{46}  S. J. Brodsky, H.-C. Pauli and S. Pinsky, Phys. Rep. \textbf{301}, 299 (1998).
\bibitem{49}  V. R. Garsevanishvili and A. N. Tavkhelidze, Phys. Part. Nucl. \textbf{30}, 258 (1999).
\bibitem{50}  G. A. Miller, Preprint NT-UW-00-04, Univ. of Washington, Seattle (2000).
\bibitem{51}  S. J. Brodsky, J. R. Hiller, D. S. Hwang and V. A. Karmanov, Phys. Rev. D \textbf{69}, 076001  (2004).
\bibitem{52}  S. J. Brodsky, Talk presented at the 58th Scottish Univ. Summer School in Physics, St. Andrews, Scotland, 2004. Preprint SLAC-PUB-10871 (2004).
\bibitem{53}  N. S. Amaglobeli et al, Eur. Phys. J. \textbf{C8}, 603 (1999).
\bibitem{54}  V. Garsevanishvili, T. Djobava, Yu. Tevzadze and L. Chkhaidze, Phys. Part. Nucl. \textbf{34}, 526 (2003).
\bibitem{55}  P.~A.~M. Dirac, Rev. Mod. Phys. \textbf{21}, 392 (1949).
\bibitem{56}  N. N. Bogolubov, D. V. Shirkov, Introduction to the Theory of Quantized Fields, 3rd Ed. Wiley, New York (1980),
\bibitem{57}  T. Suzuki et al. Progr. Theor. Phys. \textbf{56}, 922 (1976).
\bibitem{58}  T. Maskawa and K. Yamawaki, Progr. Theor. Phys. \textbf{56}, 270 (1976).
\bibitem{59}  M. Ida, Lett. Niovo Cim. \textbf{15}, 249 (1976).
\bibitem{60}  S.J. Chang, R. Root and T.-M. Yan, Phys. Rev. \textbf{D7}, 1133 (1973).
\bibitem{61}  S.J. Chang and T.-M. Yan, Phys. Rev. \textbf{D7}, 1167 (1973).
\bibitem{62}  T.-M. Yan, Phys. Rev. \textbf{D7}, 1760 (1973).
\bibitem{63}  T.-M. Yan, Phys. Rev. \textbf{D7}, 1780 (1973).
\bibitem{64}  J. H. Ten Eyck and F. Rohrlich, Phys. Rev. \textbf{D9}, 2237 (1974).
\bibitem{65}  N. M. Atakishiev, R. M. Mir-Kasimov, Sh. M. Nagiev, Theor. Mat. Phys. \textbf{32}, 34 (1977).
\bibitem{66}  L. Susskind, Phys. Rev. \textbf{165}, 1535 (1968).
\bibitem{67}  J. B. Kogut and D. E. Soper, Phys. Rev. \textbf{D1}, 2901 (1970).
\bibitem{68}  J. B. Kogut and L. Susskind, Phys. Rep. \textbf{8}, 75 (1973).
\bibitem{69}  J. F. Gunion, S. J. Brodsky, R. Blankenbecler, Phys. Rev. \textbf{D8}, 287 (1973).
\bibitem{70}  S. Weinberg, Phys. Rev. \textbf{150}, 1313 (1966).
\bibitem{71}  A. A. Khelashvili, Preprint JINR P2-4327, Dubna (1969).
\bibitem{72}  A. A. Khelashvili, in: Proc. of the 18th Intern. Conf. on High Energy Physics, Tbilisi, USSR, 1976. JINR D1,2-10400, Dubna, C111 (1977). 
\bibitem{73}  A. A. Khelashvili, A. N. Kvinikhidze, V. A. Matveev, A. N. Tavkhelidze, Theor. Mat. Phys. \textbf{29}, 3 (1976).
\bibitem{74}  N. N. Singh, Y. K. Mathur, A. N. Mitra, Few-Body Systems \textbf{1}, 47 (1986).
\bibitem{75}  D. S. Kulshreshtha, A. N. Mitra, Phys. Rev. \textbf{D37}, 1268 (1988).
\bibitem{76}  B.~H.~J. McKellar, M.~D. Scadron, R.~C. Warner, Int. J. Mod. Phys. \textbf{A3}, 203 (1988).
\bibitem{77}  H. Leutwyler, Nucl. Phys. \textbf{76B}, 413 (1974).
\bibitem{78}  M. V. Terent'ev, Sov. J. Nucl. Phys. \textbf{24}, 207 (1976).
\bibitem{79}  V. B. Berestetskii, M. V. Terent'ev, Sov. J. Nucl. Phys. \textbf{24}, 547 (1976).
\bibitem{80}  B. L. Bakker, L. A. Kondratyuk, M. V. Terent'ev, Nucl. Phys. \textbf{B15}, 497 (1979).
\bibitem{81}  S. Mandelstam, Proc. Royal Soc. \textbf{233A}, 248 (1955).
\bibitem{82}  R. N. Faustov, Ann. Phys. \textbf{78}, 176 (1973).
\bibitem{83}  N. N. Bogolubov, V. A. Matveev, A. N. Tavkhelidze, In: Proc. of the Intern. Seminar on Elementary Particle Physics, Varna, Bulgaria, 1968. JINR P2-4050, Dubna, 268 (1968). 
\bibitem{84}  C. Alabiso, G. Schierholz, Phys. Rev. \textbf{D10}, 960 (1974).
\bibitem{85}  R. N. Faustov, A. A. Khelashvili, Sov. J. Nucl. Phys. \textbf{10}, 619 (1970).
\bibitem{86}  V. A. Matveev, R. M. Muradyan, A. N. Tavkhelidze, Lett. Nuovo Cim. \textbf{7}, 719 (1973).
\bibitem{87}  S. J. Brodsky, G. R. Farrar, Phys. Rev. Lett. \textbf{31}, 1153 (1973).
\bibitem{88}  G.~H.~L. Smith, Ann. Phys. \textbf{53}, 521 (1969).
\bibitem{89}  V. R. Garsevanishvili, A. N. Kvinikhidze, V. A. Matveev, A. N. Tavkhelidze and R. N. Faustov, Sov. J. Theor. Mat. Phys. \textbf{25}, 37 (1975).
\bibitem{90}  R. P. Feynman, Phys. Rev. Lett. \textbf{23}, 1415 (1969).
\bibitem{91}  J. D. Bjorken, E. A. Paschos, Phys. Rev. \textbf{185}, 1975 (1969).
\bibitem{92}  J. D. Bjorken, E. A. Paschos, Phys. Rev. \textbf{D1}, 3151 (1970).
\bibitem{93}  R. P. Feynman, Photon-Hadron Interactions, W.~A.~Benjamin, Reading, Mass. (1972).
\bibitem{94}  A. Bodek et al, Phys. Rev. Lett. \textbf{30}, 1087 (1973).
\bibitem{95}  E. M. Riordan et al, Phys. Rev. Lett. \textbf{33}, 561 (1974).
\bibitem{96}  E. M. Riordan et al, Preprint SLAC-PUB-1634, Stanford (1975).
\bibitem{97}  A. Bodek et al, Phys. Rev. \textbf{D20}, 1471 (1979).
\bibitem{98}  B.A. Gordon et al, Phys. Rev. \textbf{D20}, 2645 (1979).
\bibitem{99}  O. Nachtman, Nucl. Phys. \textbf{B63}, 237 (1973).
\bibitem{100}  A. De Rujula, H. Georgi, H. D. Politzer, Phys. Rev. \textbf{D10}, 2141 (1974).
\bibitem{101}  H. Georgi, H. D. Politzer, Phys. Rev. \textbf{D14}, 1829 (1976).
\bibitem{102}  A. De Rujula, H. Georgi, H. D. Politzer, Ann. Phys. \textbf{103}, 315 (1977).
\bibitem{103}  A. V. Efremov, A. V. Radyushkin, In: Proc. of Intern. Seminar ``Multiple Production Processes and Inclusive Reactions at High Energies'', Serpukhov, USSR, 1976. Serpukhov,  252 (1977).
\bibitem{104}  N. N.Krasnikov, A. N. Tavkhelidze, K. G. Chetirkin, In: Proc. of the Intern. Conf. "Neutrino-77". Nauka Publ., Moscow, 189 (1978). 
\bibitem{105}  N. M. Atakishiev, R. M. Mir-Kasimov, Sh. M. Nagiev, Preprint JINR P2-80-635, Dubna (1980).
\bibitem{106}  A. M. Baldin, Short Communications in Physics \textbf{1}, 35 (1971).
\bibitem{107}  A. M. Baldin et al, Preprint JINR P1-5819, Dubna (1971).
\bibitem{108}  A. M. Baldin et al, Sov. J. Nucl. Phys. \textbf{18}, 41 (1973).
\bibitem{109}  A. M. Baldin et al, Sov. J. Nucl. Phys. \textbf{21}, 517 (1975).
\bibitem{110}  R. Arnold et al, Phys. Rev. Lett. \textbf{35}, 776 (1975).
\bibitem{111}  R. Arnold et al, Phys. Rev. Lett. \textbf{40}, 1429 (1978).
\bibitem{112}  J. J. Aubert et al, Phys. Lett. \textbf{123B}, 275 (1983).
\bibitem{113}  A. Bodek et al, Phys. Rev. Lett. \textbf{50}, 1431 (1983).
\bibitem{114}  A. Bodek et al, Phys. Rev. Lett. \textbf{51}, 534 (1983).
\bibitem{115}  R. G. Arnold et al, Phys. Rev. Lett. \textbf{52}, 727 (1984).
\bibitem{116}  G. Bari et al, Phys. Lett. \textbf{163B}, 282 (1985).
\bibitem{117}  C. H. Llewellyn Smith, Phys. Lett. \textbf{128B}, 107 (1983).
\bibitem{118}  M. Ericson, A.W. Thomas, Phys. Lett. \textbf{128B}, 112 (1983).
\bibitem{119}  F. E. Close, R. G. Roberts, G. G. Ross, Phys. Lett. \textbf{129B}, 346 (1983).
\bibitem{120}  C. E. Carlson, J. J. Havens, Phys. Rev. Lett. \textbf{51}, 261 (1983).
\bibitem{121}  S. Date, Progr. Theor. Phys. \textbf{70}, 1682 (1983).
\bibitem{122}  E. L. Berger, F. Coester, R. B. Wiringa, Phys. Rev, \textbf{D29}, 383 (1984).
\bibitem{123}  R. L. Jaffe, F.E. Close, R.G. Roberts, G.G. Ross, Phys. Lett. \textbf{134B}, 449 (1984).
\bibitem{124}  O. Nachtman, H. J. Pirner, Z. Phys. \textbf{C21}, 277 (1984).
\bibitem{125}  A. I. Titov, Sov. J. Nucl. Phys. \textbf{40}, 50 (1984).
\bibitem{126}  W. Furmanski, A. Krzywicki, Z. Phys. \textbf{C22}, 391 (1984).
\bibitem{127}  L. A. Kondratyuk, M. J. Shmatikov, JETP Lett. \textbf{39}, 389 (1984).
\bibitem{128}  J. P. Vary, Nucl. Phys. \textbf{A418}, 195c  (1984).
\bibitem{129}  V. R. Garsevanishvili, Z. R. Menteshashvili, JETP Lett. \textbf{40}, 359 (1984).
\bibitem{130}  S. V. Akulinichev, S. A. Kulagin, G. M. Vagradov, Phys. Lett. \textbf{152B}, 485 (1985).
\bibitem{131}  B. L. Birbrair, A. B. Gridnev, M. B. Zhalov, E. M. Levin, V. E. Starodubsky, Phys. Lett. \textbf{166B}, 119  (1986).
\bibitem{132}  V. V. Maryalov, Theory of Atomic Nuclei, Nauka Publ. Moscow (1967).
\bibitem{133}  S. Date, A. Nakamura, Progr. Theor. Phys. \textbf{69}, 565 (1983).
\bibitem{134}  V. A. Matveev, P. Sorba, Lett. Nuovo Cim. \textbf{20}, 443 (1977).
\bibitem{135}  P. Mulders, A. W. Thomas, Phys. Rev. Lett. \textbf{52}, 1199 (1984).
\bibitem{136}  A. M. Baldin, Preprint JINR E2-83-415, Dubna (1983). 
\bibitem{137}  I. A. Savin, In: Proc. of the 6th Intern. Conf. on High Energy Physics Problems, Dubna, USSR, 1981. JINR D1,2-81-728, Dubna, 223  (1981).   
\bibitem{138}  J. Ashman et al, Phys. Lett. \textbf{202B}, 603 (1988).
\bibitem{139}  M. Arneodo et al, Phys. Lett. \textbf{211B}, 493 (1988).
\bibitem{140}  R.~J.~M. Covolan, E. Predazzi, Preprint DFTT 21/90, Torino (1990).
\bibitem{141}  V. R. Garsevanishvili, D. G. Mirianashvili, M. S. Nioradze, Communications JINR P2-9859, Dubna (1976).
\bibitem{142}  V. R. Garsevanishvili, D. G. Mirianashvili, Rep. Math. Phys. \textbf{11}, 89 (1977).
\bibitem{143}  I. S. Shapiro, In: Proc. of the Intern. Conf. on Selected Problems of Nuclear Structure, Dubna, 1876. JINR D-9920, Dubna, 424 (1976). 
\bibitem{144}  V. V. Burov, V. K. Luk'yanov, A. I. Titov, in: Proc. of the Intern. Conf. on Selected Problems of Nuclear Structure, Dubna, 1976. JINR D-9920, Dubna, 432 (1976). 
\bibitem{145}  W. Buck, F. Gross, Phys. Lett. \textbf{63B}, 286 (1976).
\bibitem{146}  V. A. Karmanov, Sov. J. JETP \textbf{44}, 210 (1976).
\bibitem{147}  M. I. Strikman, L. L. Frankfurt, Sov. J. Nucl. Phys. \textbf{25}, 1177 (1977).
\bibitem{148}  I. Schmidt, R. Blankenbecler, Phys. Rev. \textbf{D15}, 3321 (1977).
\bibitem{149}  W. Buck, F. Gross, Phys. Rev. \textbf{D20}, 2361  (1979).
\bibitem{150}  B. S. Aladashvili, V. R. Garsevanishvili et al, Sov. J. Nucl. Phys. \textbf{33}, 681 (1981).
\bibitem{151}  J. Benecke, T. T. Chou, C. N. Yang, E. Yen, Phys. Rev. \textbf{188}, 2159 (1969).
\bibitem{152}  V. A. Matveev, R. M. Muradyan, A. N. Tavkhelidze, Lett. Nuovo Cim. \textbf{5}, 907 (1972).
\bibitem{153}  V. A. Matveev, R. M. Muradyan, A. N. Tavkhelidze, Sov. J. Theor. Mat. Phys. \textbf{15}, 332 (1973).
\bibitem{154}  V. R. Garsevanishvili, V. V. Glagolev, Z. R. Menteshashvili et al, Preprint JINR 1-81-838, Dubna (1981).
\bibitem{155}  B. S. Aladashvili et al, Nucl. Phys. \textbf{86B}, 461 (1975).
\bibitem{156}  B. S. Aladashvili et al, J. Phys. G: Nucl. Phys. \textbf{3}, 1225 (1976).
\bibitem{157}  B. S. Aladashvili et al, Sov. J. Nucl. Phys. \textbf{27}, 377 (1978).
\bibitem{158}  A. Fridman, Fortschr. Phys. \textbf{23}, 243 (1975).
\bibitem{159}  O. Benary et al, A Compilation, UCRL-20000NN, Berkeley (1970).
\bibitem{160}  V. Flaminio et al, A Compilation, CERN-HERA 79-03, Geneva (1979).
\bibitem{161}  A. Abdivaliev, K. Beshliu, V.R. Garsevanishvili et al, Sov. J. Nucl. Phys. \textbf{27}, 715 (1978).
\bibitem{162}  A. P. Gasparyan et al, Preprint JINR 1-9111, Dubna (1975).
\bibitem{163}  F. Kotorobai et al, Preprint JINR P10-9314, Dubna (1975).
\bibitem{164}  A. P. Ierusalimov et al, Preprint JINR P10-9502, Dubna (1976).
\bibitem{165}  V. R. Garsevanishvili, Z. R. Menteshashvili, D. G. Mirianashvili, M.S. Nioradze, Sov. J. Theor. Mat. Phys. \textbf{33}, 276 (1977).
\bibitem{166}  B. S. Aladashvili, V. R. Garsevanishvili, Z. R. Menteshashvili et al, Sov. J. Nucl. Phys. \textbf{34}, 591 (1981).
\bibitem{167}  V. R. Garsevanishvili, Z. R. Menteshashvili, D. G. Mirianashvili, M. S. Nioradze, Sov. J. Nucl. Phys. \textbf{41}, 587 (1985).
\bibitem{168}  V. V. Glagolev et al, Phys. Rev. \textbf{C18}, 1382 (1979).
\bibitem{169}  N. A. Buzdavina, V. V. Glagolev et al, Communications JINR 1-81-530, Dubna (1981).
\bibitem{170}  G. Bizard et al, Nucl. Phys. \textbf{A285}, 461 (1977).
\bibitem{171}  M. V. Fedoryuk, The Saddle Point Method, Nauka, Moscow (1977).
\bibitem{172}  M. M. Block, Nuovo Cim. \textbf{20}, 715 (1961).
\bibitem{173}  B. Z. Kopeliovich, I. K. Potashnikova, Sov. J. Nucl. Phys. \textbf{13}, 592 (1971).
\bibitem{174}  E. M. Levin, M. I. Strikman, Sov. J. Nucl. Phys. \textbf{23}, 216 (1976).
\bibitem{175}  J. P. Auger, J. Gillespie, R. J. Lombard, Nucl. Phys. \textbf{A262}, 372 (1976).
\bibitem{176}  G. Alberi, L. P. Rosa, Z. D. Thome, Phys. Rev. Lett. \textbf{34}, 503 (1975).

\end{thebibliography}
